\documentclass[aps,prd,preprint]{revtex4}
\usepackage{amsfonts}
\usepackage{amssymb}
\usepackage{amsmath}
\usepackage{graphicx}
\usepackage{graphics}
\usepackage{color}
\usepackage{epsfig}
\newcommand{\be}{\begin{equation}}
\newcommand{\ee}{\end{equation}}
\newcommand{\bea}{\begin{eqnarray}}
\newcommand{\eea}{\end{eqnarray}}
\newcommand{\bwt}{\begin{widetext}}
\newcommand{\ewt}{\end{widetext}}
\begin{document}
%%%%%%%%%%%%%%%%%
%%%%%%%%%%%%%%%%%
\title{Supersymmetric multi-Higgs doublet model with non-linear electroweak symmetry breaking}
\author{T.E. Clark}
\email[e-mail address:]{clarkt@purdue.edu}
\affiliation{Department of Physics,\\
 Purdue University,\\
 West Lafayette, IN 47907-2036, U.S.A.}
\author{S.T. Love}
\email[e-mail address:]{loves@purdue.edu}
\affiliation{Department of Physics,\\
 Purdue University,\\
 West Lafayette, IN 47907-2036, U.S.A.}
\author{T. ter Veldhuis}
\email[e-mail address:]{terveldhuis@macalester.edu}
\affiliation{Department of Physics \& Astronomy,\\
 Macalester College,\\
 Saint Paul, MN 55105-1899, U.S.A.}
\begin{abstract}
The electroweak symmetry is nonlinearly realized in an extension of the  minimal supersymmetric standard model (MSSM) through an additional pair of constrained Higgs doublet superfields.  The superpotential couplings of this constrained Higgs doublet pair to the MSSM Higgs doublet pair catalyze their vacuum expectation values.  The  Higgs and Higgsino-gaugino mass spectrum is presented for several choices of supersymmetry (SUSY) breaking and Higgs superpotential mass parameters.  The additional vacuum expectation values provided by the constrained fields can produce a phenomenology quite different than that of the MSSM .

\end{abstract}

\maketitle

\section{Introduction \label{intro}}

The increasing lower experimental bound on the Higgs boson mass has called into question the viability of the minimal supersymmetric standard model (MSSM) where the mass remains bounded from above by about 130 GeV even after the inclusion of radiative corrections. Augmenting the MSSM by the inclusion of an additional singlet superfield (the NMSSM) \cite{Fayet:1974pd} provides a means to raise the Higgs boson mass \cite{Espinosa:1991gr,Antoniadis:2010hs}.  Requiring the NMSSM to remain perturbative  up to the unification scale results in a Higgs mass limit of about 150 GeV \cite{Espinosa:1998re}, while permitting the singlet-Higgs doublet Yukawa coupling  to reach its Landau singularity before the unification scale allows the Higgs mass to be raised even further \cite{Barbieri:2011tw,Harnik:2003rs,Chang:2004db,Delgado:2005fq}. Taken to the extreme, the large mass limit is described by a nonlinear or chiral MSSM \cite{Clark:1993zz}. This particular nonlinear realization has been experimentally excluded by the chargino mass limits \cite{PDG}. Alternatively, a wider range of allowed tree level masses can also be achieved by the addition of families of Higgs doublets. In this case, the major model restrictions arise from  the need to suppress excessive flavor changing neutral currents (FCNC). This leads to model restrictions on the Yukawa couplings to matter superfields. The requisite safe conditions needed for the sufficient suppression of the FCNC, as well as for agreement with precision electroweak tests and anomalous magnetic moment measurements, all with perturbative Yukawa couplings,  have been extensively studied  
\cite {Skiba:2010xn,He:2001tp,Bagger:1999te,Marshall:2010qi,Gupta:2009wn,Carone:2010cp,Choi:2010an} in such extensions of the standard model and the MSSM.

The motivation for introducing additional Higgs doublet fields goes beyond the desire to alter tree level mass spectra. For example, it could be that some novel strong gauge field dynamics may be the source of the electroweak symmetry breakdown (and possibly even the supersymmetry breaking) \cite{Murayama:2003ag,Choi:1999yaa,Luty:2000fj,Carone:2006wj}, but this dynamics is not directly responsible for giving the quarks and leptons their nontrivial masses. A model independent means of characterizing the electroweak symmetry breakdown is via a nonlinear realization of the $SU(2)_L\times U(1)$. For a consistent SUSY model, this can be achieved using a constrained pair of Higgs doublet fields, where the imposition of the constraint breaks the electroweak symmetry. On the other hand, the quark and lepton superfields acquire their masses through their Yukawa coupling to an additional pair of MSSM-like Higgs doublets whose nontrivial vacuum expectation values are catalyzed by their supersymmetric coupling to the constrained Higgs doublet pairs. Thus a consistent supersymmetric version of such a picture  requires the introduction of four pairs of doublets with the additional nonlinear constraint among two of the Higgs doublet chiral superfields.  Note that in such a model, the electroweak symmetry breaking is no longer tied to the supersymmetry breaking as is the case in the MSSM. 

In this paper, we focus on such a supersymmetric model where the source for electroweak symmetry breakdown is independent of the SUSY breaking.  This is accomplished through a nonlinear realization of the $SU(2)_L\times U(1)$ symmetry.  In addition, the coupling of this sector to that of the usual MSSM, including the 
soft SUSY breaking terms, provides a rich spectrum of particle masses.  The simplest realization of the model can be expressed in terms of an additional pair of constrained doublet chiral superfields denoted $H_u^\prime$ and $H_d^\prime$ having the form
\be
H_u^\prime = \begin{pmatrix}
H_u^{+\prime}\\
H_u^{0\prime}
\end{pmatrix}=\begin{pmatrix}
i\Pi^+\\
\Sigma -i\Pi^0
\end{pmatrix} \qquad , \qquad
H_d^\prime = \begin{pmatrix}
H_d^{0\prime}\\
H_d^{-\prime}
\end{pmatrix}=\begin{pmatrix}
\Sigma+i\Pi^0\\
i\Pi^-
\end{pmatrix} ,
\label{ConstrainedDoublets}
\ee
with the vacuum expectation values
\be
<0|H_u^\prime|0> =\begin{pmatrix}
0\\
v_u^\prime /\sqrt{2}
\end{pmatrix}\qquad , \qquad
<0|H_d^\prime|0> =\begin{pmatrix}
v_d^\prime /\sqrt{2}\\ 
0
\end{pmatrix} .
\ee
These $\sigma$-model coordinates are given by the chiral superfields
$\Pi^\pm \equiv \Pi^1 \mp i \Pi^2$ and $\Pi^0 =\Pi^3$ while the superfield constraint, $H_d^\prime \epsilon H_u^\prime = v_u^\prime v_d^\prime / 2$, takes the form
\be
\Sigma = \sqrt{\frac{v_u^\prime v_d^\prime}{2}-\vec{\Pi} \cdot \vec{\Pi}~}~~ .
\ee
which allows the $\Sigma$ superfield to be eliminated in favor of the $\vec{\Pi}$ superfields.
The model action $\Gamma$ is thus given by
\be
\Gamma = \Gamma_{\rm MSSM} + \int dV \left\{ \bar{H}_u^{\prime} e^{-2g_2 A -g_1 B} H_u^\prime + 
\bar{H}_d^{\prime} e^{-2g_2 A +g_1 B} H_d^\prime \right\} +  \int dS W_{\rm Mix} + \int d \bar{S} \bar{W}_{\rm Mix} ,
\label{Action}
\ee
where $\Gamma_{\rm MSSM}$ is the action for the MSSM including soft SUSY breaking.  The electroweak gauge fields are the $SU(2)_L$ vector superfield $W=\frac{\vec{\sigma}}{2}\cdot \vec{W}$ and the $U(1)$ weak hypercharge vector superfield $B$.  The superpotential $W_{\rm Mix}$ involves the mixing of the MSSM Higgs doublets, denoted by $H_u$ and $H_d$, with the constrained coordinates $H_u^\prime$ and $H_d^\prime$
\be
W_{\rm Mix}= \mu_{12} H_u \epsilon H_d^\prime + \mu_{21} H_u^\prime \epsilon H_d  .
\ee
Note that even though the $\Sigma$ superfield is constrained, the theory remains anomaly free after its elimination.  The linear part of the $\Pi^i$-inos coupling to the $SU(2)_L$ gauge fields is in the adjoint representation and only the $\pi^\pm$-inos have a linear coupling to the $U(1)$ hypercharge gauge field.  Hence their potential contributions to the anomalies vanish.

In the MSSM, the electroweak symmetry breakdown is tied to the SUSY breaking so that without SUSY breaking there is no electroweak breaking. On the other hand, the multi-doublet sigma model can be realized in the broken electroweak symmetry phase even if SUSY remains unbroken.  In this unbroken SUSY limit, and with the global custodial $SU(2)_V$ symmetry broken only by gauging the $U(1)$ hypercharge,  the model parameters simplify to 
$v_u^\prime =v_d^\prime \equiv v^\prime$ while $\tan \beta =1$ ($v_u =v_d$) and $\mu_{12}=\mu_{21}$. Parametrizing the MSSM Higgs field doublets as 
\be
H_u = \begin{pmatrix}
H_u^+\\
H_u^0
\end{pmatrix}=\begin{pmatrix}
i\chi^+\\
H^0 -i\chi^0
\end{pmatrix} \qquad , \qquad
H_d = \begin{pmatrix}
H_d^0\\
H_d^-
\end{pmatrix}=\begin{pmatrix}
H^0+i\chi^0\\
i\chi^-
\end{pmatrix} ,
\label{MSSMDoublets}
\ee
with general vacuum expectation values $<0|H_u^0 |0> = v_u /\sqrt{2}$ and $<0|H_d^0 |0> = v_d /\sqrt{2}$,
the massless Nambu-Goldstone bosons lie in an $SU(2)_V$ triplet
\be
\vec\Pi_{\rm NG} = \vec\Pi \cos{\theta} + \vec\chi \sin{\theta},
\ee
while one of the neutral and the two charged massive Higgs chiral superfields together lie in the orthogonal $SU(2)_V$ triplet
\be
\vec{H} = -\vec\Pi \sin{\theta} + \vec\chi \cos{\theta},
\ee
with the other neutral Higgs  chiral superfield being the $SU(2)_V$ singlet $H^0$. The potential is minimized at $\mu_{12}=-\mu_{11}\tan\theta$. 
 The SUSY Higgs mechanism becomes operational with the $Z$ and $W^\pm$ vector superfields absorbing the neutral and charged Nambu-Goldstone chiral superfields to become massive with $M_Z^2 =g^2 (v_u^2 + w^2 )/2$ and $M_W^2 = M_Z^2 \cos{\theta_W}$, while  the photon vector superfield (photon and photino) remains massless. There are four additional Higgs superfields; two neutral and two charged.  The neutral chiral superfields have masses $4\mu_{11}$ and $4\mu_{11} \sec^2{\theta}$ while the charged $SU(2)_V$ partner chiral superfields have masses $4\mu_{11} \sec^2{\theta}$.  
When the SUSY breaking parameters are included and the mixing masses are chosen to be different for up and down Higgs fields, the mixing involved in forming the mass eigenstates becomes quite complicated and necessitates a numerical determination.  All told, there are two neutral pseudoscalars, three neutral Higgs scalars and three charged scalars.  In addition, the gaugino and Higgsino fields mix to yield three charginos and five neutralinos.

In section \ref{section2}, the model is expressed in terms of its component fields with the auxiliary $F-$ and $D-$ fields eliminated.  The electroweak breaking minimum of the potential is found.  The mass spectrum is extracted in section \ref{section3} for various choices of the parameters of the model.  For simplicity, the nonlinear realization of the electroweak symmetry has been taken to exhibit the custodial $SU(2)_V$ global symmetry, hence the corresponding vacuum values are chosen to satisfy: $v_u^\prime = v_d^\prime \equiv v^\prime$.  Consequently, after fixing the values of $M_Z$ and gaugino soft SUSY breaking masses $M_1$ and $M_2$, the model  spectrum  depends on five parameters: $\tan{\beta}=v_u /v_d$, $\tan{\theta}= \sqrt{(v_u^2 +v_d^2)/2v^{\prime 2}}$, the MSSM $\mu=\mu_{11}$ parameter, the $\mu_{11} B$ SUSY breaking parameter, and a mixing mass parameter $\mu_{12}$ between the MSSM Higgs and the constrained Higgs multiplets.  The K\"ahler SUSY breaking term parameters $m_u^2$, $m_d^2$ and the mixing mass parameter $\mu_{21}$  are fixed by the three electroweak symmetry breaking minimum conditions.  As usual, the $\mu-$problem still exists as a $\mu_{11}$-$\mu_{12}$ stability region of parameter space which must be determined in order to prevent $D$-flat direction runaway field values.  There is no additional $\mu$-problem tuning since the origin of field space is not an extremum of the potential as the nonlinear realization of the electroweak symmetry imposes its breakdown.

Since the quark and lepton superfield  Yukawa couplings only involve the MSSM Higgs fields, the isssue of flavor changing neutral currents (FCNC) is the same as that of the MSSM. Note that, since the $W$ and $Z$ masses are now given by the vacuum expectation value $v^2 = v_u^2 + v_d^2 + v_u^{\prime 2} +v_d^{\prime 2}=v_u^2 + v_d^2 + 2 v^{\prime 2}$,  with $M_Z=gv/2$ and $M_W = M_Z \cos{\theta_W}$ ($g^2 =g_1^2 + g_2^2$), generating the same matter masses requires that the Yukawa coupling constants  be larger than in the MSSM.  The perturbative bounds, ($ \leq 4 \pi$), for the top and bottom quarks and $\tau$ lepton  provide a further restriction on the parameter space. In section \ref{section4}, we discuss the constraints imposed by the electroweak precision tests. In addition, we consider the modifications to Higgs production and decay  due to the extra vacuum expectation values and Higgs field mixing.  Finally, note that the model has an unbroken $R$-parity which dictates the stability of the lightest supersymmetric particle (LSP) which for various regions of parameter space is the lightest neutralino and hence it is a dark matter candidate.   

\section{The Higgs-Gauge Sector Action \label{section2}}

The relevant Higgs and gauge terms in the action of Eq. (\ref{Action}) have the form
\be
\Gamma_{H-G} = \Gamma_{\rm YM} +\Gamma_{\rm K} +\Gamma_{\rm W} +\Gamma_{\rm \rlap{/}{S}} ,
\ee
where the $SU(2)_L\times U(1)$ field strength terms are
\be
\Gamma_{\rm YM} =\frac{1}{4g_2^2}\int dS {\rm Tr}[W_2 W_2] + \frac{1}{4g_1^2}\int dS W_1 W_1 +\frac{1}{4g_2^2}\int d\bar{S} {\rm Tr}[\bar{W}_2 \bar{W}_2] + \frac{1}{4g_1^2}\int d\bar{S} \bar{W}_1 \bar{W}_1 
\ee
and the two pairs of Higgs doublets have a  K\"ahler potential action given by
\be
\Gamma_{\rm K} = \int dV \left\{ \bar{H_u} e^{-2g_2 W -g_1 B} H_u + 
\bar{H}_d e^{-2g_2 W +g_1 B} H_d +\bar{H}_u^\prime e^{-2g_2 W -g_1 B} H_u^\prime + 
\bar{H}_d^\prime e^{-2g_2 W +g_1 B} H_d^\prime\right\}.
\ee
% where the MSSM Higgs doublets are given by the first equality in Eq. (\ref{MSSMDoublets}).
The Higgs doublet portion of the superpotential includes the mixing terms among the constrained and MSSM Higgs multiplets as well as the MSSM $\mu_{11}$-term so that
\be
\Gamma_{\rm W} = \int dS W + \int d\bar{S} \bar{W} 
\ee
with
\be
W=\mu_{11} H_u \epsilon H_d + W_{\rm Mix} =\mu_{11} H_u \epsilon H_d +\mu_{12} H_u \epsilon H_d^\prime + \mu_{21} H_u^\prime \epsilon H_d .
\ee
Finally the soft SUSY breaking terms for the gauginos and MSSM Higgs doublets are denoted as
\be
\Gamma_{\rm \rlap{/}{S}} = \int d^4 x {\cal L}_{\rm \rlap{/}{S}}
\ee
while, for simplicity, we take  the K\"ahler-like and $\mu_{11} B$ term type breaking to appear  only for the MSSM Higgs fields so that
\bea
{\cal L}_{\rm \rlap{/}{S}} &=& \frac{1}{2} M_1 \left( \lambda\lambda + \bar{\lambda}\bar{\lambda}\right) + \frac{1}{2} M_2 \left( \lambda^i\lambda^i + \bar{\lambda^i}\bar{\lambda^i}\right) \cr
 & & -m_u^2 H_u^\dagger H_u - m_d^2 H_d^\dagger H_d -\mu_{11} B H_u \epsilon H_d - \mu_{11} B H_u^\dagger \epsilon H_d^\dagger .
\label{LSUSYBreaking}
\eea
where $\lambda^i (\lambda)$ are the gaugino fields.

In the Wess-Zumino gauge, the component Lagrangian takes the corresponding form 
\bea
{\cal L} = {\cal L}_{\rm YM} +{\cal L}_{\rm K} +{\cal L}_{\rm W}+{\cal L}_{\rm \rlap{/}{S}} .
\eea
Here ${\cal L}_{\rm YM} = {\cal L}_{\rm SYM} + {\cal L}_{\rm DYM}$, where the individual contributions to the gauge and gaugino Lagrangian are 
\be
{\cal L}_{\rm SYM} = -\frac{1}{4}F_{\mu\nu}^i F^{i~\mu\nu} -\frac{1}{4}B_{\mu\nu} B^{\mu\nu} +i \bar{\lambda^i}\bar\sigma^\mu D_\mu \lambda^i +i \bar{\lambda}\bar\sigma^\mu \partial_\mu \lambda
\label{LSYM}
\ee
while the $D$-term contribution to the Lagrangian is simply
\be
{\cal L}_{\rm DYM} = \frac{1}{2}D^i D^i +\frac{1}{2}D D.
\ee
The field strength tensors are as usual
\bea
B_{\mu\nu} &=& \partial_\mu B_\nu - \partial_\nu B_\mu \cr
F^i_{\mu\nu} &=& \partial_\mu W^i_\nu - \partial_\nu W^i_\mu +g_2 \epsilon_{ijk} W_\mu^j W_\nu^k ,
\eea
while the $SU(2)_L$ adjoint representation gaugino covariant derivative is
\be
\left( D_\mu \lambda_\alpha  \right)^i = \partial_\mu \lambda_\alpha^i +g_2 \epsilon_{ijk} W_\mu^j \lambda_\alpha^k  .
\ee
Expanding the K\"ahler potential, the kinetic, auxiliary and gaugino-Higgsino Yukawa terms are obtained as
\bea
{\cal L}_{\rm K} &=& F_u^\dagger F_u + F_d^\dagger F_d + F^{\prime\dagger}_u F^\prime_u + F^{\prime\dagger}_d F^\prime_d \cr
 & &-\frac{g_1}{2} D \left[ H_u^\dagger H_u -H_d^\dagger H_d +H_u^{\prime\dagger} H_u^\prime -H_d^{\prime\dagger} H_d^\prime \right] \cr
 & & -\frac{g_2}{2} D^i \left[ H_u^\dagger \sigma^i H_u +H_d^\dagger \sigma^i H_d +H_u^{\prime\dagger} \sigma^i H_u^\prime +H_d^{\prime\dagger} \sigma^i H_d^\prime \right] \cr   
 & &+\left(D^\mu H_u \right)^\dagger \left(D_\mu H_u \right)+\left(D^\mu H_d \right)^\dagger \left(D_\mu H_d \right)+\left(D^\mu H_u^\prime \right)^\dagger \left(D_\mu H_u^\prime \right)+\left(D^\mu H_d^\prime \right)^\dagger \left(D_\mu H_d^\prime \right) \cr
 & &+i \bar{\tilde{H}}_u \bar\sigma^\mu D_\mu \tilde{H}_u +i \bar{\tilde{H}}_d \bar\sigma^\mu D_\mu \tilde{H}_d+i \bar{\tilde{H}}_u^\prime \bar\sigma^\mu D_\mu \tilde{H}_u^\prime +i \bar{\tilde{H}}_d^\prime \bar\sigma^\mu D_\mu \tilde{H}_d^\prime  \cr
 & &+\frac{g_1}{\sqrt{2}}\left[ H_u^\dagger \lambda \tilde{H}_u + \bar{\tilde{H}}_u \bar{\lambda} H_u -H_d^\dagger \lambda \tilde{H}_d - \bar{\tilde{H}}_d \bar{\lambda} H_d +H_u^{\prime\dagger} \lambda \tilde{H}_u^\prime + \bar{\tilde{H}}_u^\prime \bar{\lambda} H_u^\prime -H_d^{\prime\dagger} \lambda \tilde{H}_d^\prime - \bar{\tilde{H}}_d^\prime \bar{\lambda} H_d^\prime  \right]\cr
 & & +\frac{g_2}{\sqrt{2}}\left[ H_u^\dagger (\lambda^i \sigma^i) \tilde{H}_u + \bar{\tilde{H}}_u (\bar{\lambda}^i \sigma^i) H_u +H_d^\dagger (\lambda^i \sigma^i) \tilde{H}_d - \bar{\tilde{H}}_d (\bar{\lambda}^i \sigma^i) H_d \right.\cr
 & &\left. \qquad\qquad\qquad\qquad +H_u^{\prime\dagger} (\lambda^i \sigma^i) \tilde{H}_u^\prime + \bar{\tilde{H}}_u^\prime (\bar{\lambda}^i \sigma^i) H_u^\prime -H_d^{\prime\dagger} (\lambda^i \sigma^i) \tilde{H}_d^\prime - \bar{\tilde{H}}_d^\prime (\bar{\lambda}^i \sigma^i) H_d^\prime  \right] ,
\eea
with the covariant derivatives
\bea
D_\mu H_u &=& \left[ \partial_\mu -\frac{ig_2}{2} \vec\sigma \cdot \vec{W}_\mu -\frac{ig_1}{2} B_\mu \right] H_u \cr
D_\mu H_d &=& \left[ \partial_\mu -\frac{ig_2}{2} \vec\sigma \cdot \vec{W}_\mu +\frac{ig_1}{2} B_\mu \right] H_d ,
\eea
and likewise for $H_u^\prime$ and $H_d^\prime$ and the associated Higgsino partners $\tilde{H}_u$, 
$\tilde{H}_u^\prime$, $\tilde{H}_d$ and $\tilde{H}_d^\prime$. 
The superpotential contribution to the Lagrangian takes its familiar doublet auxiliary field  and Higgsino mass term form
\bea
{\cal L}_{\rm W} &=& -4 F^a \frac{\partial W}{\partial A^a} +2 \lambda^a \frac{\partial^2 W}{\partial A^a \partial A^b} \lambda^b  + ~ {\rm h.c.} \cr
 &=& -4\mu_{11} F_u \epsilon H_d -4 \mu_{12} F_u \epsilon H_d^\prime -4\mu_{11} H_u \epsilon F_d -4 \mu_{21} H_u^\prime \epsilon F_d -4 \mu_{12} H_u \epsilon F_d^\prime -4 \mu_{21} F_u^\prime \epsilon H_d \cr
 & & \qquad\qquad + 4\mu_{11} \tilde{H}_u \epsilon \tilde{H}_d + 4\mu_{12} \tilde{H}_u \epsilon \tilde{H}_d^\prime + 4\mu_{21} \tilde{H}_u^\prime \epsilon \tilde{H}_d  +{\rm h.c.}  .
\eea
The soft SUSY breaking Lagrangian is given by Eq. (\ref{LSUSYBreaking}).

The chiral superfields have the  component expansion 
\bea
\Sigma (x,\theta, \bar\theta) &= &e^{-i\theta \rlap{/}{\partial}\bar\theta} \left[ \sigma (x) +\sqrt{2} \theta^\alpha \tilde\sigma_\alpha (x) + \theta^2 F_\sigma (x) \right]\cr
\Pi^i (x,\theta, \bar\theta) &= &e^{-i\theta \rlap{/}{\partial}\bar\theta} \left[ \pi^i (x) +\sqrt{2} \theta^\alpha \tilde\pi^i_\alpha (x) + \theta^2 F_\pi^i (x) \right] .
\eea
Applying the constraint to the $H_u^\prime$ and $H_d^\prime$ doublets, $H_d^\prime \epsilon H_u^\prime = v_u^\prime v_d^\prime / 2$, the  component fields take the form
\bea
\sigma &=& \sqrt{\frac{v_u^\prime v_d^\prime}{2} -{\vec{\pi}}^2~} \cr
\tilde\sigma_\alpha &=& -\frac{\vec\pi \cdot \vec{\tilde{\pi}}_\alpha}{\sqrt{\frac{v_u^\prime v_d^\prime}{2} -{\vec{\pi}}^2}~} \cr
F_\sigma &=& \frac{-\vec{F_\pi}\cdot \vec{\pi}+\frac{1}{2}\vec{\tilde{\pi}} \cdot \vec{\tilde{\pi}}}{\sqrt{\frac{v_u^\prime v_d^\prime}{2} -{\vec{\pi}}^2}~} ~~.
\eea
The auxiliary fields can now be eliminated through field equations.  Focusing on the relevant $D$- and $F$-terms, the Lagrangian for $D$-terms has contributions from ${\cal L}_{\rm DYM}$ and ${\cal L}_{\rm K}$ and is given by
\bea
{\cal L}_{\rm D} &=& \frac{1}{2} D^i D^i  + \frac{1}{2} D D -\frac{1}{2} \left[ H_u^{\prime\dagger} \left( 2 g_2 \frac{\sigma^i}{2} D^i +2 g_1 \frac{1}{2}D \right) H_u^\prime + H_d^{\prime\dagger} \left( 2 g_2 \frac{\sigma^i}{2} D^i - 2 g_1 \frac{1}{2}D \right) H_d^\prime  \right. \cr
 & &\left.  +H_u^\dagger \left( 2 g_2 \frac{\sigma^i}{2} D^i +2 g_1 \frac{1}{2}D \right) H_u + H_d^\dagger \left( 2 g_2 \frac{\sigma^i}{2} D^i - 2 g_1 \frac{1}{2}D \right) H_d  \right]  \cr
 &\equiv& \frac{1}{2} D^A Z_{AB} D^B -\frac{1}{2} D^A {\cal J}_A ,
\eea
with 
\be
Z^{-1~AB} = \begin{pmatrix}
(2g_2)^2 \delta^{ij} &0\\
0&(2g_1)^2 
\end{pmatrix}_{AB}
\ee
and where $D^A = (2g_2 D^i~, 2g_1 D)$, with $A=1,2,3,4$.  The $D$-term contribution is given by the Killing potentials  
\be
{\cal J}_A = J_A + H_u^\dagger T_u^A H_u  + H_d^\dagger  T_d^A  H_d 
\label{JA}
\ee
which are the $\theta$-$\bar\theta$ independent components of the gauge superfield Noether currents. Here the representation matrices are combined according to $T_u^A = (\vec{\sigma}, 1)/2$ and $T_d^A =(\vec{\sigma}, -1)/2$ while the nonlinear sigma model Killing potential \cite{CL} is found to be
\bea
J_A &=& H_u^{\prime\dagger}  T_u^A H_u^\prime + H_d^{\prime\dagger}  T_d^A  H_d^\prime \cr
 &=& \frac{i}{2} H^{\prime\dagger}_u  \frac{\partial H_u^\prime}{\partial \pi^i} A_A^i -\frac{i}{2} \frac{\partial H^{\prime\dagger}_u}{\partial \pi^{\bar{i}\dagger}} A_A^{\bar{i}\dagger} H_u^\prime  +\frac{i}{2} H^{\prime\dagger}_d  \frac{\partial H_d^\prime}{\partial \pi^i} A_A^i -\frac{i}{2} \frac{\partial H^{\prime\dagger}_d}{\partial \pi^{\bar{i}\dagger}} A_A^{\bar{i}\dagger} H_d^\prime \cr
 &=& \frac{i}{2} \frac{\partial K}{\partial \pi^i} A_A^i -\frac{i}{2} \frac{\partial K}{\partial \pi^{\bar{i}\dagger}} A_A^{\bar{i}\dagger} ,
\eea
with 
\bea
K= H^{\prime\dagger}_u H_u^\prime + H^{\prime\dagger}_d H_d^\prime = 2( \sigma^\dagger \sigma +\vec{\pi}^\dagger \cdot \vec\pi ).
\eea
The (anti-)chiral Killing vectors ($A_A^{\bar{i}\dagger} (\pi^\dagger))~A_A^i (\pi)$ are given according to the $\sigma$-model realization through the variation of the constrained doublets $H_u^\prime$ and $H_d^\prime$. They are secured as the $\theta-\bar\theta$ independent components of the defining superfield relations
\bea
\delta (\Lambda)H_u^\prime &=& -i \Lambda^A T_u^A H_u^\prime = \frac{\partial H_u^\prime}{\partial \Pi^i} \delta (\Lambda) \Pi^i = \frac{\partial H_u^\prime}{\partial \pi^i} \Lambda^A A_A^i (\Pi) \cr
\delta (\Lambda)H_d^\prime &=& -i \Lambda^A T_d^A H_d^\prime = \frac{\partial H_d^\prime}{\partial \Pi^i} \delta (\Lambda) \Pi^i = \frac{\partial H_d^\prime}{\partial \pi^i} \Lambda^A A_A^i (\Pi) ,
\label{transf}
\eea
where, analogously to the gauge fields, $V^A =(2g_2 \vec{W} , 2g_1 B)$, the four  chiral gauge transformation parameters are defined as $ \Lambda^A = (2g_2 \vec\Lambda_2, 2g_1 \Lambda_1 )$.  Recalling the expression for the constrained doublets in terms of the $\sigma$-model coordinates, equation (\ref{ConstrainedDoublets}), the Killing vectors are obtained
\be
A_A^i =\left\{
\begin{matrix}
\frac{1}{2} \epsilon^{ikj} \Pi^j -\frac{1}{2}\delta_k^i \Sigma & , ~A=k \\
\frac{1}{2} \epsilon^{i3j} \Pi^j +\frac{1}{2}\delta_3^i \Sigma & , ~A=4 \\
\end{matrix} \right.,
\ee
with the constraint $\Sigma = \sqrt{v_u^\prime v_d^\prime /2 -\vec\Pi^2~}$.  The superfield Killing vectors are given in terms of the derivative of the Killing potentials.  As seen from above 
\bea
\frac{\partial}{\partial \bar\pi^{\bar{i}}} J_A &=& iA_A^i g_{\bar{i}i} \cr
\frac{\partial}{\partial \pi^{i}} J_A &=& -i\bar{A}_A^{\bar{i}} g_{\bar{i}i} 
\eea
with
\be
g_{\bar{i}i} = \frac{\partial \bar{H^\prime}_u}{\partial \bar\Pi^{\bar{i}}} \frac{\partial H_u^\prime}{\partial \Pi^i} +\frac{\partial \bar{H^\prime}_d}{\partial \bar\Pi^{\bar{i}}} \frac{\partial H_d^\prime}{\partial \Pi^i} .
\label{metric}
\ee
Expanding Eqs. (\ref{transf})-(\ref{metric}) in powers of $\theta$ and $\bar\theta$ allows for the extraction of  the various component relations.

Hence, by the straightforward application of the auxiliary $D$-field equation of motion, the $D$-term (component) Lagrangian becomes
\be
{\cal L}_{\rm D} = -\frac{1}{8} {\cal J}_A Z^{-1~ AB} {\cal J}_B .
\ee
where here ${\cal J}_A$ denotes the $\theta-\bar\theta$ independent component of the defining superfield relation as given in Eqs. (\ref{JA})-(\ref{transf}).

The $F$-terms are contained in ${\cal L}_{\rm K}$ and ${\cal L}_{\rm W}$.  For the unconstrained MSSM doublets, they have the combined form
\bea
{\cal L}_{\rm F} &=& F_u^\dagger F_u +F_d^\dagger F_d -4 F_u \epsilon \left(  \mu_{11} H_d + \mu_{12} H_d^\prime \right)-4 \left( \mu_{11} H_u + \mu_{21} H_u^\prime  \right) \epsilon F_d \cr
 & &-4 F_u^\dagger \epsilon \left(  \mu_{11} H_d^\dagger + \mu_{12} H_d^{\prime\dagger} \right)-4 \left( \mu_{11} H_u^\dagger + \mu_{21} H_u^{\prime\dagger}  \right) \epsilon F_d^\dagger .
\eea
Eliminating the $F_u$ and $F_d$ doublet auxiliary fields yields
\be
{\cal L}_{\rm F} = -16 | \mu_{11} H_d +\mu_{12} H_d^\prime |^2 -16 |\mu_{11} H_u +\mu_{21} H_u^\prime |^2  .
\ee
The constrained auxiliary fields couple to the scalar and fermion fields through the K\"ahler potential as well as the $\mu$-term superpotential.  Their combined Lagrangian is
\bea
{\cal L}_{\rm F^\prime} &=& \left[F_\pi^{\bar{i}^\dagger} - \frac{1}{2} \Gamma^{\bar{i}\dagger}_{\bar{m} \bar{n}} \bar{\tilde{\pi}}^{\bar{m}} \bar{\tilde{\pi}}^{\bar{n}} \right] g_{\bar{i}~i} \left[ F_\pi^i -\frac{1}{2} \Gamma^i_{rs} \tilde{\pi}^r \tilde{\pi}^s \right] \cr
 & &\qquad-4 \left\{ \left[ \mu_{12} H_u \epsilon \frac{\partial H_d^\prime}{\partial \pi^i} + \mu_{21} \frac{\partial H_u^\prime}{\partial \pi^i} \epsilon H_d \right] F_\pi^i  +{\rm h.c.} \right\} ,
\eea
where the K\"ahler metric is obtained from the K\"ahler potential to be
\be
g_{\bar{i}i}= 2\left(\delta_{\bar{i}i} + \frac{\pi^{\bar{i}\dagger}\pi^i}{\sigma^\dagger \sigma}  \right)
\ee
and the associated Christoffel symbols are
\be
\Gamma^i_{jk} = g^{i\bar{i}} g_{\bar{i} j,k}
\ee
and similarly for $\Gamma^{\bar{i}\dagger}_{\bar{m} \bar{n}}$.
Employing the $F_\pi$ Euler-Lagrange equations then gives
\bea
{\cal L}_{\rm F^\prime} &=& -16 \left[ \mu_{12} H_u \epsilon \frac{\partial H_d^\prime}{\partial \pi^i} + \mu_{21} \frac{\partial H_u^\prime}{\partial \pi^i} \epsilon H_d \right] g^{i~\bar{i}} \left[ \mu_{12} H_u^\dagger \epsilon \frac{\partial H^{\prime\dagger}_d}{\partial \pi^{\bar{i}\dagger}} + \mu_{21} \frac{\partial H^{\prime\dagger}_u}{\partial \pi^{\bar{i}\dagger}} \epsilon H_d^\dagger \right]\cr
 & & -2 \left[ \mu_{12} H_u \epsilon \frac{\partial H_d^\prime}{\partial \pi^i} + \mu_{21} \frac{\partial H_u^\prime}{\partial \pi^i} \epsilon H_d \right] \Gamma^i_{rs} \tilde\pi^r \tilde\pi^s \cr
 & & -2 \Gamma^{\bar{i}\dagger}_{\bar{m} \bar{n}} \bar{\tilde{\pi}}^{\bar{m}} \bar{\tilde{\pi}}^{\bar{n}} \left[ \mu_{12} H_u^\dagger \epsilon \frac{\partial H^{\prime\dagger}_d}{\partial \pi^{\bar{i}\dagger}} + \mu_{21} \frac{\partial H^{\prime\dagger}_u}{\partial \pi^{\bar{i}\dagger}} \epsilon H_d^\dagger \right]  .
\eea

Hence the Lagrangian with auxiliary fields eliminated has the form ${\cal L} = {\cal L}_{\rm SYM} + {\cal L}_{\rm \rlap{/}{S}} + {\cal L}_{\sigma}$ where the $\sigma$-model Lagrangian, ${\cal L}_{\sigma}$, consists of all the terms coming from ${\cal L}_{\rm D}, {\cal L}_{\rm K} $ and ${\cal L}_{\rm W}$ and takes the form
\bea
{\cal L}_{\sigma} &=&   {\cal L}_{\rm D}  + {\cal L}_{\rm F} +{\cal L}_{\rm F^\prime} \cr
 & &+D_\mu \pi^{\bar{i}\dagger} g_{\bar{i}~ i} D^\mu \pi^i + i \bar{\tilde{\pi}}^{\bar{i}} \bar\sigma^\mu g_{\bar{i}~i} D_\mu \tilde{\pi}^i  +\frac{1}{4}R_{r\bar{m} s \bar{n}} \bar{\tilde{\pi}}^{\bar{m}} \bar{\tilde{\pi}}^{\bar{n}} \tilde{\pi}^r \tilde{\pi}^s \cr
 & &+\left(D^\mu H_u \right)^\dagger \left(D_\mu H_u \right)+\left(D^\mu H_d \right)^\dagger \left(D_\mu H_d \right)+i \bar{\tilde{H}}_u \bar\sigma^\mu D_\mu \tilde{H}_u +i \bar{\tilde{H}}_d \bar\sigma^\mu D_\mu \tilde{H}_d \cr
 & &+\frac{1}{\sqrt{2}}\left[ H_u^\dagger \lambda^A T_u^A \tilde{H}_u + \bar{\tilde{H}}_u \bar{\lambda}^A T_u^A H_u + H_d^\dagger \lambda^A T_d^A \tilde{H}_d + \bar{\tilde{H}}_d \bar{\lambda}^A T_d^A H_d \right. \cr
 & &\left. \qquad\qquad -iA_A^{\bar{i}\dagger} g_{\bar{i}~i} \lambda^A \tilde\pi^i + i \bar{\tilde{\pi}}^{\bar{i}} g_{\bar{i}~i} \bar{\lambda}^A  A_A^i \right]\cr
 & & +4 \mu_{12} H_u \epsilon \frac{\partial^2 H_d^\prime}{\partial \pi^i \partial \pi^j} \tilde{\pi}^i \tilde{\pi}^j +2\mu_{21} \frac{\partial^2 H_u^\prime}{\partial \pi^i \partial \pi^j}\tilde{\pi}^i \tilde{\pi}^j \epsilon H_d \cr
 & &\qquad\qquad  +4 \mu_{12} H_u^\dagger \epsilon \frac{\partial^2 H_d^{\prime\dagger}}{\partial \pi^{\bar{i}^\dagger} \partial \pi^{\bar{j}^\dagger}} \bar{\tilde{\pi}}^{\bar{i}} \bar{\tilde{\pi}}^{\bar{j}} +2\mu_{21} \frac{\partial^2 H^{\prime\dagger}_u}{\partial \pi^{\bar{i}\dagger} \partial \pi^{\bar{j}\dagger}}\bar{\tilde{\pi}}^{\bar{i}} \bar{\tilde{\pi}}^{\bar{j}} \epsilon H_d^\dagger \cr
 & & + 4\mu_{11} \tilde{H}_u \epsilon \tilde{H}_d + 4\mu_{12} \tilde{H}_u \epsilon \tilde{H}_d^\prime + 4\mu_{21} \tilde{H}_u^\prime \epsilon \tilde{H}_d \cr
 & &\qquad\qquad + 4\mu_{11} \bar{\tilde{H}}_u \epsilon \bar{\tilde{H}}_d + 4\mu_{12} \bar{\tilde{H}}_u \epsilon \bar{\tilde{H}}_d^\prime + 4\mu_{21} \bar{\tilde{H}}_u^\prime \epsilon \bar{\tilde{H}}_d ,
\label{Lsigma}
\eea
where the Riemann tensor is given by 
\be
R^m_{r\bar{m}s} = \frac{\partial}{\partial \pi^{\bar{m}\dagger}} \Gamma^m_{rs} 
\ee
with
\be
R_{r\bar{m}s\bar{n}}= g_{\bar{n}m} R^m_{r\bar{m}s} = \frac{\partial}{\partial \pi^{\bar{m}\dagger}} \Gamma_{\bar{n}rs} - \Gamma^{\bar{i}\dagger}_{\bar{m}\bar{n}}\Gamma_{\bar{i}rs} .
\ee
The covariant derivatives are found by expressing the K\"ahler kinetic energy terms for the constrained doublets in terms of the unconstrained $\sigma$-model $\pi$ fields so that
\be
|D_\mu H_u^\prime |^2 + |D_\mu H_d^\prime |^2 = D_\mu \pi^{\bar{i}\dagger} g_{\bar{i}~i} D^\mu \pi^i  ,
\ee
with
\be
D_\mu \pi^i = \partial_\mu \pi^i +\frac{1}{2} V_\mu^A  A_A^i (\pi) .
\ee
Similarly for the Higgsino fields
\be
i \bar{\tilde{H}}_u^\prime \bar\sigma^\mu D_\mu \tilde{H}_u^\prime +i \bar{\tilde{H}}_d^\prime \bar\sigma^\mu D_\mu \tilde{H}_d^\prime  = i\bar{\tilde{\pi}}^{\bar{i}} \bar\sigma^\mu g_{\bar{i}~i} D_\mu \tilde{\pi}^i  ,
\ee
with
\be
D_\mu \tilde{\pi}^i = \partial_\mu \tilde{\pi}^i +\frac{1}{2}V_\mu^A \frac{\partial A_A^i}{\partial \pi^j} \tilde{\pi}^j  +\Gamma^i_{jk} (D_\mu \pi^j ) \tilde{\pi}^k  .
\ee

From the Lagrangian the scalar potential $V$ can be read off as
\bea
V &=& m_u^2 H_u^\dagger H_u  + m_d^2 H_d^\dagger H_d -\mu_{11} B H_u \epsilon H_d - \mu_{11} B H_u^\dagger \epsilon H_d^\dagger \cr
 & & +\frac{1}{8} {\cal J}_A Z^{-1~ AB} {\cal J}_d +16 | \mu_{11} H_d +\mu_{12} H_d^\prime |^2 +16 |\mu_{11} H_u +\mu_{21} H_u^\prime |^2  \cr
 & &+16 \left[ \mu_{12} H_u \epsilon \frac{\partial H_d^\prime}{\partial \pi^i} + \mu_{21} \frac{\partial H_u^\prime}{\partial \pi^i} \epsilon H_d \right] g^{i~\bar{i}} \left[ \mu_{12} H_u^\dagger \epsilon \frac{\partial H^{\prime\dagger}_d}{\partial \pi^{\bar{i}\dagger}} + \mu_{21} \frac{\partial H^{\prime\dagger}_u}{\partial \pi^{\bar{i}\dagger}} \epsilon H_d^\dagger \right]  .
\eea
Taking the derivatives of the potential with respect to the shifted scalar fields $(H^{0\dagger}_u +H_u^0)$, $(H_d^{0\dagger} +H_d^0)$ and $(\pi^{0\dagger} -\pi^0)$ and evaluating it at the vacuum expectation values
$<0|H_u^0 |0> = v_u /\sqrt{2}$ and $<0|H_d^0 |0> = v_d /\sqrt{2}$, yields the three electroweak symmetry breaking minima equations
\bea
1)~~&0=&m_u^2 v_u -\mu_{11} B v_d +16 \mu_{11} \left( \mu_{11} v_u + \mu_{21} v_u^\prime  \right) \cr
 & &\qquad +16 \mu_{12}\frac{v_d^\prime}{(v_u^{\prime 2} +v_d^{\prime 2})} \left( \mu_{12} v_u v_d^\prime-\mu_{21} v_d v_u^\prime \right) \cr
 & &\qquad +\frac{(g_1^2 + g_2^2 )}{8} \left(v_u^{\prime 2} -v_d^{\prime 2} +v_u^2 -v_d^2\right) v_u  \cr
2)~~&0=& m_d^2 v_d -\mu_{11} B v_u +16 \mu_{11} \left( \mu_{11} v_d + \mu_{12} v_d^\prime  \right) \cr
 & &\qquad -16 \mu_{21}\frac{v_u^\prime}{(v_u^{\prime 2} +v_d^{\prime 2})} \left( \mu_{12} v_u v_d^\prime-\mu_{21} v_d v_u^\prime \right) \cr
 & &\qquad -\frac{(g_1^2 + g_2^2 )}{8}  \left(v_u^{\prime 2} -v_d^{\prime 2} +v_u^2 -v_d^2 \right) v_d  \cr
3)~~&0=&\frac{(g_1^2 + g_2^2 )}{8} \left(v_u^{\prime 2} -v_d^{\prime 2} +v_u^2 -v_d^2\right) \left(  \frac{v_u^{\prime 2} +v_d^{\prime 2}}{v_u^\prime +v_d^\prime}   \right)  \cr
 & &\qquad -16 \mu_{12} \left( \frac{ v_d^\prime}{v_u^\prime +v_d^\prime} \right) \left( \mu_{11} v_d +\mu_{12} v_d^\prime    \right) \cr
& &\qquad +16 \mu_{21} \left( \frac{ v_u^\prime}{v_u^\prime +v_d^\prime} \right) \left( \mu_{11} v_u +\mu_{21} v_u^\prime    \right) \cr
 & &\qquad -32 \frac{v_u^\prime v_d^\prime}{(v_u^{\prime 2} v_d^{\prime 2}) (v_u^\prime +v_d^\prime)^2} \left(\mu_{12} v_u +\mu_{21} v_d  \right) \left( \mu_{12} v_u v_d^\prime -\mu_{21} v_d v_u^\prime \right) \cr
 & &\qquad-32  \frac{v_u^\prime v_d^\prime (v_u^\prime -v_d^\prime )}{(v_u^{\prime 2} v_d^{\prime 2})^2  (v_u^\prime +v_d^\prime)^2}  \left( \mu_{12} v_u v_d^\prime -\mu_{21} v_d v_u^\prime \right)^2  .
\eea
Note that these equations admit no non-trivial solutions for $v_u, v_d$ in the limit $v_u^\prime=v_d^\prime=0$ and $\mu_{12}^2=\mu_{21}^2=0$ and the good SUSY limit $B=m_u^2=m_d^2=0$. Consequently, it is the non-trivial vacuum expectation values of the constrained Higgs doublets which catalyze the vacuum expectation values of the MSSM Higgs doublets through their bilinear superpotential coupling with coefficients $\mu_{12}, \mu_{21}$.

In order to simplify the parameter space the nonlinearly realized symmetry breakdown is taken to respect the custodial $SU(2)_V$ symmetry hence, $v_u^\prime =v_d^\prime \equiv v^\prime$.   The $Z$ and $W$ vector boson masses are then given by the vacuum value
$v^2 = v_u^2 + v_d^2 +2 v^{\prime 2}$ with $M_Z = g v/2$ and $M_W = M_Z \cos{\theta_W}$.  The 3 potential minimum equations simplify to 
\bea
1)~~&0=&\frac{M_Z^2}{2} \frac{v_u^2 - v_d^2}{v^2} v_u +m_u^2 v_u +16 \mu_{11}^2 v_u -\mu_{11} B v_d +16 \mu_{11} \mu_{21} v^\prime +8 \mu_{12} \left( \mu_{12} v_u -\mu_{21} v_d \right) \cr
2)~~&0=&-\frac{M_Z^2}{2} \frac{v_u^2 - v_d^2}{v^2} v_d +m_d^2 v_d +16 \mu_{11}^2 v_d -\mu_{11} B v_u +16 \mu_{11} \mu_{12} v^\prime -8 \mu_{21} \left( \mu_{12} v_u -\mu_{21} v_d \right) \cr
3)~~&0=&\frac{M_Z^2}{2} \frac{v_u^2 - v_d^2}{v^2}  -8 \mu_{12}^2 \left( 1+\frac{v_u^2}{2 v^{\prime 2}} \right)
+8 \mu_{21}^2 \left( 1+\frac{v_d^2}{2 v^{\prime 2}} \right) -8 \mu_{11} \left( \mu_{12} \frac{v_d}{v^\prime} -\mu_{21} \frac{v_u}{v^\prime}  \right)  .
\eea
Introducing spherical polar coordinates for the 3 vacuum values
\bea
\sqrt{2}~~v^\prime &=& v \cos{\theta} \cr
v_u &=& v \sin{\theta} \sin{\beta} \cr
v_d &=& v \sin{\theta} \cos{\beta} ,
\eea
where  $ \tan{\beta} =v_u /v_d$ and $\tan{\theta} = \sqrt{(v_u^2 +v_d^2 )/(2 v^{\prime 2})}$, the minimum conditions take the form
\bea
1)~~&&m_u^2 +16 \mu_{11}^2 -\frac{M_Z^2}{2} \sin^2{\theta} \cos{2\beta}  =\mu_{11} B \cot{\beta} -8 \sqrt{2} \mu_{11} \mu_{21}  \cot{\theta} \csc{\beta} -8 \mu_{12} \left( \mu_{12} -\mu_{21} \cot{\beta} \right)\cr
2)~~&&m_d^2 +16 \mu_{11}^2 +\frac{M_Z^2}{2} \sin^2{\theta} \cos{2\beta}  =\mu_{11} B \tan{\beta} -8 \sqrt{2} \mu_{11} \mu_{12}  \cot{\theta} \sec{\beta}  +8 \mu_{21} \left( \mu_{12} \tan{\beta} -\mu_{21} \right)\cr
3)~~&&\frac{M_Z^2}{2}  \sin^2{\theta} \cos{2\beta}  = -8 \mu_{12}^2 \left( 1+\tan^2{\theta} \sin^2{\beta} \right) +8 \mu_{21}^2 \left( 1+\tan^2{\theta} \cos^2{\beta}   \right)\cr
 & &\qquad\qquad -8 \sqrt{2}\mu_{11}  \tan{\theta} \left( \mu_{12} \cos{\beta} -\mu_{21} \sin{\beta}  \right)  .
\eea
The first two conditions are used to eliminate $m_u^2$ and $m_d^2$ from the parameters of the model while the third condition is used to express $\mu_{21}$ in terms of the remaining parameters.  Thus the five variables upon which the potential depends are the MSSM parameters $\tan{\beta}$, $\mu_{11}$ and $b=-\mu_{11} B$, as well as the independent electroweak symmetry breaking vacuum angle $\tan{\theta}$ and the Higgs doublet mixing mass coupling $\mu_{12}$.  The tuning of the $\mu_{11}$ and $\mu_{12}$ parameters is required as can be seen by expressing the first two minimum conditions as
\bea
 & &16 \mu_{11}^2 -8\sqrt{2} \mu_{11} \frac{\cot{\theta}}{\tan^2{\beta}-1} \left[\mu_{12} -\mu_{21} \tan{\beta}\right] \sec{\beta}= \frac{m_d^2 -m_u^2 \tan^2{\beta}}{\tan^2{\beta} -1} -\frac{M_Z^2}{2} \sin^2{\theta} \cr
 & &\qquad\qquad\qquad\qquad -8 \frac{\mu_{12}^2 \tan^2{\beta} -\mu_{21}^2}{\tan^2{\beta} -1} \cr
 & &2\mu_{11} B =\left[ m_u^2 +m_d^2 + 32 \mu_{11}^2 + 8\left( \mu_{12}^2 + \mu_{21}^2 \right)
+8 \sqrt{2} \mu_{11} \cot{\theta} \left( \mu_{12} \sec{\beta}
 + \mu_{21} \csc{\beta} \right) \right] \sin{2\beta}\cr
 & &\qquad\qquad -16 \mu_{12} \mu_{21}.
\eea

%\iffalse
\begin{figure*}
\begin{center}
$\begin{array}{cc}
\includegraphics[scale=1.00]{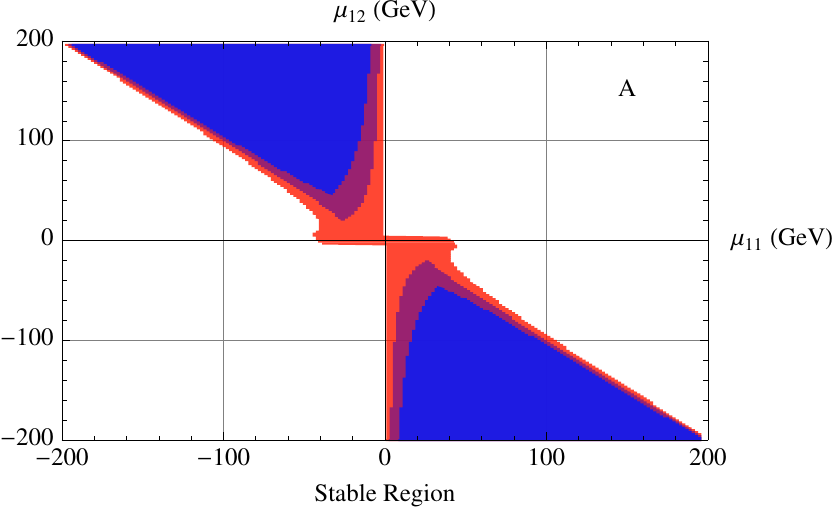} &
\includegraphics[scale=1.00]{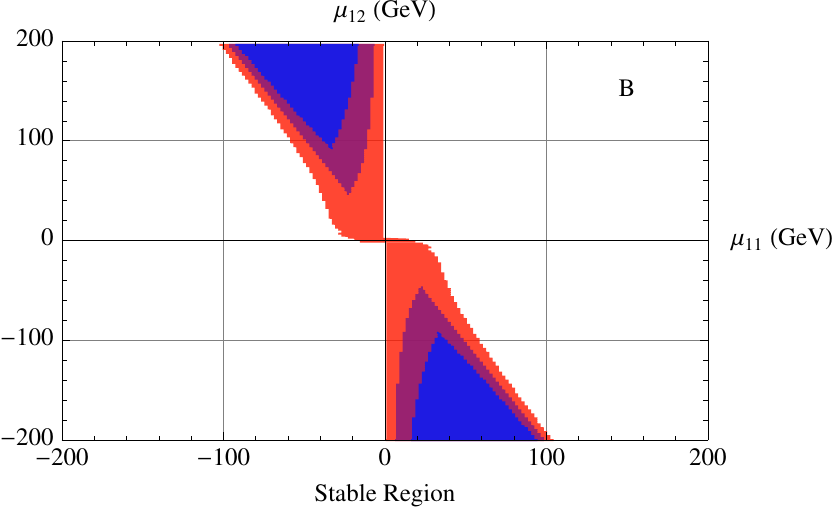} \\
\includegraphics[scale=1.00]{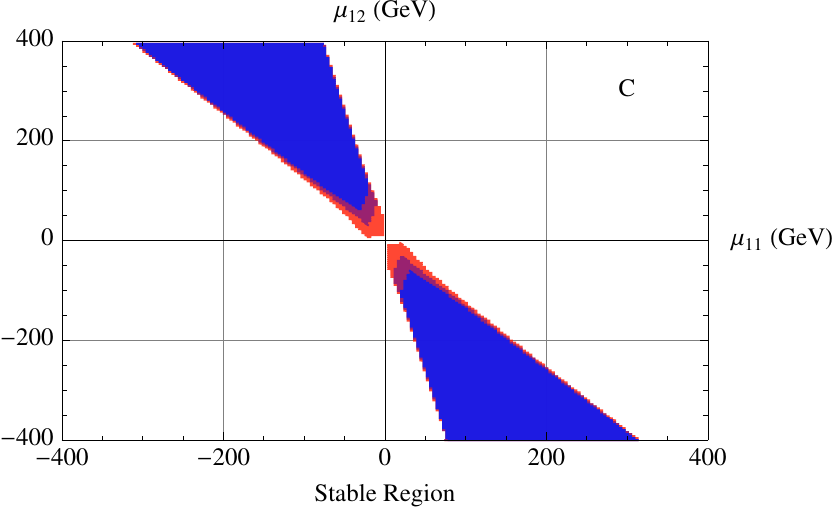} &
\includegraphics[scale=1.00]{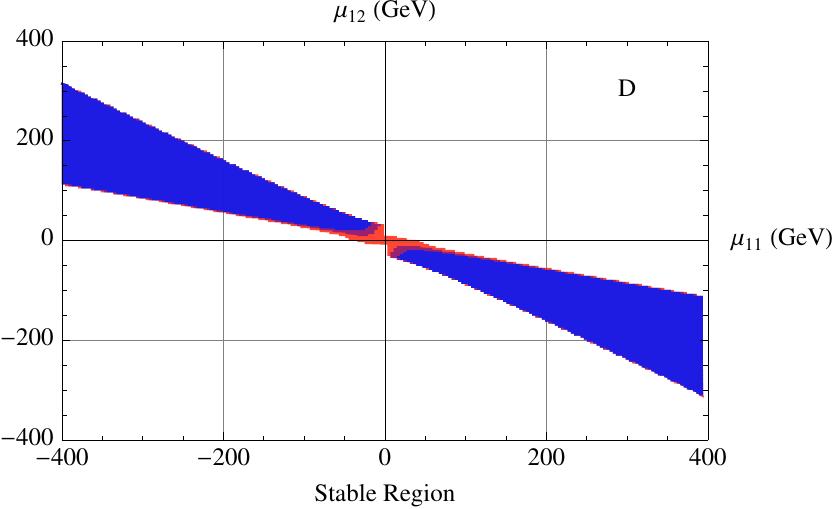} 
\end{array}$
\caption{Stability of the potential against $D$-flat direction runaway field values is determined in the $\mu_{11}$-$\mu_{12}$ parameter plane.  Each region of SUSY breaking parameter $b=-4,000,~4,000,~12,000$ ${\rm GeV^2}$ is depicted by the overlapping orange, violet, blue regions, respectively.  Finally, stability region A has $\tan{\beta}=1$, $\tan{\theta}=1$, region B has $\tan{\beta}=1$, $\tan{\theta}=2$, region C has $\tan{\beta}=2$, $\tan{\theta}=2$, and region D has $\tan{\beta}=10$, $\tan{\theta}=2$.}
\label{StabilityRegion}
\end{center}
\end{figure*}

\section{Mass Spectrum \label{section3}}

In order to determine the mass spectrum of the model, the Lagrangian must be expanded about the non-trivial vacuum expectation values. We focus on the case $v_u^\prime =v_d^\prime = v^\prime$.  In the neutral Higgs field sector, the scalar, $S$, and pseudoscalar, $P$, fields with canonically normalized kinetic terms are introduced in terms of the shifted Higgs fields as
\bea
P_\pi = \left( \pi^{0^\dagger} +\pi^0 \right)\qquad &,&\qquad S_\pi = -i \left(\pi^{0\dagger} -\pi^0\right)\cr
P_u = \frac{i}{\sqrt{2}} \left( H_u^{0\dagger} - H_u^0 \right) \qquad &,&\qquad S_u= \frac{1}{\sqrt{2}} \left( H_u^{0\dagger} + H_u^0 \right) \cr
P_d = \frac{i}{\sqrt{2}} \left( H_d^{0\dagger} - H_d^0 \right) \qquad &,&\qquad S_d= \frac{1}{\sqrt{2}} \left( H_d^{0\dagger} + H_d^0 \right).
\eea
The pseudoscalar and scalar mass squared matrices are determined from the second derivatives of the potential evaluated at the minimum
\bea
\left( M^2_{\rm PS} \right)_{ij} &=& \frac{\partial^2 V}{\partial P_i \partial P_j}\vert_{\rm minimum}~~;~~\left( M^2_{\rm S} \right)_{ij} = \frac{\partial^2 V}{\partial S_i \partial S_j}\vert_{\rm minimum} .
\eea
The pseudoscalar mass squared matrix is given in the $(P_u , P_d , P_\pi )$ basis as 
\be
M_{\rm PS}^2 = \begin{pmatrix}
M_{uu}^2 & M_{ud}^2 &M_{u\pi}^2 \\
M_{du}^2 & M_{dd}^2 &M_{d\pi}^2 \\
M_{\pi u}^2 & M_{\pi d}^2 &M_{\pi\pi}^2
\end{pmatrix}
\ee
with
\bea
M_{uu}^2 &=&\left( \mu_{11} B + 8\mu_{12} \mu_{21} \right)\cot{\beta} -8 \sqrt{2} \mu_{11} \mu_{21} \cot{\theta} \csc{\beta} \cr
M_{dd}^2 &=&\left( \mu_{11} B + 8\mu_{12} \mu_{21} \right)\tan{\beta} -8 \sqrt{2} \mu_{11} \mu_{12} \cot{\theta} \sec{\beta} \cr
M_{\pi\pi}^2 &=&16\mu_{12} \mu_{21} \tan^2{\theta} \sin{2\beta} -8 \sqrt{2}\mu_{11} \tan{\theta} \left( \mu_{12} \cos{\beta} +\mu_{21} \sin{\beta}\right) \cr
M_{ud}^2 &=& \mu_{11} B -8 \mu_{12} \mu_{21} =M_{du}^2 \cr
M_{u\pi}^2 &=& -8\sqrt{2} \mu_{11} \mu_{21} +16 \mu_{12} \mu_{21} \tan{\theta} \cos{\beta} = M_{\pi u}^2 \cr
M_{d\pi}^2 &=& +8\sqrt{2} \mu_{11} \mu_{12} -16 \mu_{12} \mu_{21} \tan{\theta} \sin{\beta} = M_{\pi d}^2.
\eea

In the $SU(2)_V$ limit, where $\mu_{12} =\mu_{21}$ , $m_u^2 = m_d^2 = m^2$ and $\tan{\beta} =1$, the potential minimum condition reduces to $[m^2 +16\mu_{11}^2 -\mu_{11} B ] = -16 \mu_{11} \mu_{12} \cot{\theta}$. In this case, the mass matrix has eigenvalues corresponding to the massless Nambu-Goldstone boson which is absorbed by the $Z$ vector field and two physical massive pseudoscalars with values
\bea
m_{a}^2 &=& 2\mu_{11} B -16 \mu_{11} \mu_{12} \cot{\theta} \cr
m_{A}^2 &=& \left(16 \mu_{12}^2 -16 \mu_{11} \mu_{12} \cot{\theta}\right)\sec^2 {\theta} .
\eea
For $D$-flat direction stability of the potential, it is required that $m_a^2 >0$. As shall be seen,  the scalar sector stability condition requires that $\mu_{11} \mu_{12}<0$. Hence, as long as $m_A^2 -m_a^2 = 16\mu_{12}^2\sec^2 \theta +2b -16 \mu_{11}\mu_{12}\tan\theta >0$, the mass $m_a$ corresponds to the lightest pseudoscalar in this limit.  
The scalar Higgs mass squared matrix in the $(S_u , S_d , S_\pi )$ basis can be written as
\be
M_{\rm S}^2 = M_{\rm PS}^2 + \Delta M_{\rm S}^2 
\ee
with
\be
\Delta M_{\rm S}^2 = \begin{pmatrix}
\Delta M_{uu}^2 & \Delta M_{ud}^2 &\Delta M_{u\pi}^2 \\
\Delta M_{du}^2 & \Delta M_{dd}^2 & \Delta M_{d\pi}^2 \\
\Delta M_{\pi u}^2 & \Delta M_{\pi d}^2 &\Delta M_{\pi\pi}^2
\end{pmatrix}
\ee
where
\bea
\Delta M_{uu}^2 &=&M_Z^2 \sin^2{\theta} \sin^2{\beta} \cr 
\Delta M_{dd}^2 &=&M_Z^2 \sin^2{\theta} \cos^2{\beta} \cr 
\Delta M_{\pi\pi}^2 &=&M_Z^2 \cos^2{\theta} +16\left( \mu_{12}^2 +  \mu_{21}^2 \right)+ 16 \sqrt{2}\mu_{11} \tan{\theta} \left( \mu_{12} \cos{\beta} +\mu_{21} \sin{\beta}\right) \cr
\Delta M_{ud}^2 &=&-\frac{1}{2} M_Z^2 \sin^2{\theta} \sin{2\beta}-2\mu_{11} B =\Delta M_{du}^2 \cr
\Delta M_{u\pi}^2 &=& -\frac{1}{2} M_Z^2 \sin{2\theta} \sin{\beta}+ 16 \mu_{12} \tan{\theta} \left( \mu_{12} \sin{\beta} -\mu_{21} \cos{\beta}\right) = \Delta M_{\pi u}^2 \cr
\Delta M_{d\pi}^2 &=&\frac{1}{2} M_Z^2 \sin{2\theta} \cos{\beta}+ 16 \mu_{21} \tan{\theta} \left( \mu_{12} \sin{\beta} -\mu_{21} \cos{\beta}\right)  = \Delta M_{\pi d}^2  .
\eea

\begin{figure*}
\begin{center}
$\begin{array}{cc}
\includegraphics[scale=1.00]{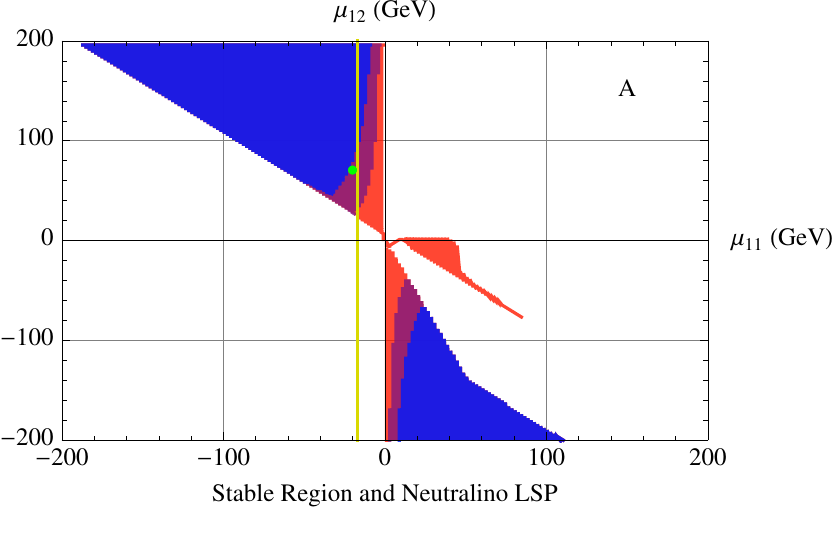} &
\includegraphics[scale=1.00]{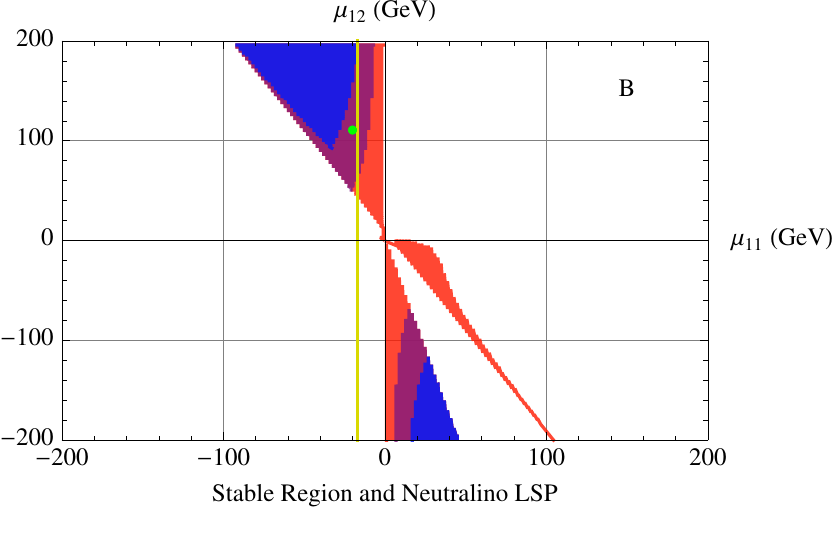} \\
\includegraphics[scale=1.00]{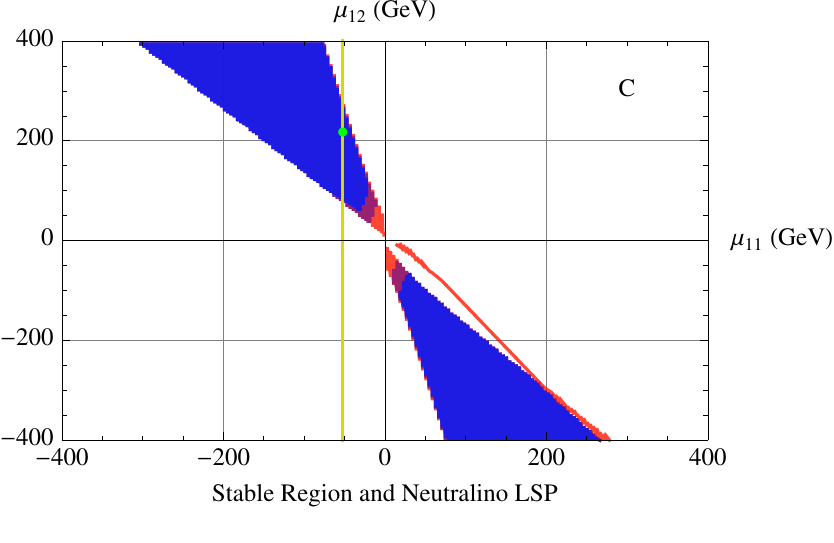} &
\includegraphics[scale=1.00]{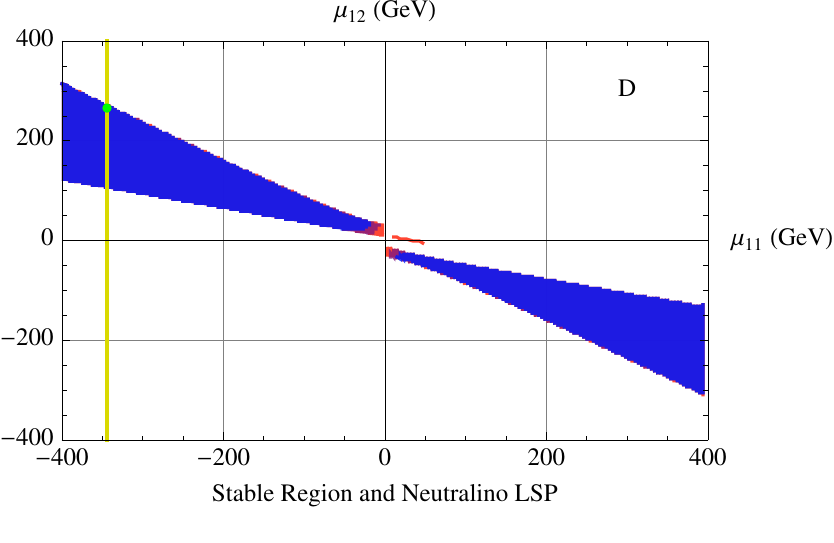} 
\end{array}$
\caption{The requirement that a neutralino is the LSP further delineates the stability regions of Fig.~\ref{StabilityRegion} as shown here for the same slices of parameter space. The green dots indicate the points in parameters space associated with the detailed mass spectrum in Fig. \ref{MassSpectrum}. The yellow lines indicate the value of $\mu_{11}$ along which the parameter $\mu_{12}$ is scanned in  the subsequent mass spectrum plots. For each plot the value of the gaugino SUSY breaking masses are $M_1 =200$ GeV and $M_2=800$ GeV.  }
\label{StabilityandLSPRegion}
\end{center}
\end{figure*}

In the $SU(2)_V$ limit, stability requires that $\mu_{11} \mu_{12}<0$. The smallest eigenvector of this matrix corresponds to an $SU(2)_V$ singlet which can be identified as the lightest Higgs scalar boson  with mass squared
\be
m_h^2 =-16 \mu_{11} \mu_{12} \cot{\theta}=m_a^2 + 2b = m_A^2\cos^2\theta -16\mu_{12}^2 .
\ee
In this limit, the mass of the lightest Higgs is lighter than the heaviest pseudoscalar but heavier or lighter than the lightest pseudoscalar depending on the sign of $b$. After extracting the contribution of this singlet, the remainder of the scalar mass squared matrix can be combined into a $2\times 2$ matrix denoted as $m_s^2$.  Since $\tan{\beta} =1$ is a $D$-flat direction, the stability of the potential against runaway moduli is guaranteed by the mass squared (second derivatives of the potential) matrix having positive eigenvalues.  Since the eigenvalues are given by 
\be
m_\pm^2 = \frac{1}{2} \left[ {\rm Tr}~m_s^2 \pm \sqrt{\left( {\rm Tr}~m_s^2 \right)^2 -4{\rm det}~m_s^2 ~}\right],
\ee
their reality requires $({\rm Tr}~m_s^2)^2 >4 {\rm det}~m_s^2$ and their positivity leads to ${\rm det}~m_s^2 >0$.  The expressions for the trace and determinant are readily extracted as
\bea
{\rm Tr}~m_s^2 &=& M_Z^2 -2b +16 \mu_{12}^2 [3 + \tan^2{\theta} ]-32 \mu_{11} \mu_{12} \cot{2\theta} \cr
{\rm det}~ m_s^2 &=& 16 M_Z^2 \left( \mu_{12}^2 -\mu_{11} \mu_{12} \cot{\theta} \right) \sec^2{\theta} + 2 \mu_{11} B \left[ M_Z^2 \cos^2{\theta} +16 \left( \mu_{12}^2 -\mu_{11} \mu_{12} \cot{\theta} \right) \tan^2{\theta} \right] \cr
 & &\qquad\qquad +32 \left( \mu_{12}^2 +\mu_{11} \mu_{12} \tan{\theta} \right)\left[ M_Z^2 \sin^2{\theta} +2\mu_{11} B +16 \left( \mu_{12}^2 -\mu_{11} \mu_{12} \cot{\theta} \right)  \right] .
\eea
The region of stability can be mapped out for various parameters. If $\mu_{12}^2$ corresponds to the largest mass squared parameter, the trace and determinant simplify to 
\bea
{\rm Tr}~m_s^2 &\approx& 16 \mu_{12}^2 [ 3+\tan^2{\theta} ]\cr
({\rm Tr}~m_s^2 )^2 > 4 ~{\rm det} m_s^2 &\approx& 8 (16 \mu_{12}^2)^2 >0 ,
\eea
with the heavier 2 neutral Higgs fields having mass squares (with $m_A^2 \approx 16\mu_{12}^2\sec^2\theta$)
\bea
m_{H1}^2 &\approx& \frac{1}{2}m_A^2\cos^2\theta \left[3+\tan^2{\theta} + \sqrt{(3+\tan^2{\theta})^2 -8~}  \right] \cr
m_{H2}^2 &\approx& \frac{1}{2}m_A^2 \cos^2\theta \left[3+\tan^2{\theta} - \sqrt{(3+\tan^2{\theta})^2 -8~}  \right] .
\eea

In an analogous fashion, the charged Higgs mass squared matrix, denoted $M_{Ch}^2$, can also be obtained from the potential curvature at the minimum.  The matrix and its elements in the $(H_u^+ ,H_d^{-\dagger} , \pi^+ ,\pi^{-\dagger} )$ basis are given by
\be
M_{Ch}^2 =\begin{pmatrix}
M_{u^+ \bar{u}^+}^2 & M_{u^+ d^-}^2 & M_{u^+ \bar\pi^+}^2 & M_{u^+ \pi^-}^2\\
M_{\bar{d}^- \bar{u}^+}^2 & M_{\bar{d}^- d^-}^2 & M_{\bar{d}^- \bar\pi^+}^2 & M_{\bar{d}^- \pi^-}^2\\
M_{\pi^+ \bar{u}^+}^2 & M_{\pi^+ d^-}^2 & M_{\pi^+ \bar\pi^+}^2 & M_{\pi^+ \pi^-}^2\\
M_{\bar\pi^- \bar{u}^+}^2 & M_{\bar\pi^- d^-}^2 & M_{\bar\pi^- \bar\pi^+}^2 & M_{\bar\pi^- \pi^-}^2\\
\end{pmatrix}
\ee
where
\bea
M_{u^+ \bar{u}^+}^2 &=& M_W^2 \sin^2{\theta} \cos^2{\beta} +\mu_{11} B \cot{\beta} +8\mu_{12} \mu_{21} \cot{\beta} +8\mu_{12}^2 -8\sqrt{2}\mu_{11} \mu_{21} \cot{\theta} \csc{\beta} \cr
M_{u^+ d^-}^2 &=& \frac{1}{2} M_W^2 \sin^2{\theta} \sin{2\beta} + \mu_{11} B = M_{\bar{d}^- \bar{u}^+}^2 \cr
M_{u^+ \bar\pi^+}^2 &=&-\frac{1}{2\sqrt{2}} M_W^2 \sin{2\theta} \sin{\beta} -16i \mu_{11} \mu_{21} + 8i \sqrt{2} \mu_{12} \tan{\theta} \left( \mu_{12} \sin{\beta} + \mu_{21} \cos{\beta}  \right)= - M_{\pi^+ \bar{u}^+}^2 \cr
M_{u^+ \pi^-}^2 &=&\frac{i}{2\sqrt{2}} M_W^2 \sin{2\theta} \sin{\beta} = -M_{\bar\pi^- \bar{u}^+}^2  \cr
M_{\bar{d}^- d^-}^2 &=&M_W^2 \sin^2{\theta} \sin^2{\beta} +\mu_{11} B \tan{\beta} +8\mu_{12} \mu_{21} \tan{\beta} +8\mu_{21}^2 -8\sqrt{2}\mu_{11} \mu_{12} \cot{\theta} \sec{\beta} \cr
M_{\bar{d}^- \bar\pi^+}^2 &=& -\frac{i}{2\sqrt{2}} M_W^2 \sin{2\theta} \cos{\beta} = -M_{\pi^+ d^-}^2  \cr
M_{\bar{d}^- \pi^-}^2 &=&\frac{i}{2\sqrt{2}} M_W^2 \sin{2\theta} \cos{\beta} +16i \mu_{11} \mu_{12} - 8i \sqrt{2} \mu_{21} \tan{\theta} \left( \mu_{12} \sin{\beta} + \mu_{21} \cos{\beta}  \right)=-M_{\bar\pi^- d^-}^2 \cr
M_{\pi^+ \bar\pi^+}^2 &=& \frac{1}{2}M_W^2 \left[ \cos^2{\theta} +\sin^2{\theta} \left( 1-\tan^2{\theta_W} \right)\cos{2\beta} \right] +16\mu_{21}^2 + 8\tan^2{\theta} \left( \mu_{12} \sin{\beta} + \mu_{21}\cos{\beta} \right)^2  \cr
M_{\bar\pi^- \pi^-}^2 &=& \frac{1}{2}M_W^2 \left[ \cos^2{\theta} -\sin^2{\theta} \left( 1-\tan^2{\theta_W} \right)\cos{2\beta} \right] +16\mu_{12}^2 + 8\tan^2{\theta} \left( \mu_{12} \sin{\beta} + \mu_{21}\cos{\beta} \right)^2  \cr
M_{\pi^+ \pi^-}^2 &=& -\frac{1}{2}M_W^2 \cos^2{\theta} -8\left( \mu_{12}^2 + \mu_{21}^2 \right) - 8\sqrt{2} \mu_{11} \tan{\theta} \left( \mu_{12} \cos{\beta} + \mu_{21}\sin{\beta} \right) = M_{\bar\pi^- \bar\pi^+}^2  .
\eea

The sfermion mass matrices are obtained directly from the Lagrangian, Eqs.~ (\ref{LSUSYBreaking}), (\ref{LSYM}) and (\ref{Lsigma}).  The chargino mass matrix, denoted $M_{\rm Chino}$, in the $(\tilde{W}^+ , \tilde{H}_u^+ , \tilde{\pi}^+ )$ basis is found to be
\be
M_{\rm Chino} =\begin{pmatrix}
M_{W^+ W^-} & M_{W^+ d^-} & M_{W^+ \pi^-}\\
M_{u^+ W^-} & M_{u^+ d^-} & M_{u^+ \pi^-}\\
M_{\pi^+ W^-} & M_{\pi^+ d^-} & M_{\pi^+ \pi^-}\\
\end{pmatrix}
\ee
where
\bea
M_{W^+ W^-}&=& M_2 ~;~M_{W^+ d^-}= M_W \sqrt{2} \sin{\theta} \cos{\beta} ~;~M_{W^+ \pi^-}=iM_W \cos{\theta} \cr
M_{u^+ W^-} &=& M_W \sqrt{2} \sin{\theta} \sin{\beta} ~;~M_{u^+ d^-} = 4\mu_{11}  ~;~M_{u^+ \pi^-} = 4i\mu_{12}  \cr
M_{\pi^+ W^-} &=& iM_W \cos{\theta} ~;~M_{\pi^+ d^-}  = 4i \mu_{21} ~;~M_{\pi^+ \pi^-} = 2\sqrt{2} \tan{\theta} \left( \mu_{12} \sin{\beta} + \mu_{21} \cos{\beta} \right)  .
\eea

\begin{figure*}
\begin{center}
$\begin{array}{cc}
\includegraphics[scale=0.65]{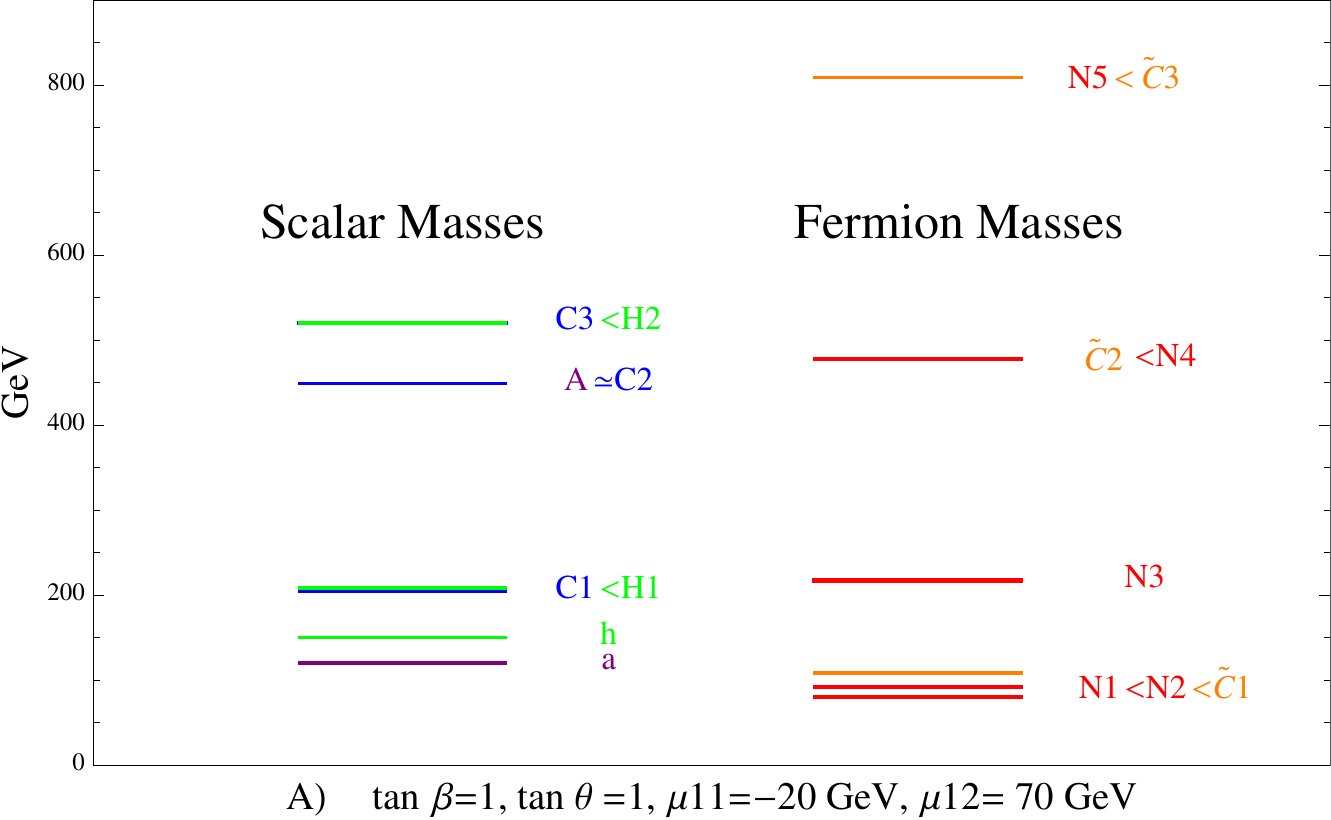} &
\includegraphics[scale=0.65]{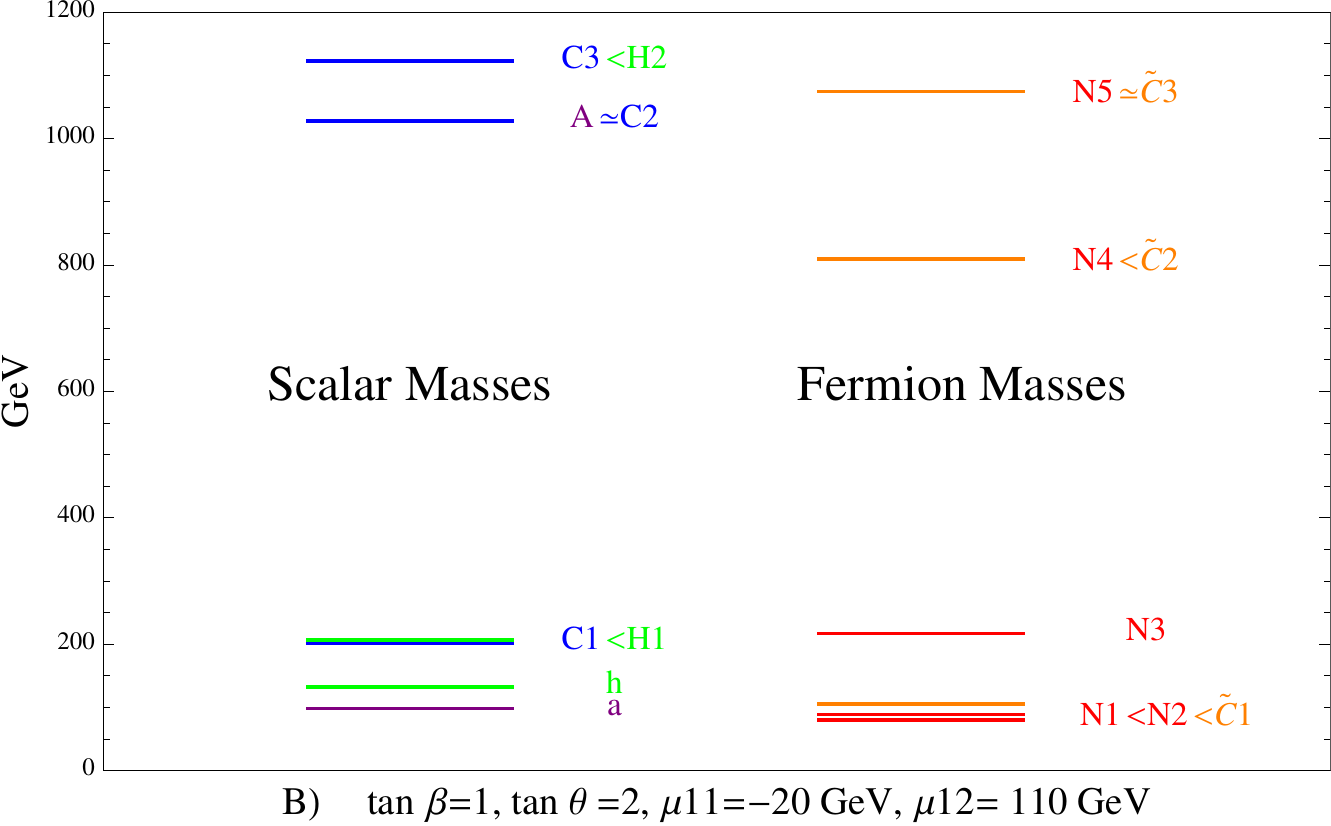} \\
\includegraphics[scale=0.65]{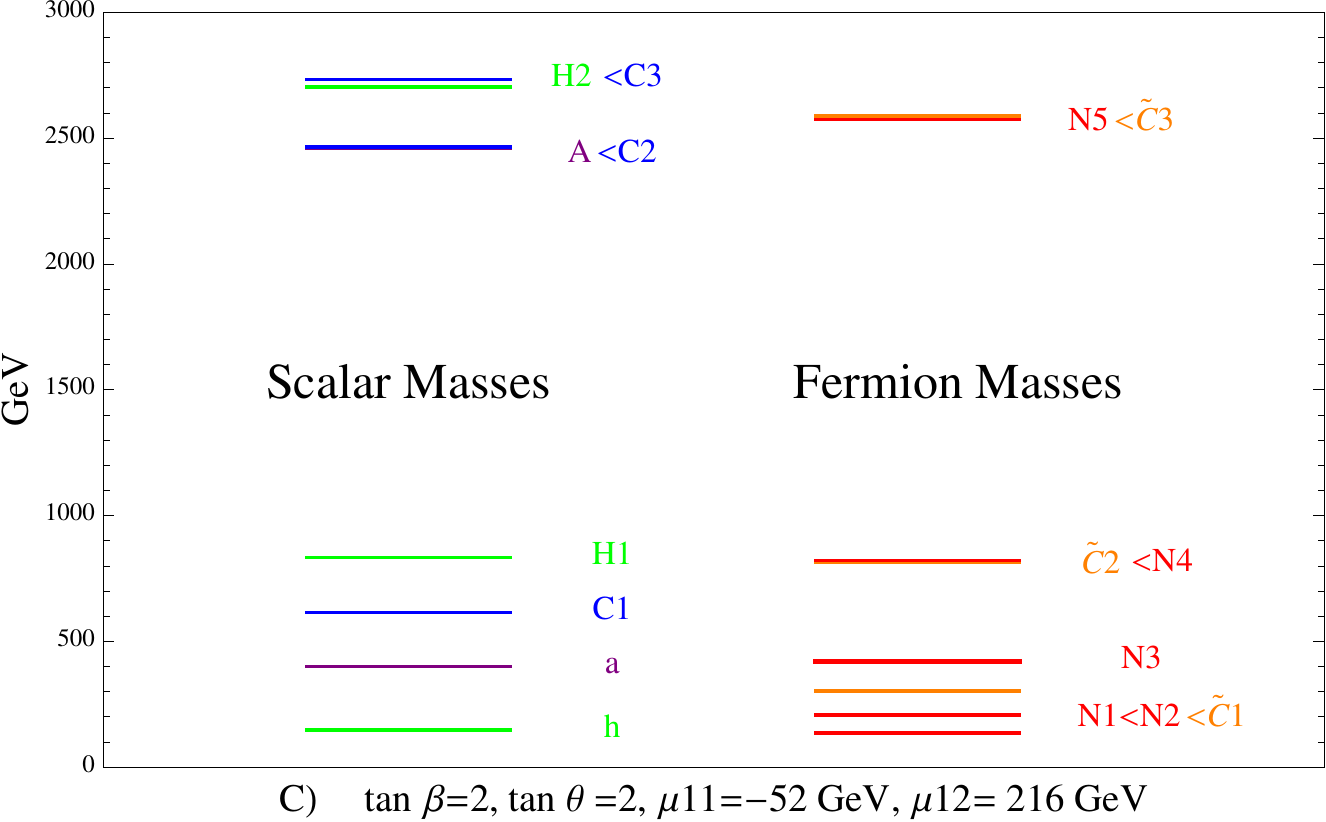} &
\includegraphics[scale=0.65]{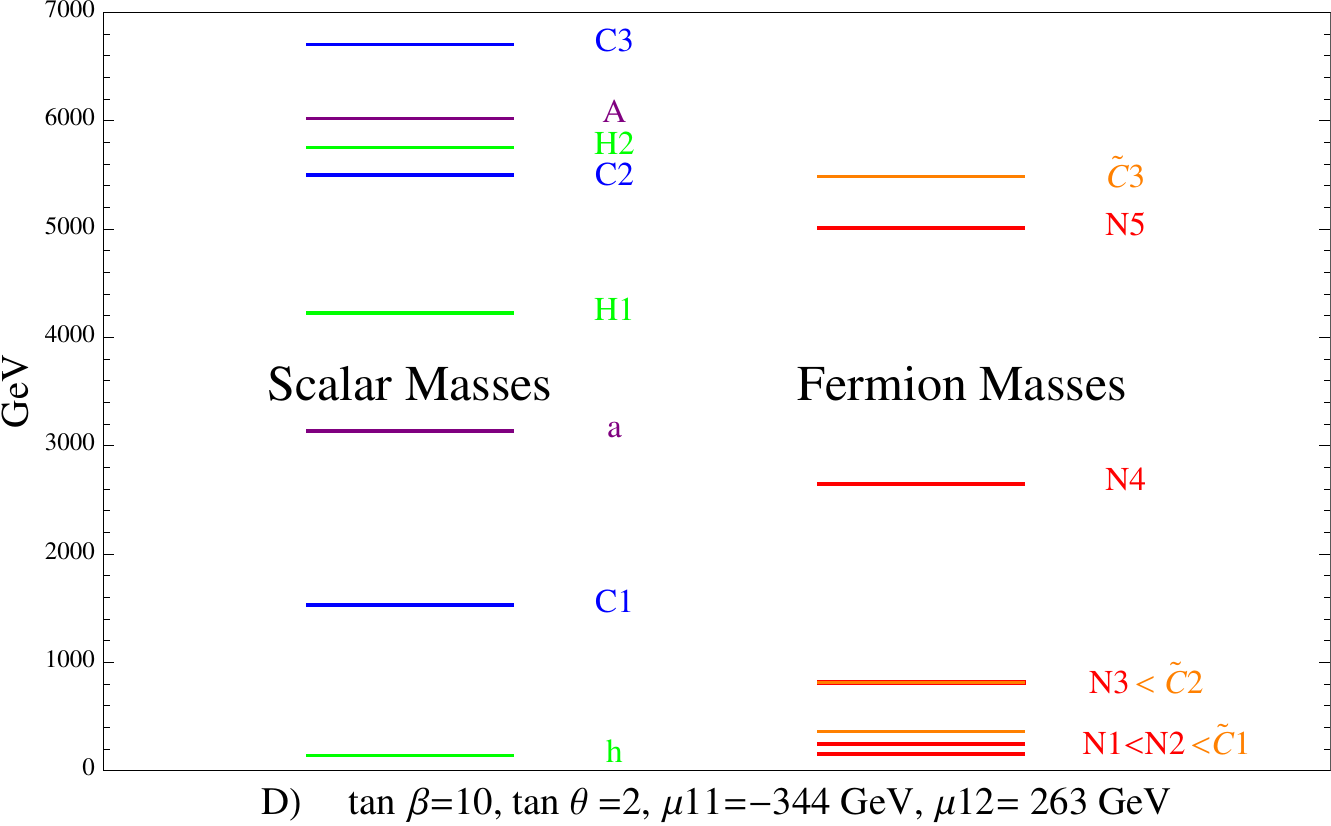} 
\end{array}$
\caption{The Higgs (pseudo-) scalars and gaugino-Higgsino mass spectrum for a point in the LSP-stability regions indicated by the green dot in 
Fig.~\ref{StabilityandLSPRegion}.  The gaugino soft SUSY breaking masses are $M_1 =200$ GeV and $M_2=800$ GeV, and $b=4,000 $ ${\rm GeV}^2$ for all regions.}
\label{MassSpectrum}
\end{center}
\end{figure*}

There are five neutralino fields with their mass matrix in the $(\lambda , \tilde{Z} , \tilde{H}_u^0 , \tilde{H}_d^0 , \tilde\pi^0 )$ basis given by
\be
M_{\rm Nino} =\begin{pmatrix}
\tilde{M}_{\gamma \gamma} & \tilde{M}_{\gamma Z} & \tilde{M}_{\gamma u} & \tilde{M}_{\gamma d} & \tilde{M}_{\gamma \pi}\\
\tilde{M}_{Z \gamma} & \tilde{M}_{Z Z} & \tilde{M}_{Z u} & \tilde{M}_{Z d} & \tilde{M}_{Z \pi}\\
\tilde{M}_{u \gamma} & \tilde{M}_{u Z} & \tilde{M}_{u u} & \tilde{M}_{u d} & \tilde{M}_{T \pi}\\
\tilde{M}_{d \gamma} & \tilde{M}_{d Z} & \tilde{M}_{d u} & \tilde{M}_{d d} & \tilde{M}_{d \pi}\\
\tilde{M}_{\pi \gamma} & \tilde{M}_{\pi Z} & \tilde{M}_{\pi u} & \tilde{M}_{\pi d} & \tilde{M}_{\pi \pi}\\
\end{pmatrix}
\ee
where
\bea
\tilde{M}_{\gamma \gamma} &=& m_{\gamma \gamma} ~;~\tilde{M}_{\gamma Z}=m_{\gamma Z} ~;~\tilde{M}_{\gamma u}=0\cr
\tilde{M}_{\gamma B} &=&0 ~;~\tilde{M}_{\gamma \pi}=0 ~;~\tilde{M}_{Z \gamma} = m_{Z \gamma } \cr
\tilde{M}_{Z Z} &=&m_{ZZ} ~;~\tilde{M}_{Z u} =-M_Z \sin{\theta} \sin{\beta} ~;~\tilde{M}_{Z d} =M_Z \sin{\theta} \cos{\beta} ~;~\tilde{M}_{Z \pi}=i M_Z \cos{\theta} \cr
\tilde{M}_{u \gamma} &=&0~;~ \tilde{M}_{u Z} =-M_Z \sin{\theta} \sin{\beta} ~;~\tilde{M}_{u u} =0\cr
\tilde{M}_{u d} &=&-4\mu_{11} ~;~ \tilde{M}_{T \pi}=-2i\sqrt{2} \mu_{12} ~;~ \tilde{M}_{d \gamma} = 0\cr
\tilde{M}_{d Z} &=&+M_Z \sin{\theta} \cos{\beta} ~;~ \tilde{M}_{d u} =-4\mu_{11} ~;~ \tilde{M}_{d d} =0\cr
\tilde{M}_{d \pi}&=&i2\sqrt{2} \mu_{21} ~;~ \tilde{M}_{\pi \gamma} = 0 ~;~ \tilde{M}_{\pi Z} =iM_Z \cos{\theta} \cr
\tilde{M}_{\pi u} &=&-2i\sqrt{2} \mu_{12} ~;~ \tilde{M}_{\pi d} = 2i\sqrt{2} \mu_{21} ~;~ \tilde{M}_{\pi \pi}=2\sqrt{2} \tan{\theta} \left( \mu_{12} \sin{\beta} + \mu_{21} \cos{\beta} \right) ,
\eea
with the SUSY breaking gaugino masses defined as
\bea
m_{\gamma \gamma} &=& M_1 \cos^2{\theta_W} + M_2 \sin^2{\theta_W} \cr
m_{ZZ} &=& M_1 \sin^2{\theta_W} + M_2 \cos^2{\theta_W} \cr
m_{\gamma Z} &=& \left( M_2 -M_1 \right) \sin{\theta_W}\cos{\theta_W} =m_{Z \gamma}  .
\eea

\begin{figure*}
\begin{center}
$\begin{array}{cc}
\includegraphics[scale=1.00]{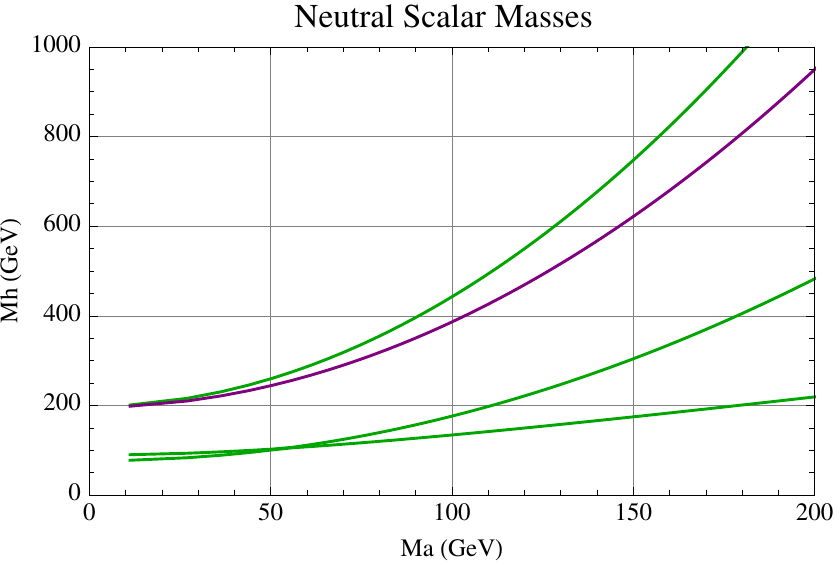} &
\includegraphics[scale=1.00]{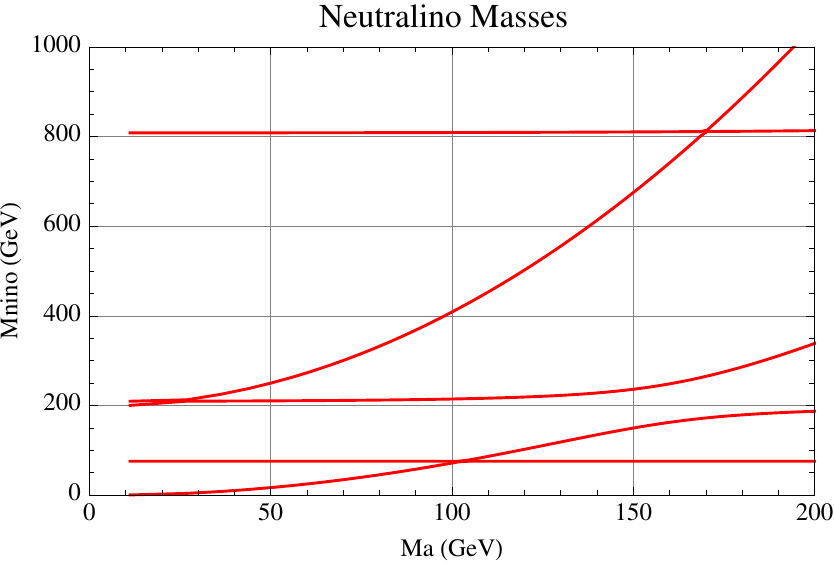} \\
\includegraphics[scale=1.00]{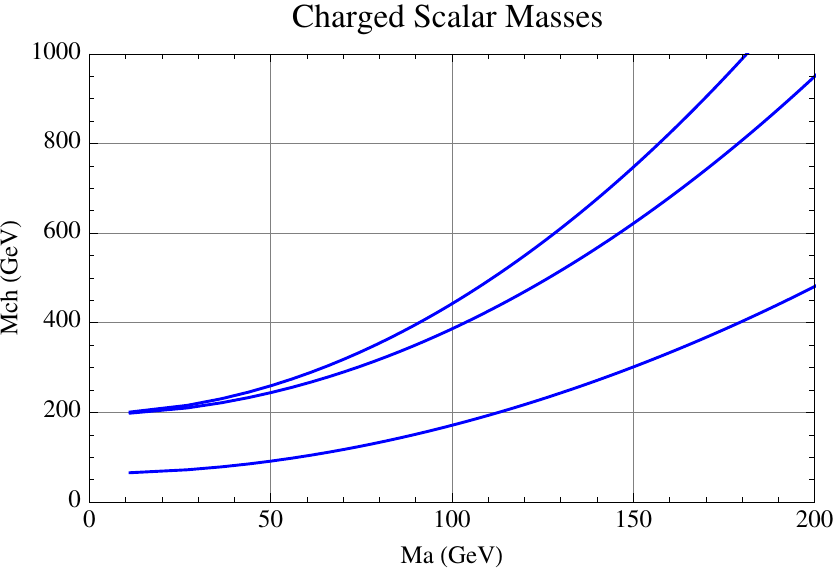} &
\includegraphics[scale=1.00]{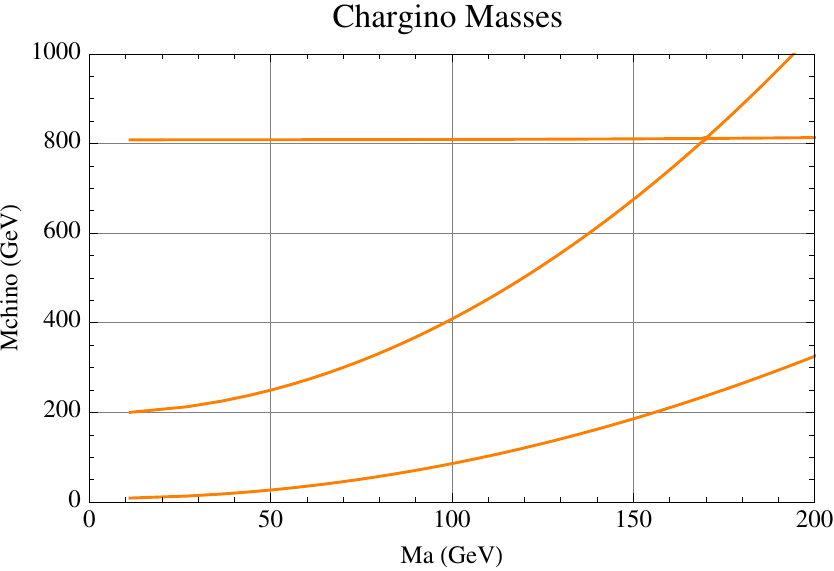} 
\end{array}$
\caption{Masses  as a function of the lightest pseudoscalar mass $m_a$ for a $\mu_{12}$ scan along the yellow line across region A in Fig. \ref{StabilityandLSPRegion}.  The parameters for the plots are $\tan{\beta}=1$, $\tan{\theta}=1$, $b=4,000$ ${\rm GeV^2}$ and $\mu_{11} =-12$ GeV. In the top left panel green curves correspond to scalar $h, H1, H2$ masses, while the purple curve corresponds to the  pseudoscalar 
$A$ mass. In the bottom left panel, the blue curves correspond to the charged Higgs C1, C2, C3 masses. In the top right panel, the red curves correspond to the neutralino $N1-N5$ masses, while the orange curves in the lower right panel correspond to the chargino $\tilde{C1}, \tilde{C2}, \tilde{C3}$ masses.}
\label{MassesTanBeta1}
\end{center}
\end{figure*}

\begin{figure*}
\begin{center}
$\begin{array}{cc}
\includegraphics[scale=1.00]{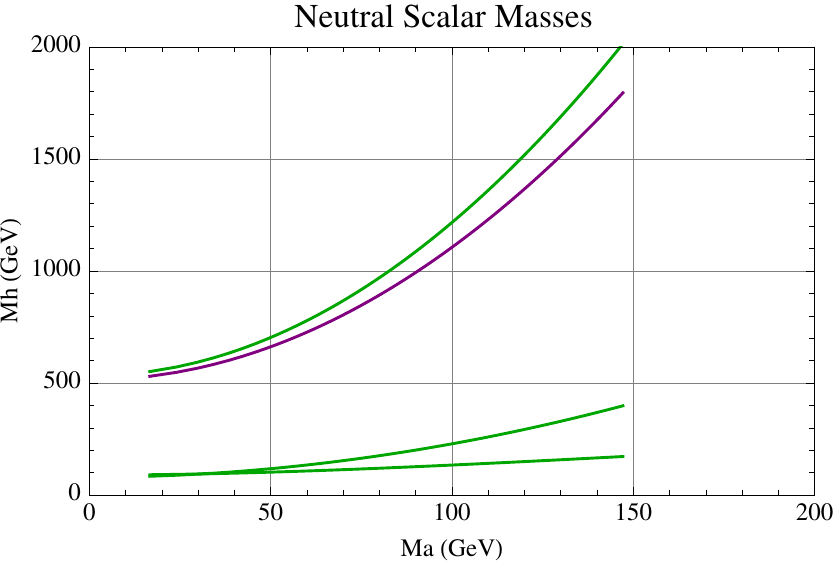} &
\includegraphics[scale=1.00]{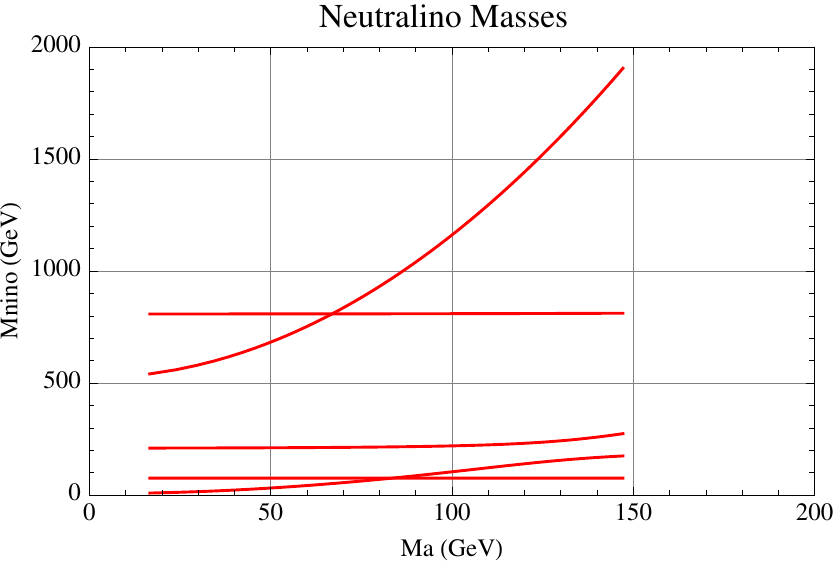} \\
\includegraphics[scale=1.00]{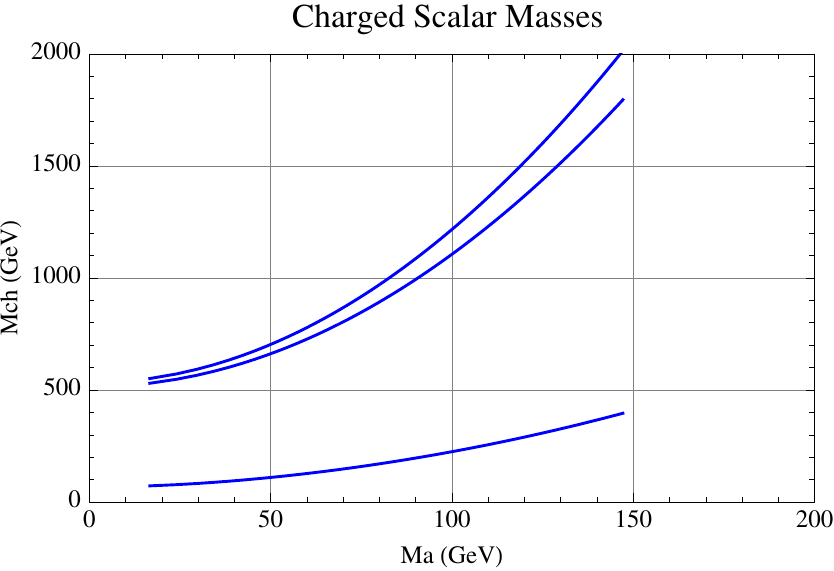} &
\includegraphics[scale=1.00]{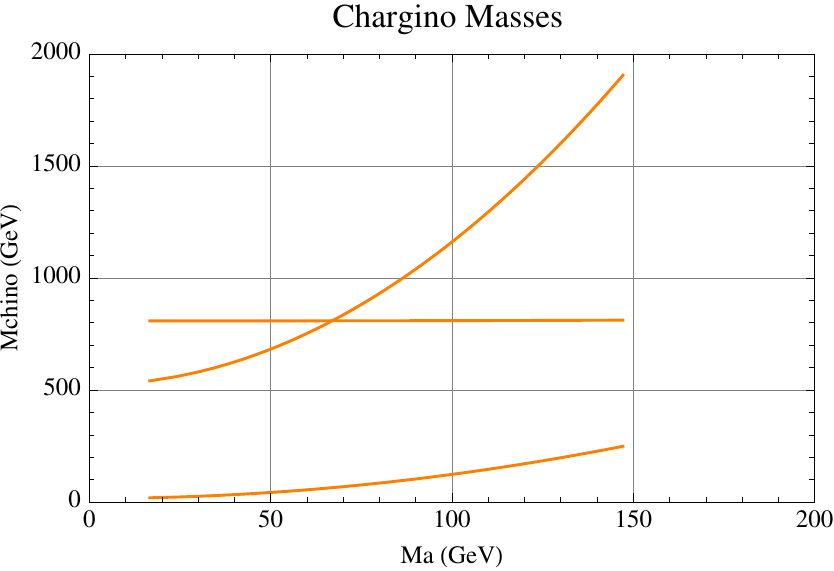} 
\end{array}$
\caption{Masses as a function of the lightest pseudoscalar mass $m_a$  for a $\mu_{12}$ scan along the yellow line across region B in Fig. \ref{StabilityandLSPRegion}.  The parameters for the plots are $\tan{\beta}=1$, $\tan{\theta}=2$, $b=4,000$ ${\rm GeV^2}$ and $\mu_{11} =-16$ GeV. The  curves correspond to the various particles  just as described in the caption to Fig. \ref{MassesTanBeta1}. }
\label{MassesTanBeta2}
\end{center}
\end{figure*}

\begin{figure*}
\begin{center}
$\begin{array}{cc}
\includegraphics[scale=1.00]{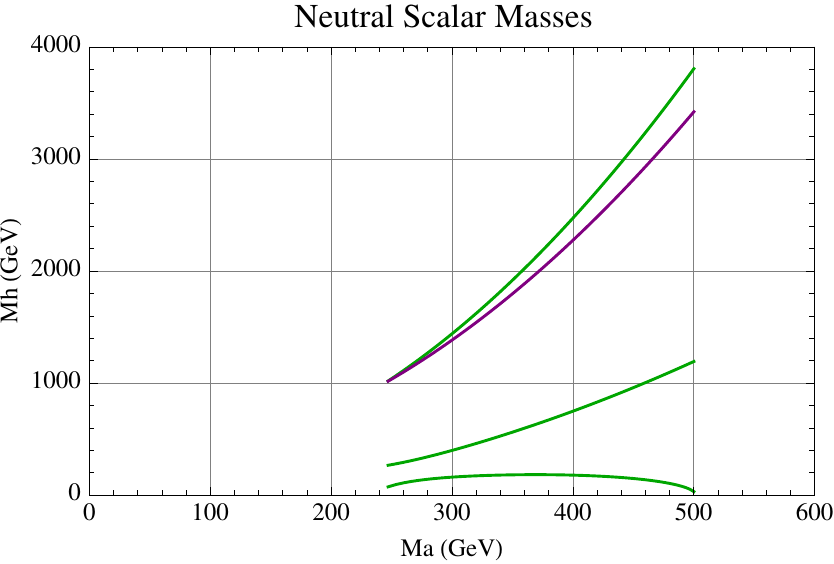} &
\includegraphics[scale=1.00]{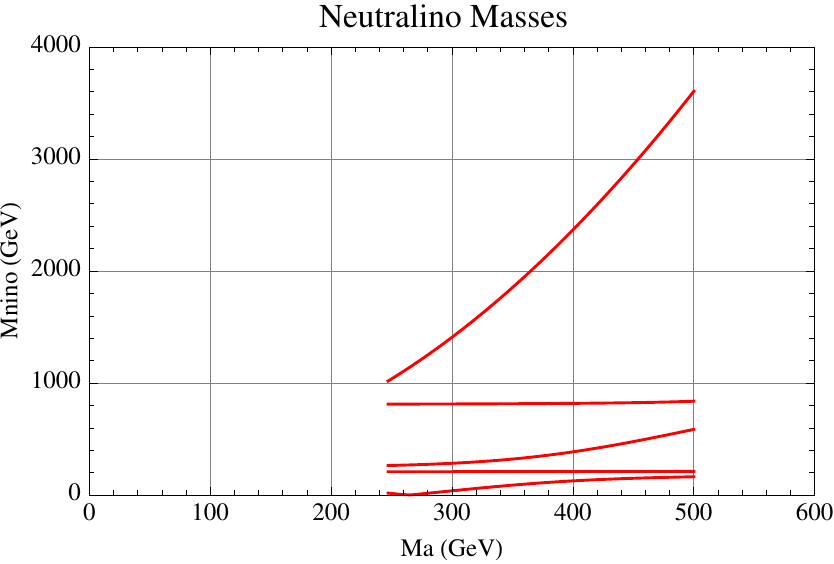} \\
\includegraphics[scale=1.00]{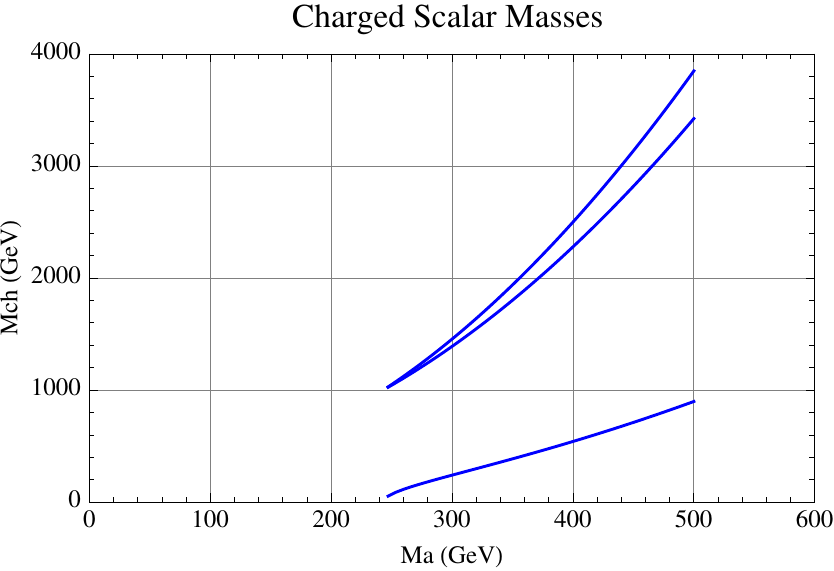} &
\includegraphics[scale=1.00]{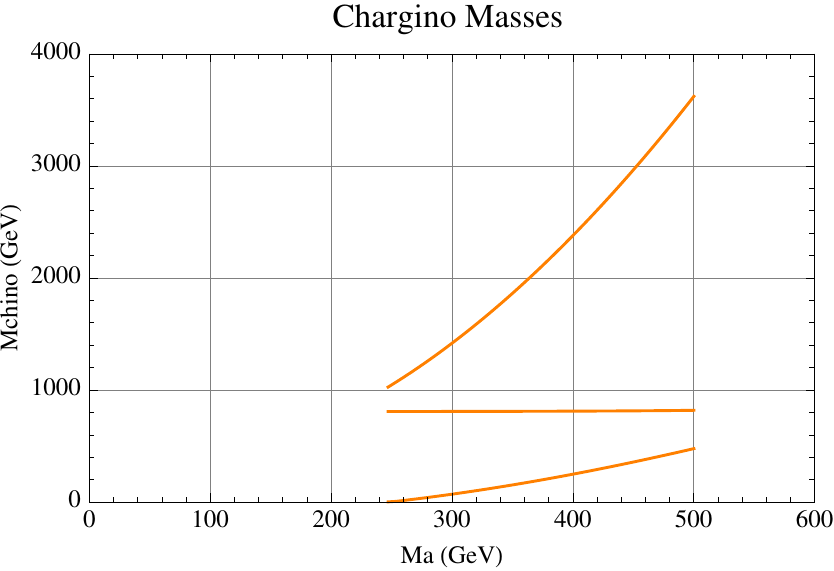} 
\end{array}$
\caption{Masses as a function of the lightest pseudoscalar mass $m_a$ for a $\mu_{12}$ scan along the yellow line across region C in Fig. \ref{StabilityandLSPRegion} .  The parameters for the plots are $\tan{\beta}=2$, $\tan{\theta}=2$, $b=4,000$ ${\rm GeV^2}$ and $\mu_{11} =-52$ GeV. The  curves correspond to the various particles  just as described in the caption to Fig. \ref{MassesTanBeta1}.}
\label{MassesTanBeta10}
\end{center}
\end{figure*}

\begin{figure*}
\begin{center}
$\begin{array}{cc}
\includegraphics[scale=1.00]{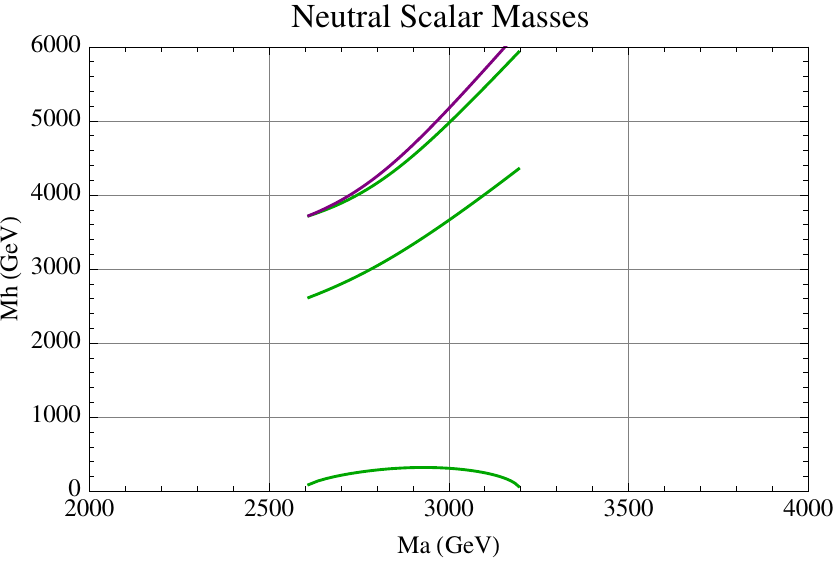} &
\includegraphics[scale=1.00]{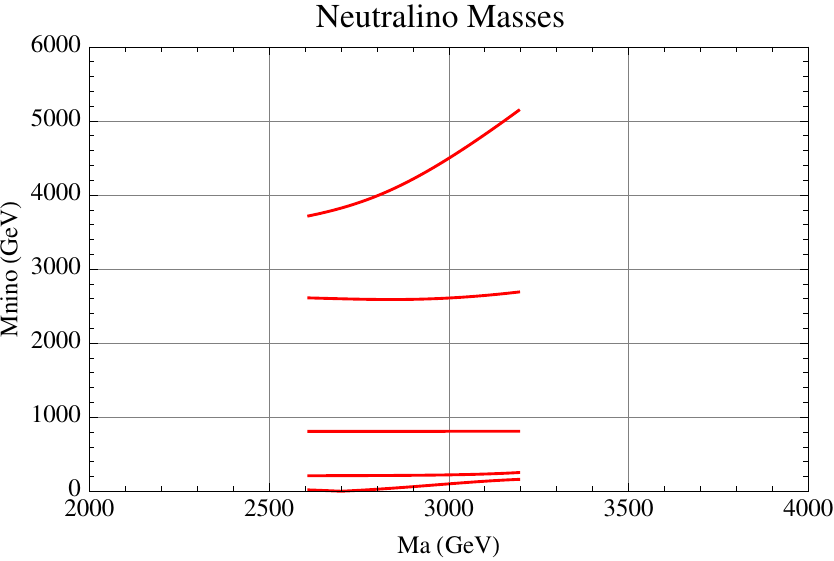} \\
\includegraphics[scale=1.00]{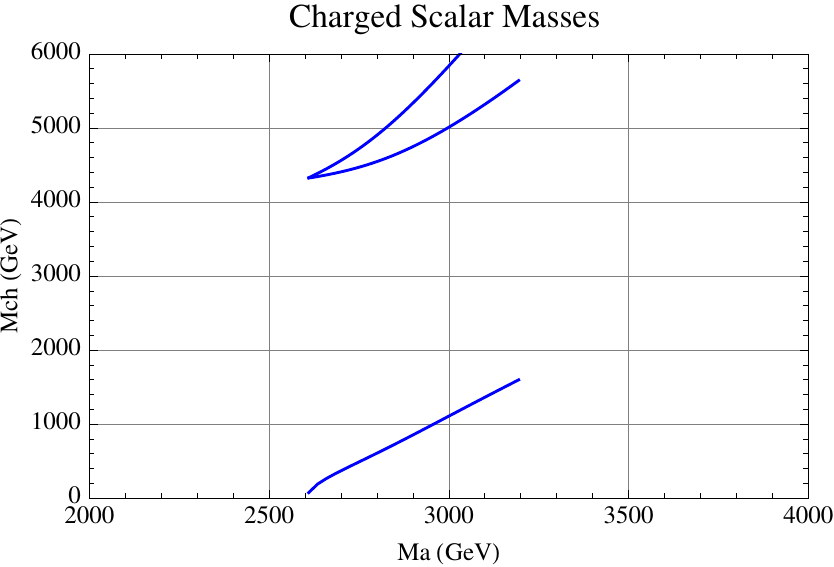} &
\includegraphics[scale=1.00]{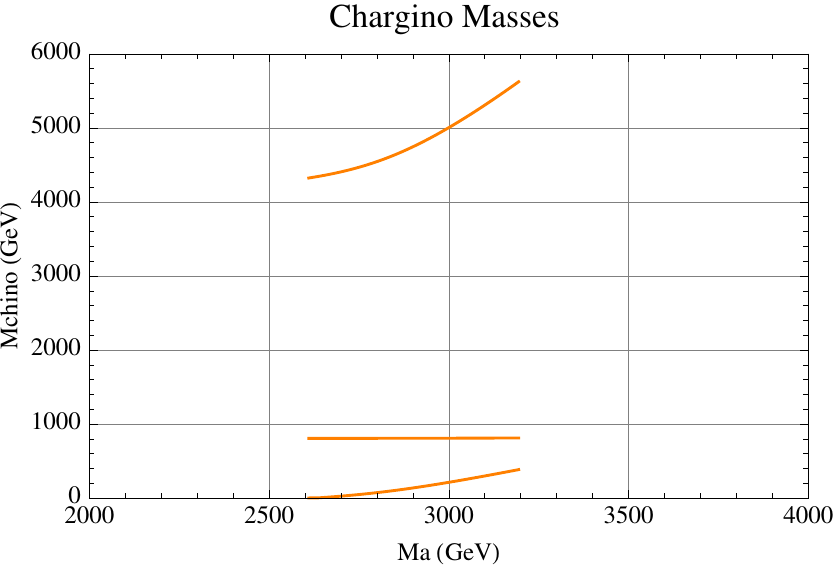} 
\end{array}$
\caption{Masses  as a function of the lightest pseudoscalar mass $m_a$ for a $\mu_{12}$ scan along the yellow line across region D in Fig. \ref{StabilityandLSPRegion}.  The parameters for the plots are $\tan{\beta}=10$, $\tan{\theta}=2$, $b=4,000$ ${\rm GeV^2}$ and $\mu_{11} =-344$ GeV. The  curves correspond to the various particles  just as described in the caption to Fig. \ref{MassesTanBeta1}.}
\label{MassesNegativeb}
\end{center}
\end{figure*}

The stability region in parameter space is determined by requiring all scalar squared masses to be positive.  Four typical stability regions, denoted as  A, B, C, and D, are exhibited in Fig.~\ref{StabilityRegion} in the $\mu_{11}$ -- $\mu_{12}$ plane. For each panel in the figure the value of the gaugino SUSY breaking masses are $M_1 =200$ GeV and $M_2=800$ GeV.  Stability region A has $\tan{\beta}=1$, $\tan{\theta}=1$, region B has $\tan{\beta}=1$, $\tan{\theta}=2$, region C has $\tan{\beta}=2$, $\tan{\theta}=2$, and region D has $\tan{\beta}=10$, $\tan{\theta}=2$. Each region is considered for three values of the SUSY breaking parameter $b=-4,000,~4,000,~12,000$ ${\rm GeV^2}$.  
Additional delineation in parameter space is obtained when a neutralino is required to be the LSP as illustrated in Fig.~\ref{StabilityandLSPRegion} for the same four regions of parameter space. In general, the eigenvalues of the mass matrices must be determined numerically. Detailed mass spectra for specific points in parameter space indicated by  green dots in Fig. \ref{StabilityandLSPRegion} are displayed in Fig. \ref{MassSpectrum}. Note that the lightest spin zero particle can be either the neutral pseudoscalar $a$ (panels A,B) or the neutral scalar $h$ (panels C, D). The next heaviest neutral pseudoscalar is denoted by $A$, while the remaining neutral scalars in order of increasing mass  are denoted as $H1, H2$. Adapting a similar notation, the neutralinos in order of increasing mass are denoted as ${N1}, {N2}, {N3}, {N4}, {N5}$, while  the charged scalars (charginos) are  $C1, C2, C3$ ($\tilde{C1}, \tilde{C2}, \tilde{C3}$).

To further explore the mass spectra, the neutral (pseudo-)scalar, charged scalar, neutralino, and chargino masses as a function of the lightest pseudoscalar mass are exhibited in  Figs. \ref{MassesTanBeta1} -- \ref{MassesNegativeb}. The various curves in the figures follow the parameter scans from left to right for fixed $\mu_{11}$ with increasing $\mu_{12}$ over the range  indicated by the yellow lines in  Fig.~\ref{StabilityandLSPRegion} for each of the four regions A, B, C, and D. The left endpoint of all the curves in each of the figures is dictated by the stability bounds as is the right endpoint of the curves in Figs. \ref{MassesTanBeta10}-\ref{MassesNegativeb}. On the other hand the right endpoints of the curves in Fig. \ref{MassesTanBeta2} corresponds to the maximum value for $\mu_{12}$ plotted in Fig. \ref{StabilityandLSPRegion}. Note that in regions A and B $\tan \beta =1$. In these regions the $U(1)$ gauge coupling  forms the only breaking of the global $SU(2)_V$ symmetry, and as a consequence some near degeneracies in the  mass spectra occur. Appendix A includes the explicit form of certain masses  and eigenvectors in the $SU(2)_V$ limit. 
All four panels allow for a lightest Higgs boson, $h$, with mass greater than $130$ GeV. Using the experimental bound\cite{PDG} on the lightest MSSM pseudo-scalar of $m_a > 94.3$ GeV as the bound for the current model, we see that  region $A$ allows a lightest Higgs boson tree level mass in the range $130 ~{\rm GeV}~ <m_h< 200$ GeV which corresponds to the range $94.3 ~{\rm GeV} ~<m_a < 180$ GeV, while for region $B$, the lightest Higgs boson mass varies from  $130 ~{\rm GeV}~ <m_h< 172$ GeV as $m_a$ ranges from  $94.3 ~{\rm GeV} ~<m_a < 148$ GeV over the scanned region.
A lightest Higgs scalar with a mass in the range $115 ~{\rm GeV}~ <m_h< 130 $ GeV is also allowed provided  different (SUSY breaking) parameters are employed. For the scans considered, region $C$ admits a lightest Higgs boson mass in a range from $182 ~{\rm GeV}~> m_h > 115$ GeV as $m_a$ varies from $370 ~{\rm GeV}~ < m_a < 475$ GeV. For $m_a$ less than around 350 GeV, there is some conflict with the current experimental limit on the mass of the lightest chargino. Finally region $D$  admits a lightest Higgs boson mass in a range from $200 ~{\rm GeV}~>m_h> 115$ GeV as $m_a$ varies from $3140 ~{\rm GeV}~ < m_a < 3180$ GeV. For $m_a$ less than around 3000 GeV, there is some tension with the current experimental limit on the mass of the lightest chargino and/or neutralino.

\begin{figure*}
\begin{center}
$\begin{array}{cc}
\includegraphics[scale=0.75]{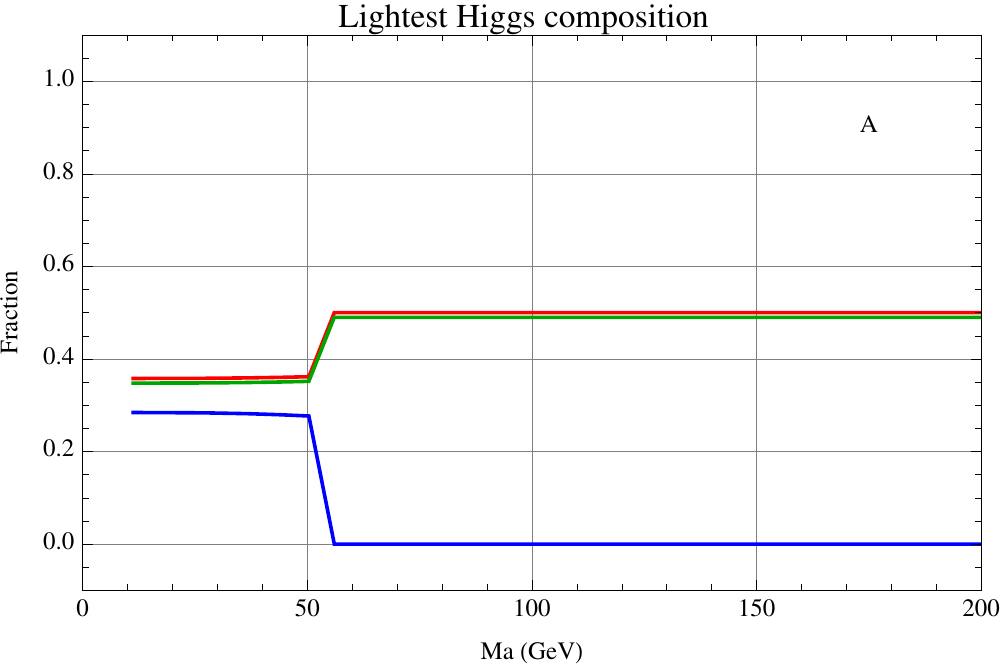} &
\includegraphics[scale=0.75]{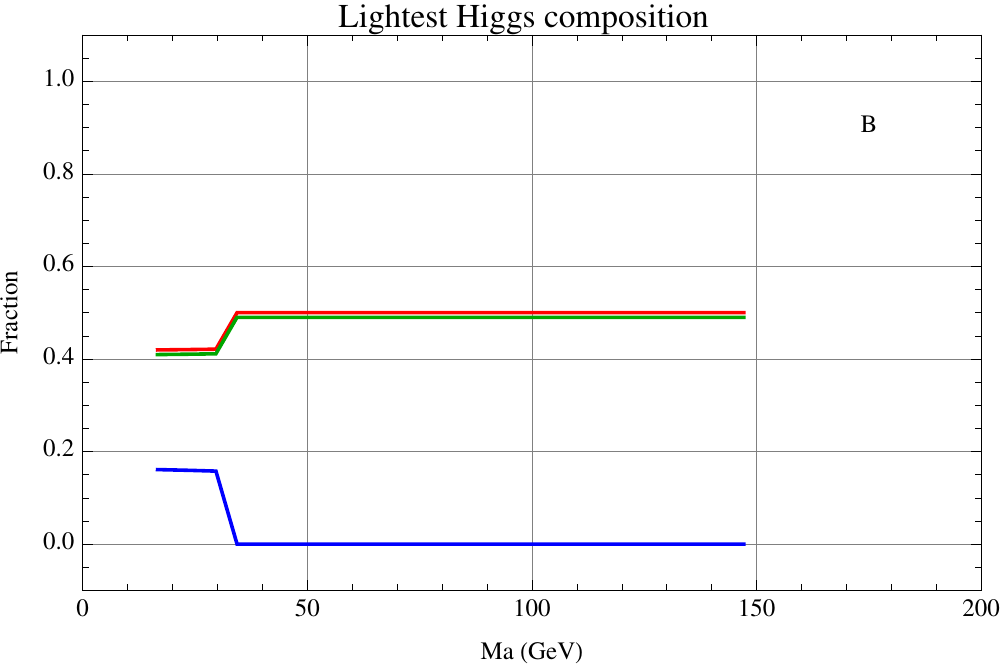} \\
\includegraphics[scale=0.75]{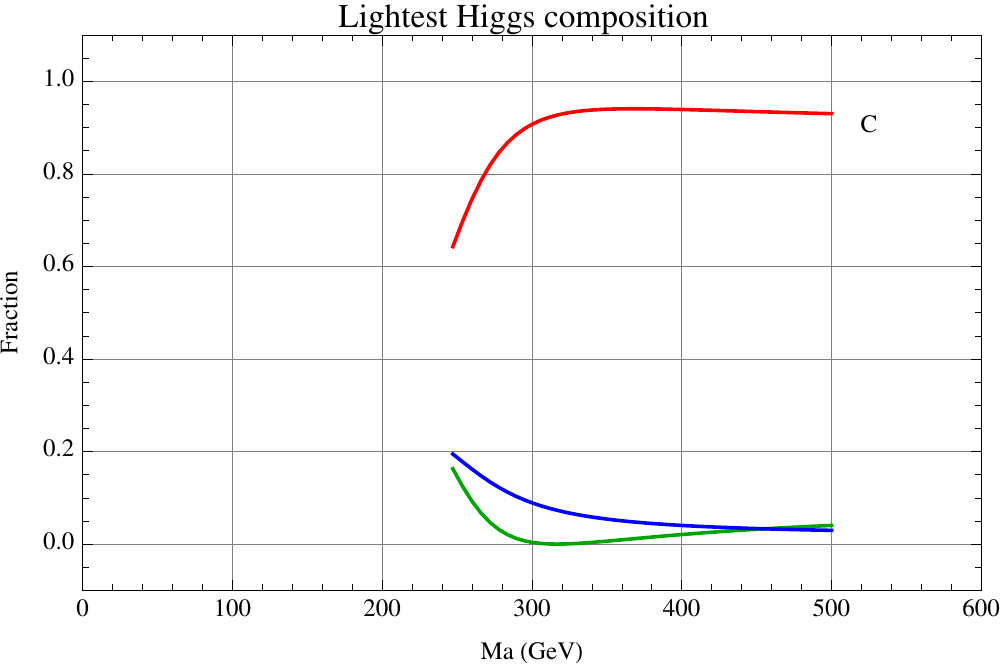} &
\includegraphics[scale=0.75]{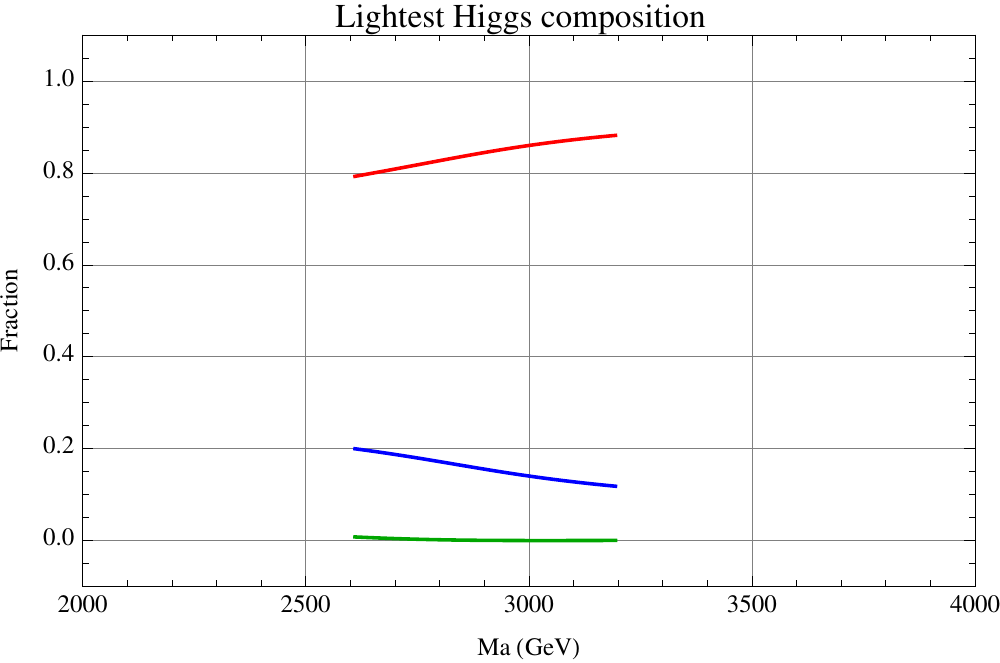} 
\end{array}$
\caption{Lightest neutral Higgs boson, $h$, content as a function of the lightest pseudoscalar mass for a $\mu_{12}$ scan corresponding to the yellow lines across the A, B, C, and D regions in Fig. \ref{StabilityandLSPRegion}. For each plot the values of the gaugino SUSY breaking masses are $M_1 =200$ GeV and $M_2=800$ GeV, and $b=4,000$ ${\rm GeV^2}$.  The scan through region A has $\tan{\beta}=1$, $\tan{\theta}=1$, and $\mu_{11}=-12$ GeV, the one through region B has $\tan{\beta}=1$, $\tan{\theta}=2$ and $\mu_{11}=-16$ GeV, the one through region C has $\tan{\beta}=2$, $\tan{\theta}=2$, and $\mu_{11}=-52$  GeV, and the one through region D has $\tan{\beta}=10$, $\tan{\theta}=2$, and $\mu_{11}=-344$ GeV. The red curve corresponds to the $S_u$ fraction, the green curve to the $S_d$ fraction, and the blue curve to the $S_\pi$ fraction. }
\label{HiggsContent}
\end{center}
\end{figure*}

\begin{figure*}
\begin{center}
$\begin{array}{cc}
\includegraphics[scale=0.75]{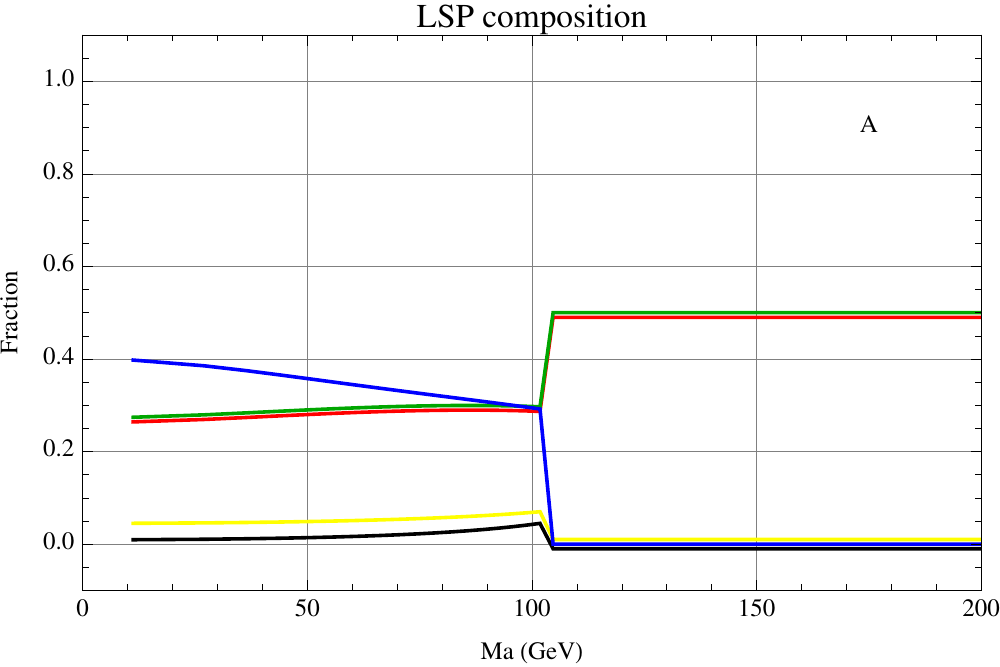} &
\includegraphics[scale=0.75]{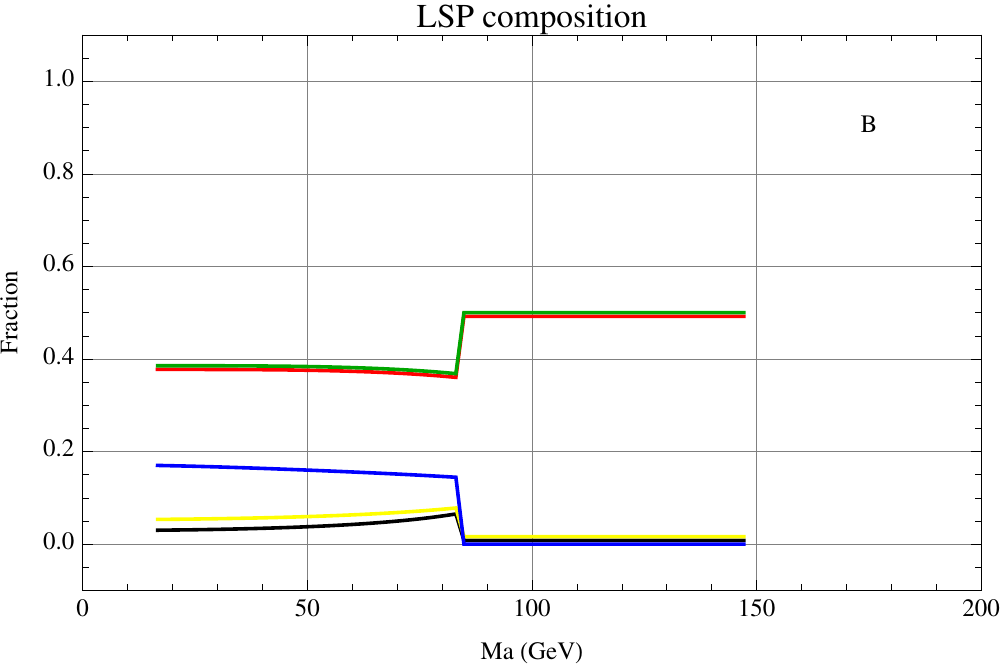} \\
\includegraphics[scale=0.75]{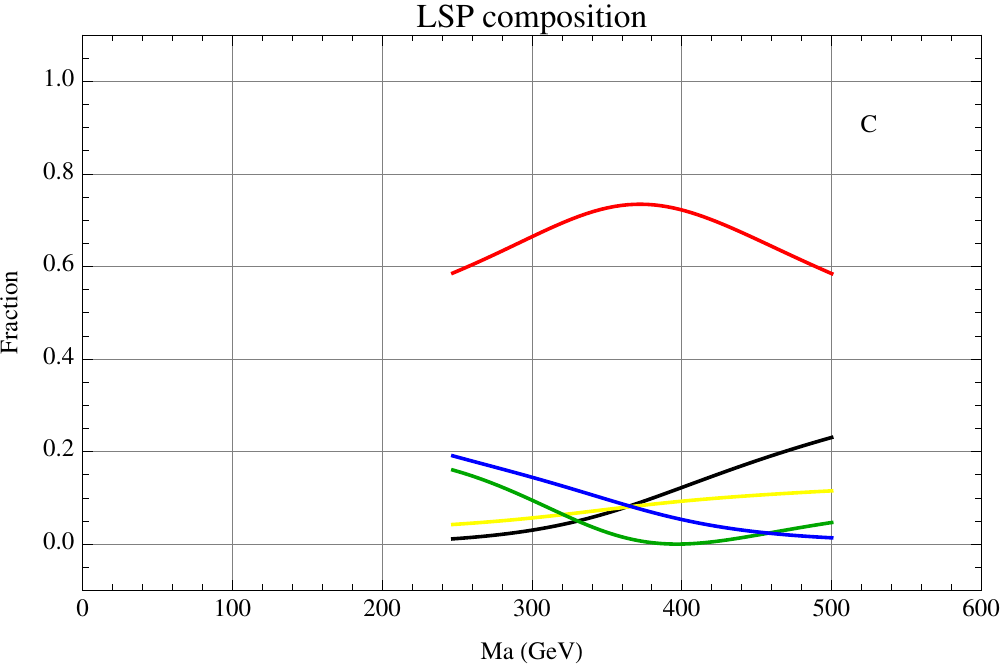} &
\includegraphics[scale=0.75]{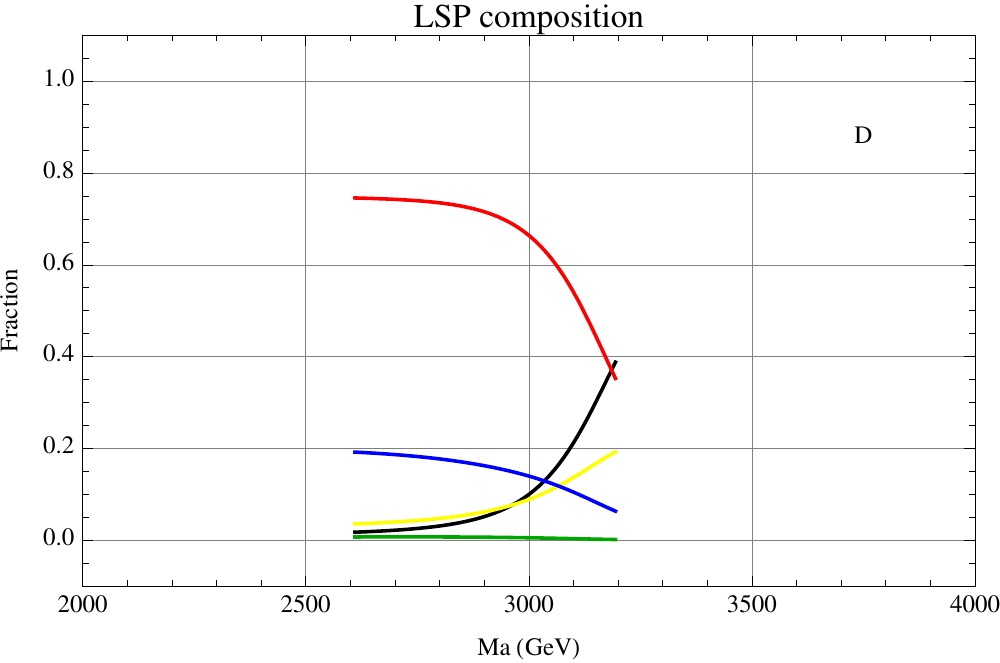} 
\end{array}$
\caption{LSP-neutralino, $N1$, content as a function of the lightest pseudoscalar mass for a $\mu_{12}$ scan  corresponding to the yellow lines across the A, B, C, and D regions in Fig. \ref{StabilityandLSPRegion}. For each plot the values of the gaugino SUSY breaking masses are $M_1 =200$ GeV and $M_2=800$ GeV, and $b=4,000$ ${\rm GeV^2}$.  The scan through region A has $\tan{\beta}=1$, $\tan{\theta}=1$, and $\mu_{11}=-12$ GeV, the one through region B has $\tan{\beta}=1$, $\tan{\theta}=2$ and $\mu_{11}=-16$ GeV, the one through region C has $\tan{\beta}=2$, $\tan{\theta}=2$, and $\mu_{11}=-52$  GeV, and the one through region D has $\tan{\beta}=10$, $\tan{\theta}=2$, and $\mu_{11}=-344$ GeV. The black curve corresponds to the $\lambda_\gamma$ fraction, the yellow curve to the $ \lambda_Z$ fraction, the red curve to the $\tilde{H}_u^0 $ fraction, the green curve to the $\tilde{H}_d^0$ fraction, and the blue curve to the $\tilde\pi^0 $ fraction.}
\label{NeutralinoContent}
\end{center}
\end{figure*}

It is instructive to quantify the contribution of the components of the  constrained Higgs doublet multiplets to the lightest Higgs neutral (pseudo-) scalar  and charged scalars as well as the lightest neutralino and chargino fermions. The fractions of the lightest neutral Higgs scalar $h$ in a decomposition in terms of the MSSM neutral scalars $S_u, S_d$ and the  scalar $S_\pi$ arising from the constrained doublets are displayed in Fig. \ref{HiggsContent}  as a function of $m_a$. For regions $A$ and $B$,  a lightest Higgs scalar is essentially devoid of the nonlinearly transforming scalar $S_\pi$ over the entire range  $ 94 ~{\rm GeV}~  <m_a$. As such, the composition of the Higgs scalar is thus almost identical to that of the MSSM.  In  region $C$, the $S_\pi$ fraction of is less than  $6-4\%$ for a lightest Higgs scalar mass in the range $182 ~{\rm GeV}~> m_h > 115 $ Gev which corresponds to $370 ~{\rm GeV}~ < m_a < 475$ GeV. 
 While not completely negligible, the Higgs scalar is still predominately composed of the MSSM fields. Finally, for region $D$, the $S_\pi$ content in the lightest Higgs scalar is about $13-12\%$ for the mass range $200 ~{\rm GeV}~ > m_h  > 115 $ GeV which corresponds to $3140 ~{\rm GeV}~ < m_a < 3180$ GeV. The modification to this lightest Higgs production and decay due to the admixture of the non-MSSM content will be addressed in the next section. The discontinuity in the slope appearing in the plots for regions $A$ and $B$ is a consequence  of the crossover in the particle content of  the lightest mass eigenvalue and the $m_a$ step size used in the numerical calculation. Note that this slope discontinuity occurs at a value of $m_a$ which is less than $94.3$ GeV and hence excluded by the current experimental bound.

\begin{figure*}
\begin{center}
$\begin{array}{cc}
\includegraphics[scale=0.75]{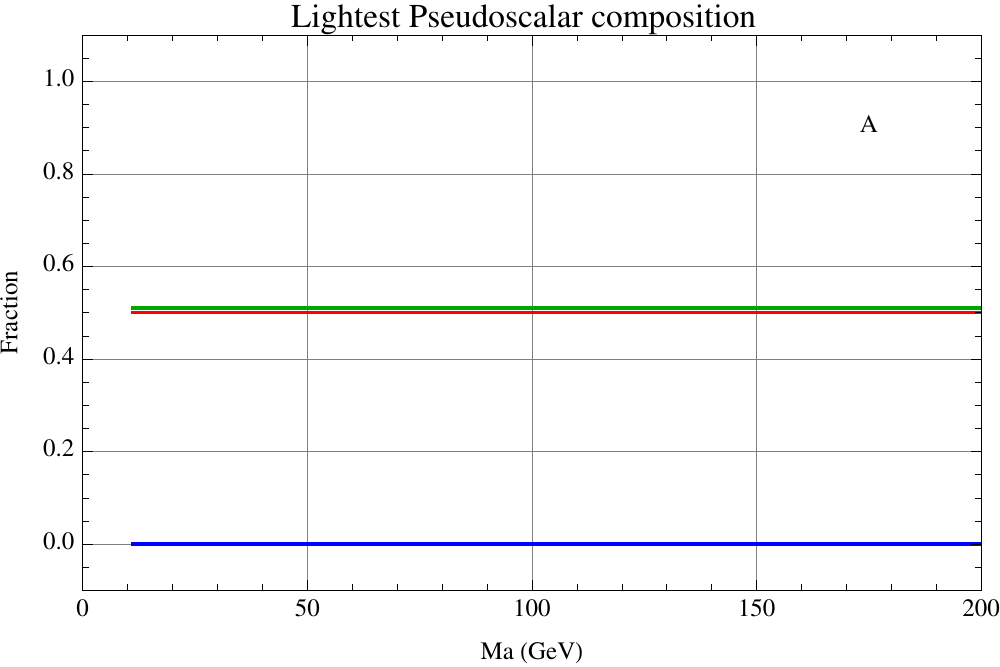} &
\includegraphics[scale=0.75]{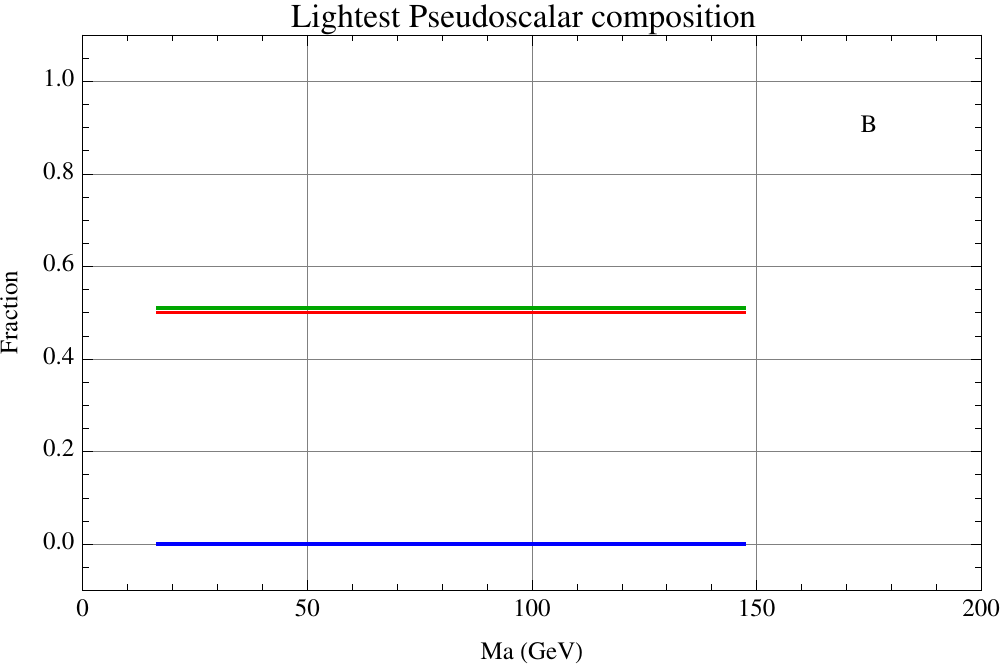} \\
\includegraphics[scale=0.75]{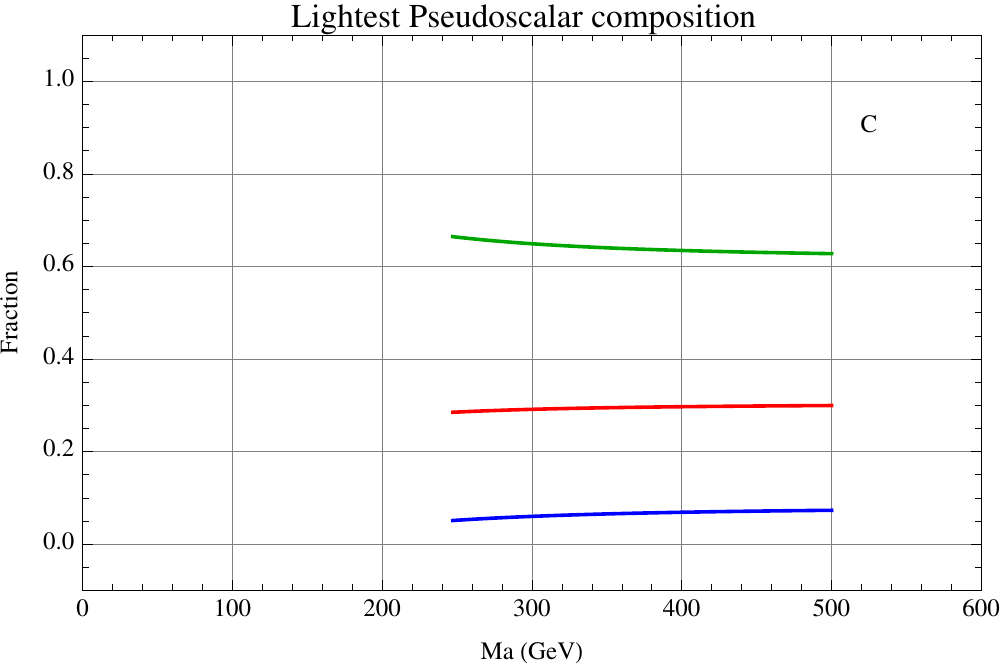} &
\includegraphics[scale=0.75]{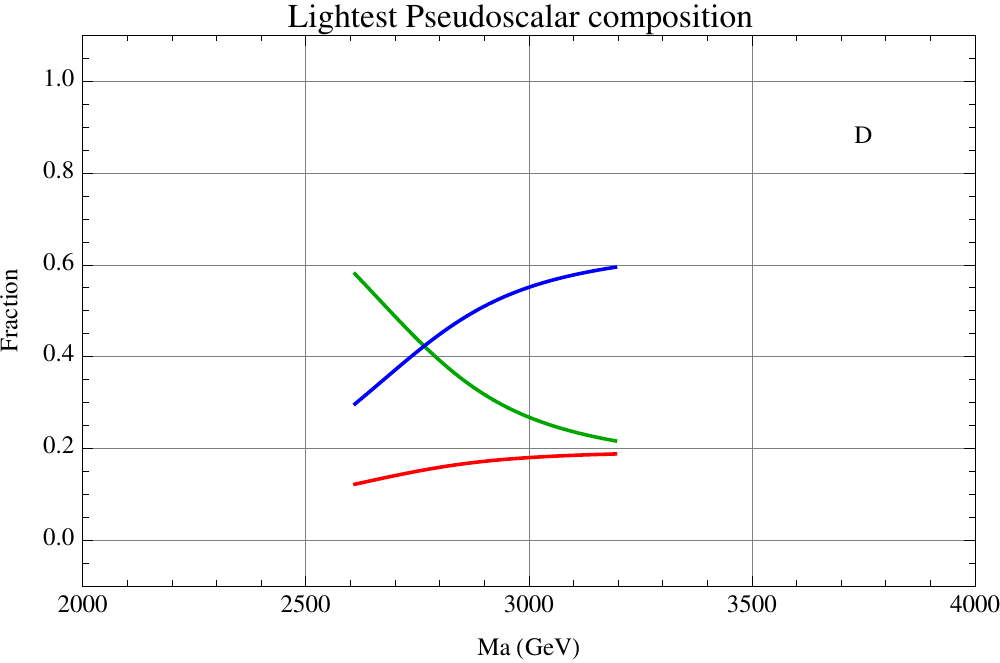} 
\end{array}$
\caption{Lightest Pseudoscalar, $a$, content as a function of the lightest pseudoscalar mass for a $\mu_{12}$ scan  corresponding to the yellow lines across the A, B, C, and D regions in Fig. \ref{StabilityandLSPRegion}. For each plot the values of the gaugino SUSY breaking masses are $M_1 =200$ GeV and $M_2=800$ GeV, and $b=4,000$ ${\rm GeV^2}$.  The scan through region A has $\tan{\beta}=1$, $\tan{\theta}=1$, and $\mu_{11}=-12$ GeV, the one through region B has $\tan{\beta}=1$, $\tan{\theta}=2$ and $\mu_{11}=-16$ GeV, the one through region C has $\tan{\beta}=2$, $\tan{\theta}=2$, and $\mu_{11}=-52$  GeV, and the one through region D has $\tan{\beta}=10$, $\tan{\theta}=2$, and $\mu_{11}=-344$ GeV. The red curve corresponds to the $P_u$ fraction, the green curve to the $P_d$ fraction, and the blue curve to the $P_\pi$ fraction..} 
\label{PseudoContent}
\end{center}
\label{fig11b}
\end{figure*}

\begin{figure*}
\begin{center}
$\begin{array}{cc}
\includegraphics[scale=0.75]{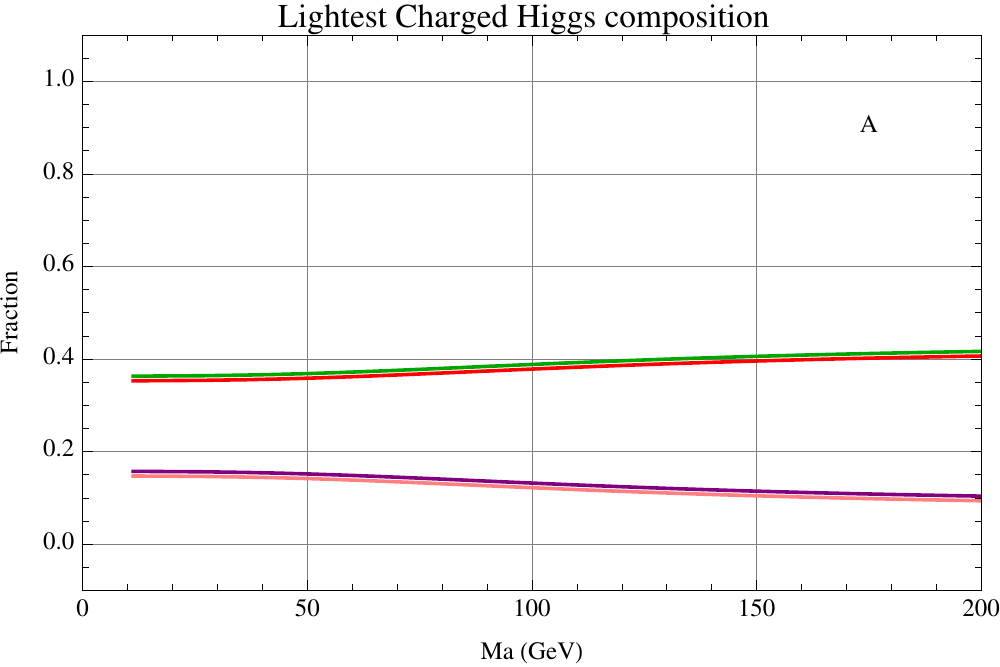} &
\includegraphics[scale=0.75]{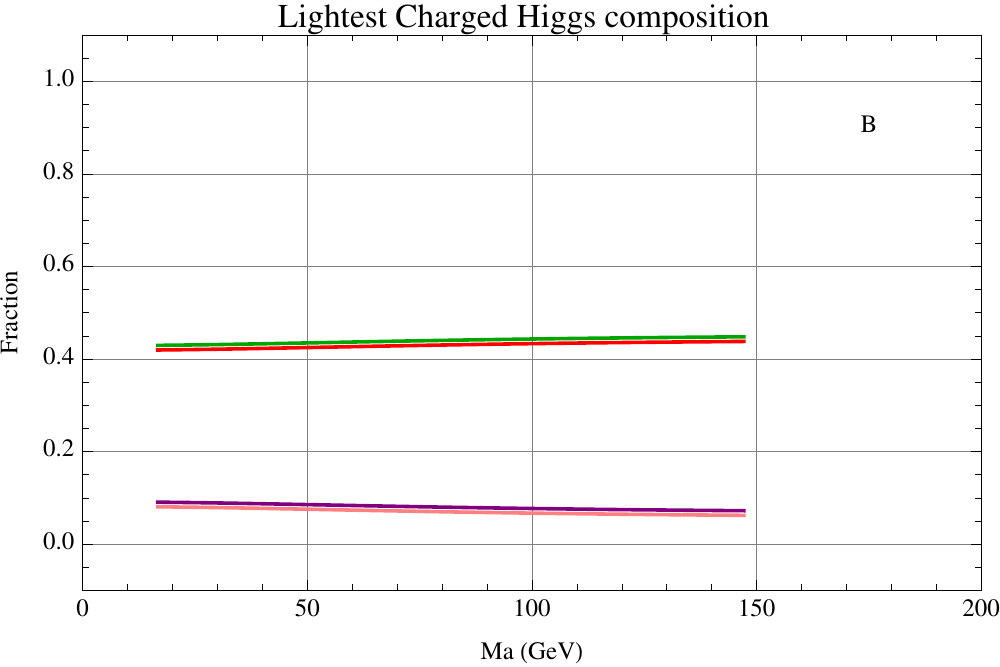} \\
\includegraphics[scale=0.75]{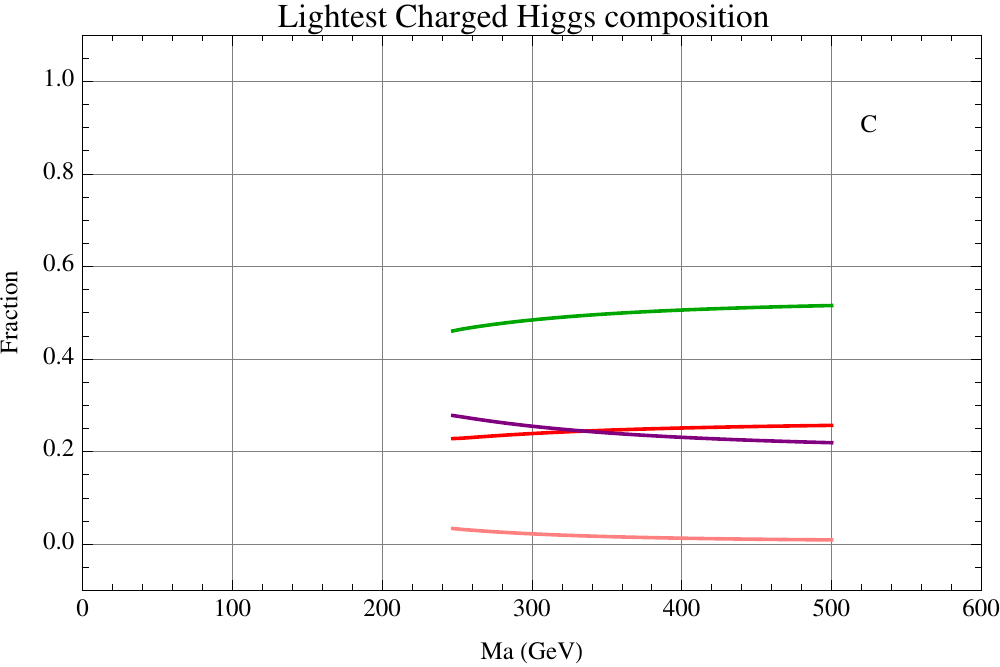} &
\includegraphics[scale=0.75]{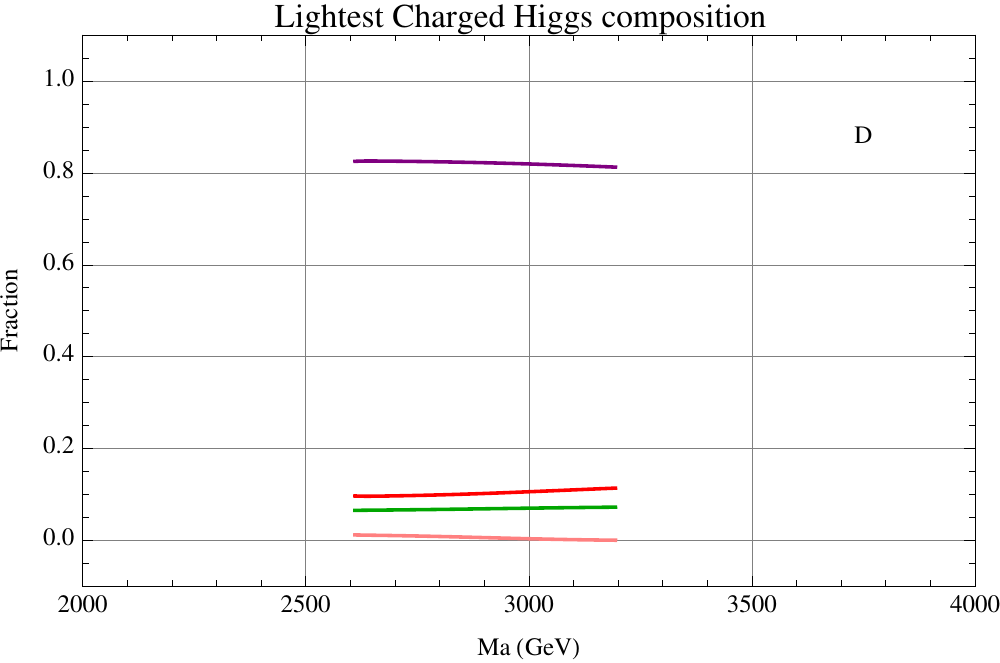} 
\end{array}$
\caption{Lightest charged Higgs boson, $C1$, content as a function of the lightest pseudoscalar mass for a $\mu_{12}$ scan  corresponding to the yellow lines across the A, B, C, and D regions in Fig. \ref{StabilityandLSPRegion}. For each plot the values of the gaugino SUSY breaking masses are $M_1 =200$ GeV and $M_2=800$ GeV, and $b=4,000$ ${\rm GeV^2}$.  The scan through region A has $\tan{\beta}=1$, $\tan{\theta}=1$, and $\mu_{11}=-12$ GeV, the one through region B has $\tan{\beta}=1$, $\tan{\theta}=2$ and $\mu_{11}=-16$ GeV, the one through region C has $\tan{\beta}=2$, $\tan{\theta}=2$, and $\mu_{11}=-52$  GeV, and the one through region D has $\tan{\beta}=10$, $\tan{\theta}=2$, and $\mu_{11}=-344$ GeV. The red curve corresponds to the $H_u^+$ fraction, the green curve to the $\bar{H}_d^-$ fraction, the pink curve to the $\pi^+$ fraction, and the purple curve to the $\bar{\pi}^- $ fraction.}
\label{ChargedHiggsContent}
\end{center}
\label{fig10}
\end{figure*}

\begin{figure*}
\begin{center}
$\begin{array}{cc}
\includegraphics[scale=0.75]{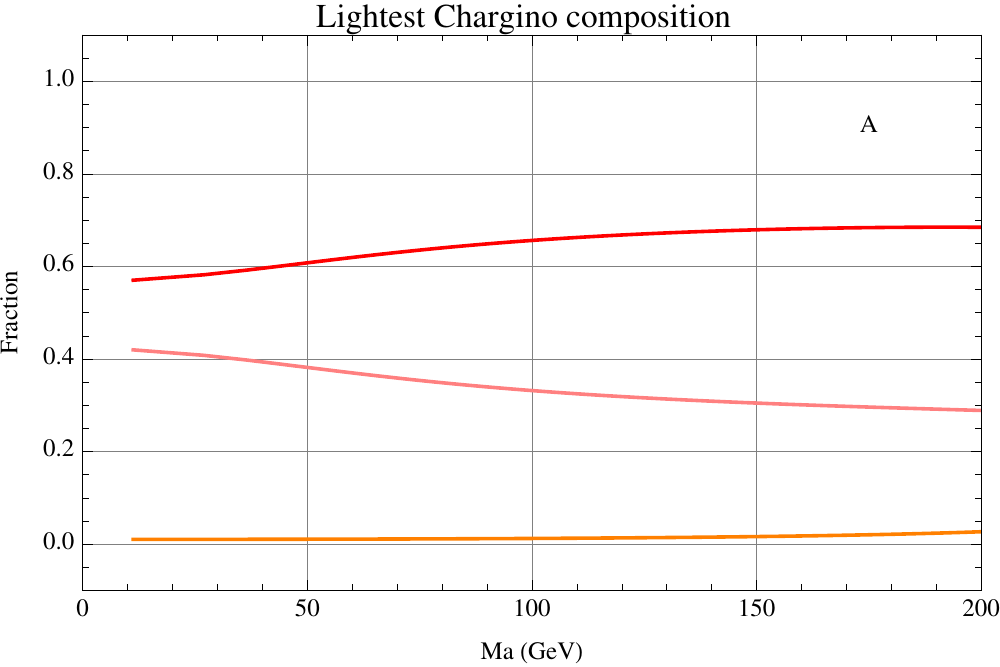} &
\includegraphics[scale=0.75]{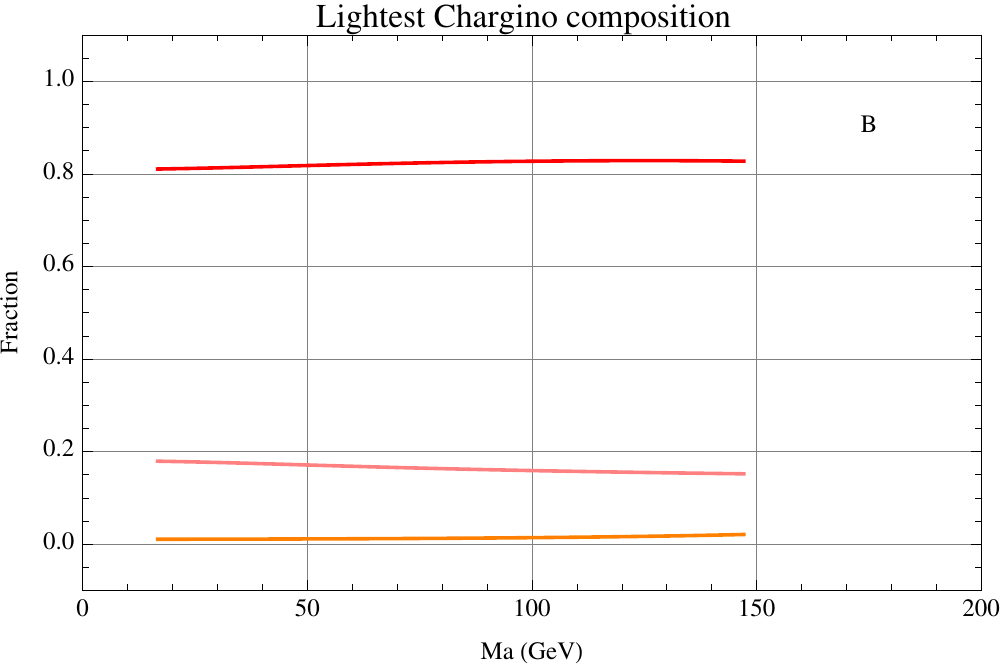} \\
\includegraphics[scale=0.75]{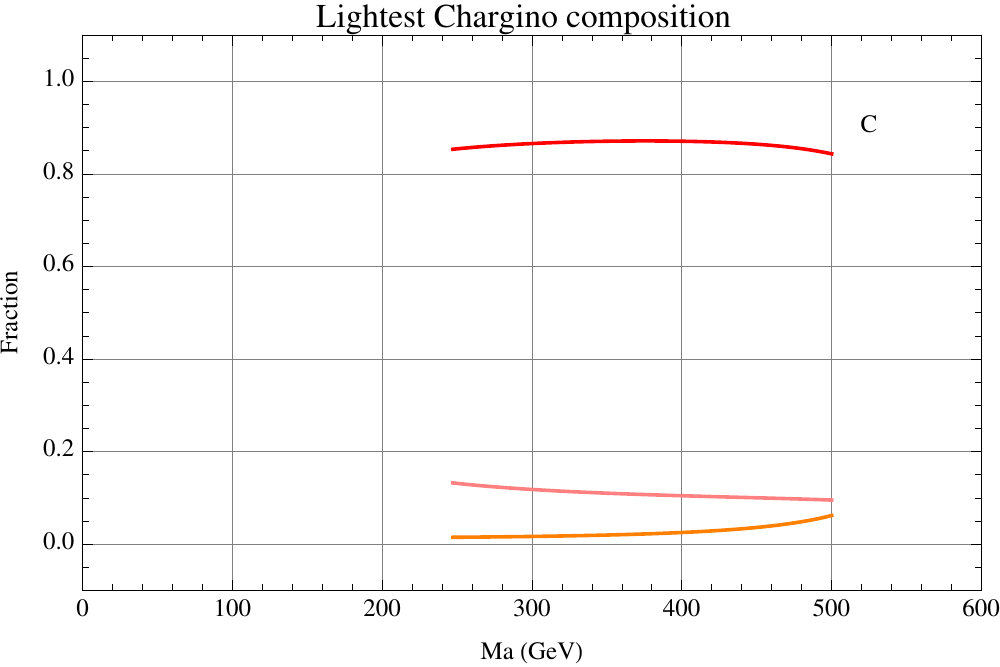} &
\includegraphics[scale=0.75]{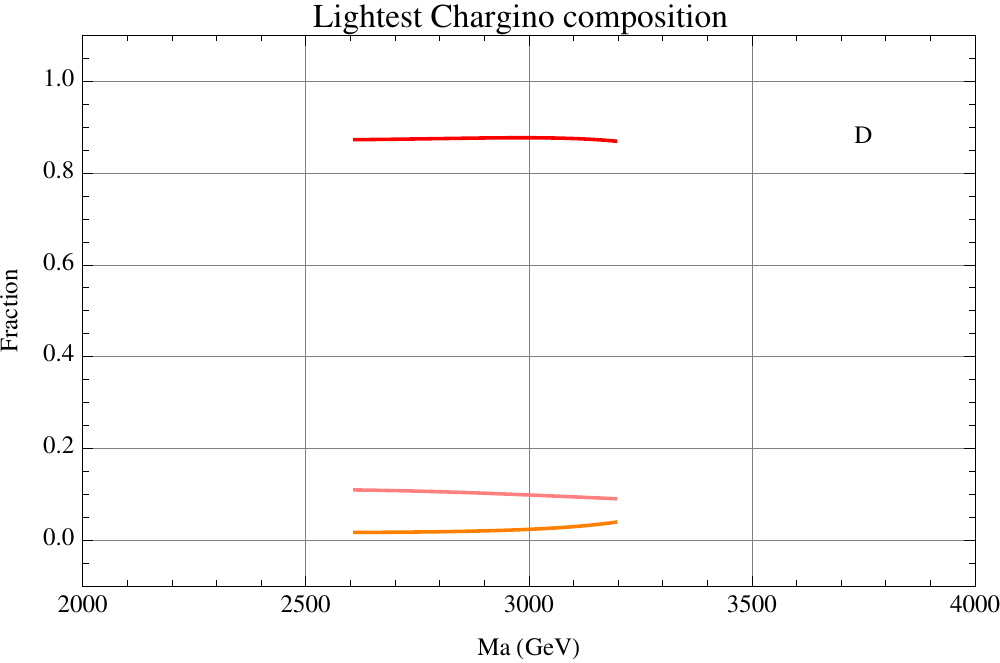} 
\end{array}$
\caption{Lightest Chargino, $\tilde{C1}$, content as a function of the lightest pseudoscalar mass for a $\mu_{12}$ scan  corresponding to the yellow lines across the A, B, C, and D regions in Fig. \ref{StabilityandLSPRegion}. For each plot the values of the gaugino SUSY breaking masses are $M_1 =200$ GeV and $M_2=800$ GeV, and $b=4,000$ ${\rm GeV^2}$.  The scan through region A has $\tan{\beta}=1$, $\tan{\theta}=1$, and $\mu_{11}=-12$ GeV, the one through region B has $\tan{\beta}=1$, $\tan{\theta}=2$ and $\mu_{11}=-16$ GeV, the one through region C has $\tan{\beta}=2$, $\tan{\theta}=2$, and $\mu_{11}=-52$  GeV, and the one through region D has $\tan{\beta}=10$, $\tan{\theta}=2$, and $\mu_{11}=-344$ GeV. The orange curve corresponds to the $\tilde{W}^+$ fraction, the red curve to the $ \tilde{H}_u^+$ fraction, and the pink curve to the $\tilde{\pi}^+ $ fraction.} 
\label{CharginoContent}
\end{center}
\label{fig11}
\end{figure*}

\begin{figure*}
\begin{center}
$\begin{array}{cc}
\includegraphics[scale=1.00]{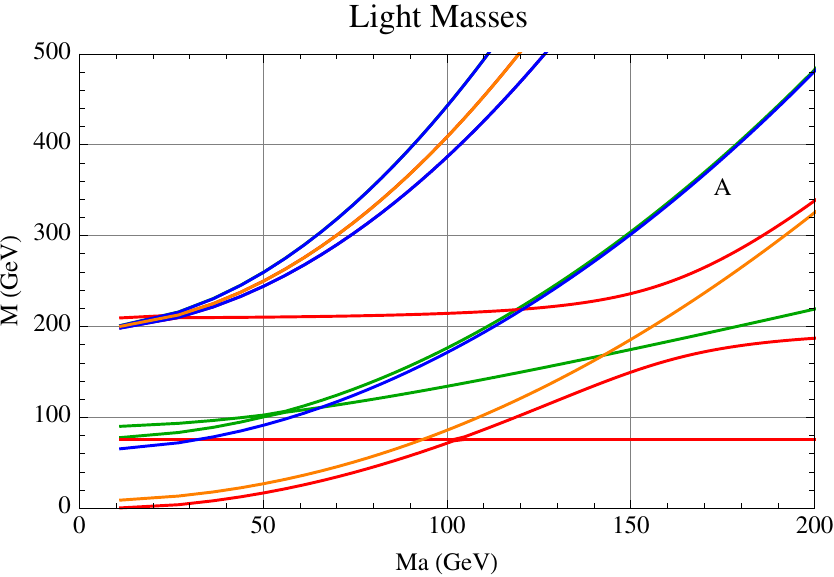} &
\includegraphics[scale=1.00]{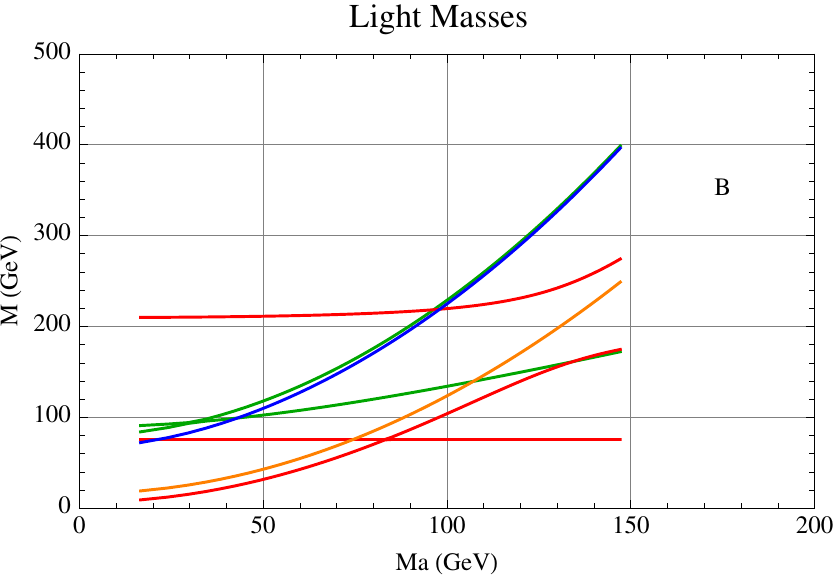} \\
\includegraphics[scale=1.00]{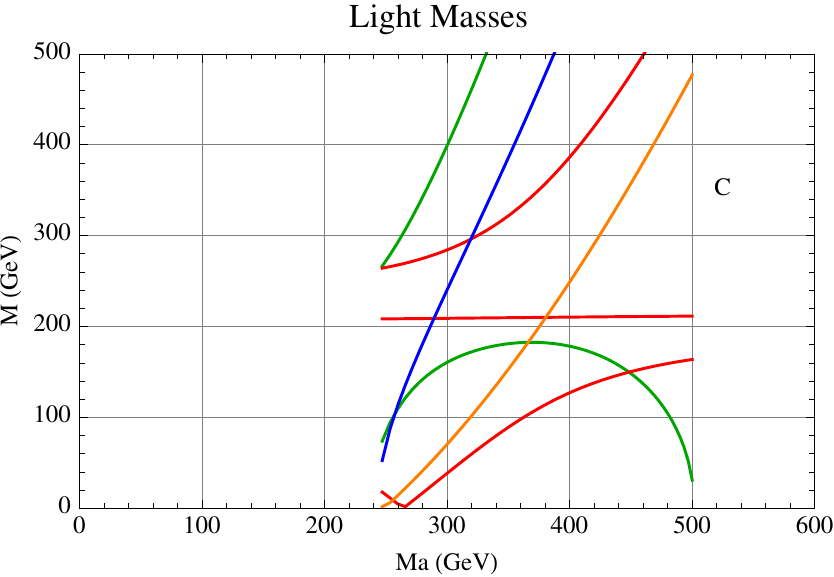} &
\includegraphics[scale=1.00]{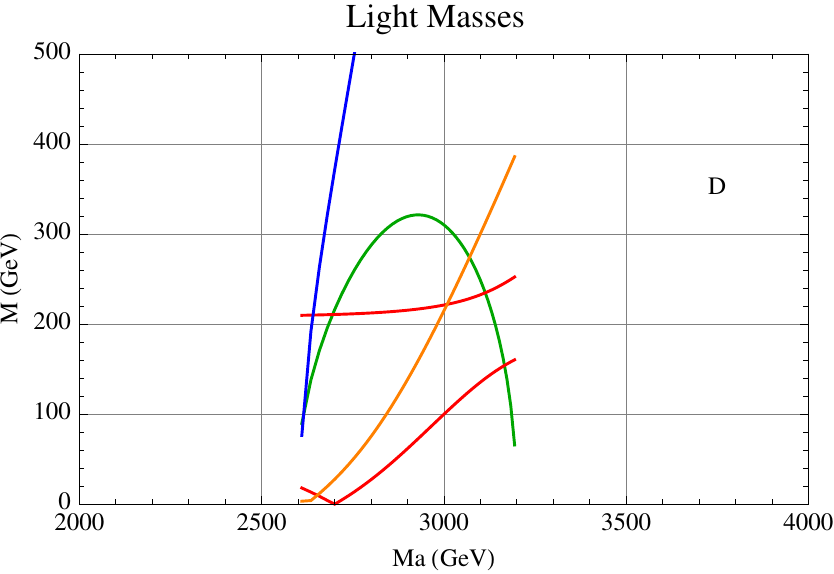} 
\end{array}$
\caption{Detailed light spectra as a function of the lightest pseudoscalar mass for a $\mu_{12}$ scan  corresponding to the yellow lines across the A, B, C, and D regions in Fig. \ref{StabilityandLSPRegion}. For each plot the values of the gaugino SUSY breaking masses are $M_1 =200$ GeV and $M_2=800$ GeV, and $b=4,000$ ${\rm GeV^2}$.  The scan through region A has $\tan{\beta}=1$, $\tan{\theta}=1$, and $\mu_{11}=-12$ GeV, the one through region B has $\tan{\beta}=1$, $\tan{\theta}=2$ and $\mu_{11}=-16$ GeV, the one through region C has $\tan{\beta}=2$, $\tan{\theta}=2$, and $\mu_{11}=-52$  GeV, and the one through region D has $\tan{\beta}=10$, $\tan{\theta}=2$, and $\mu_{11}=-344$ GeV. Green curves correspond to neutral scalar masses, blue curves to charged scalar masses, red curves to neutralino masses, and orange curves to chargino masses.}
\label{Lightspectrum}
\end{center}
\label{fig12}
\end{figure*}

The fractions of the lightest neutralino ${N1}$, the LSP,  in its decomposition in terms of the  photino $\lambda_\gamma$, zino $\lambda_Z$, the MSSM neutral Higgsinos $\tilde{H}_u^0, \tilde{H}_d^0$ and the neutral $\pi$-ino  originating from the constrained multiplets are displayed in Fig. \ref{NeutralinoContent} for these scans. For the considered regions in parameter space, the  nonlinearly transforming $\pi-$ino field composition of the neutralino LSP is very similar to the nonlinearly transforming Higgs field composition of the lightest neutral scalar detailed above for regions $A, B, C$. Consequently, its  identification  with  dark matter can proceed just as in the MSSM. For region $D$,  the fraction of $\pi-$ino is somewhat larger being of order $10-5\%$ for $3100 ~{\rm GeV}~ < m_a < 3150$ GeV. Fig. \ref{PseudoContent} displays the fractions of the lightest pseudoscalar, $a$, in its decomposition in terms of MSSM pseudoscalars, $P_u, P_d$, and the nonlinearly transforming $P_\pi$. The contribution of $P_\pi$ in regions $A$ and $B$ is completely negligible, while for region $C$, $P_\pi$ contributes at roughly a $5-10\% $. On the other hand, for region $D$, the lightest pseudoscalar is predominately composed of $P_\pi$ for the larger scanned $m_a$ values. The fractions of the lightest charged scalar $C1$ in its decomposition in terms of the  MSSM charged scalars $H_u^+, H_d^{-\dagger}$ and the charged scalars $\pi^+, \pi^{-\dagger}$ arising from the nonlinearly transforming Higgs multiplets is displayed in Fig. \ref{ChargedHiggsContent}. In this case, each of the nonlinearly transforming scalars contribute a fraction which is a decreasing function of $m_a$. This time, the largest fraction, which is still  $\sim 15 \%$, occurs for panel A, while panels B, C, D have successively smaller nonlinear transforming field content over the entire scanned range. Finally, the fractions of the lightest chargino $\tilde{C1}$ in its decomposition in terms of the wino $\lambda_{W_+}$,  the MSSM charged Higgsino $\tilde{H}_u^+$,and the Higgsino $\tilde{\pi}^+$ originating from the constrained multiplets are displayed in Fig. \ref{CharginoContent} for these scans. In this case, the contribution of nonlinearly transforming Higgsino $\tilde{\pi}^+$ is consistently larger than in the previously considered cases, although it is still subdominant. Detailed plots of the light mass spectra including only particles with a mass less than $500$ GeV are presented in Fig. \ref{Lightspectrum} for the scans through each of the four regions.

%\fi
\newpage
\section{Electroweak Precision Tests and Lightest Higgs Boson Production and Decay\label{section4}}

Since only the MSSM Higgs fields couple directly to the standard model matter fields, one anticipates that the flavor physics in this model should be quite similar to that of the MSSM. The only difference arises due to the fact that the MSSM Higgs field vacuum expectation values only partially contribute to the electroweak vacuum value $v=246$ GeV. Consequently, the matter field Yukawa couplings must be proportionately larger in order to compensate for the smaller $v_u$ and $v_d$ values. For the top and bottom quarks and tau lepton the masses are related to the Yukawa coupings as 
\bea
\frac{m_t}{v} &=& \frac{1}{\sqrt{2}} y_{t} \sin{\theta} \sin{\beta} \cr
\frac{m_b}{v} &=& \frac{1}{\sqrt{2}} y_{b} \sin{\theta} \cos{\beta} \cr
\frac{m_\tau}{v} &=& \frac{1}{\sqrt{2}} y_{\tau} \sin{\theta} \cos{\beta} .
\eea
Comparing with the MSSM values, we have the effective replacements $y^{MSSM}=y \sin\theta$. Thus the Yukawa couplings will differ significantly from their MSSM values for small $\tan\theta$. Placing a perturbative bound on the size of the Yukawa coupling constants so that $y< 4 \pi$ translates to  bounds on $\tan{\beta}$ and $\tan{\theta}$ given by
\bea
\left[ 1 +\frac{1}{\tan^2{\theta}}\right] \left[ 1 +\frac{1}{\tan^2{\beta}}\right] &=& \frac{y_t^2 v^2}{2 m_t^2} \leq \frac{8 \pi^2 v^2}{m_t^2} \approx 160 \cr
\left[ 1 +\frac{1}{\tan^2{\theta}}\right] \left[ 1 +\tan^2{\beta}\right] &=& \frac{y_b^2 v^2}{ 2m_b^2} \leq \frac{8 \pi^2 v^2}{m_b^2} \approx 2\times 10^5\cr
\left[ 1 +\frac{1}{\tan^2{\theta}}\right] \left[ 1 +\tan^2{\beta}\right] &=& \frac{y_\tau^2 v^2}{2 m_\tau^2} \leq \frac{8 \pi^2 v^2}{m_\tau^2} \approx 1.5 \times 10^6 .
\eea
In addition to the very small $\tan\theta$ values, this also excludes regions corresponding to fractionally small values of  $\tan{\theta}$ and $\tan{\beta}$ (e.g. $\tan{\theta} =0.1$ and $\tan{\beta}=1$) as well as excessively large values of $\tan{\beta}$.

The $W$ and $Z$ masses satisfy the  $\rho = M_W^2/M_Z^2 \cos{\theta_W}=1$ relation at tree level. The effects of radiative corrections to the gauge field vacuum polarizations can be encapsulated in the electroweak precision parameters $S$ and $T$. One source of contributions to $S$ and $T$ can arise from loop effects in the effective model under consideration here. The precise form of their 1-loop contribution is beyond the scope of this paper.  However, one anticipates a contribution of  the form 
\be
\Delta S = \frac{c}{16 \pi^2} \ln{\frac{\Lambda}{M_Z}} \quad , \quad \Delta T = \frac{d}{16 \pi^2} \ln{\frac{\Lambda}{M_Z}} ,
\ee
where $\Lambda$ is the mass scale above which the effective theory no longer accurately describes the dynamics and $c, d$ are the specific values obtained from the 1-loop Feynman diagrams. In addition, there are contributions to $S$ and $T$ arising from the underlying theory responsible for the electroweak symmetry breaking and the resulting nonlinear sigma model. Although we  do not specify a particular theory, we can parametrize its effects by the inclusion of 
additional supersymmetric higher dimensional operators, albeit suppressed by powers of the effective action cutoff $\Lambda$.  There are four lowest dimension effective operators contributing to the electroweak precision parameter $S$.  The action for each is given by
\bea
\Gamma_{S11} &=& \frac{-s_{11}}{128 g_1 g_2 \Lambda^2} \left( \int dS H_u \epsilon W_2 W_1 H_d +\int d \bar{S} \bar{H}_u \epsilon \bar{W}_2 \bar{W}_1 \bar{H}_d \right) \cr
 &=&\frac{s_{11} v_u v_d}{8 \Lambda^2}\int d^4 x \left[ \sin{2\theta_W}\left( Z_{\mu\nu} Z^{\mu\nu} -A_{\mu\nu} A^{\mu\nu} \right) - 2\cos{2\theta_W} Z_{\mu\nu} A_{\mu\nu} + \cdots\right] \cr
\Gamma_{S12} &=& \frac{-s_{12}}{128 g_1 g_2 \Lambda^2} \left( \int dS H_u \epsilon W_2 W_1 H_d^\prime +\int d \bar{S} \bar{H}_u \epsilon \bar{W}_2 \bar{W}_1 \bar{H}_d^\prime \right) \cr
 &=&\frac{s_{12} v_u v^\prime}{8 \Lambda^2}\int d^4 x \left[ \sin{2\theta_W}\left( Z_{\mu\nu} Z^{\mu\nu} -A_{\mu\nu} A^{\mu\nu} \right) - 2\cos{2\theta_W} Z_{\mu\nu} A_{\mu\nu} + \cdots\right] \cr
\Gamma_{S21} &=& \frac{-s_{21}}{128 g_1 g_2 \Lambda^2} \left( \int dS H_u^\prime \epsilon W_2 W_1 H_d +\int d \bar{S} \bar{H^\prime}_u \epsilon \bar{W}_2 \bar{W}_1 \bar{H}_d \right) \cr
 &=&\frac{s_{21} v_d v^\prime}{8 \Lambda^2}\int d^4 x \left[ \sin{2\theta_W}\left( Z_{\mu\nu} Z^{\mu\nu} -A_{\mu\nu} A^{\mu\nu} \right) - 2\cos{2\theta_W} Z_{\mu\nu} A_{\mu\nu} + \cdots\right] \cr
\Gamma_{S22} &=& \frac{-s_{22}}{128 g_1 g_2 \Lambda^2} \left( \int dS H_u^\prime \epsilon W_2 W_1 H_d^\prime +\int d \bar{S} \bar{H^\prime}_u \epsilon \bar{W}_2 \bar{W}_1 \bar{H^\prime}_d \right) \cr
 &=&\frac{s_{22} v^{\prime 2}}{8 \Lambda^2}\int d^4 x \left[ \sin{2\theta_W}\left( Z_{\mu\nu} Z^{\mu\nu} -A_{\mu\nu} A^{\mu\nu} \right) - 2\cos{2\theta_W} Z_{\mu\nu} A_{\mu\nu} + \cdots\right], \cr
 & & 
\eea
with the ellipses denoting the higher dimensional terms.  The contribution of these operators to $S$ is given by
\be
\alpha S / \sin{2\theta_W} = \frac{(s_{11} v_u v_d + s_{12} v_u v^\prime +s_{21} v_d v^\prime +  s_{22} w^2)}{\Lambda^2} ,
\ee
while they do not contribute to $T$.

Likewise their are several effective operators that contribute  to $T$ but not to $S$. These are higher dimensional contributions to the  K\"ahler potential 
%\bea
%K_{uu} &=& \bar{H}_u e^{-2g_2 W -g_1 B}H_u = \frac{v_u^2}{2}\left[ 1+ gZ + g_2^2 W^+ W^- +\frac{1}{2}g^2 Z^2 + \cdots  \right]\cr
%K_{dd} &=& \bar{H}_d e^{-2g_2 W +g_1 B}H_d = \frac{v_d^2}{2}\left[ 1- gZ + g_2^2 W^+ W^- +\frac{1}{2}g^2 Z^2 + \cdots  \right]\cr
%K_{uu}^{\prime\prime} &=& \bar{H^\prime}_u e^{-2g_2 W -g_1 B}H_u^\prime = \frac{w^2}{2}\left[ 1+ gZ + g_2^2 W^+ W^- +\frac{1}{2}g^2 Z^2 + \cdots  \right]\cr
%K_{dd}^{\prime\prime} &=& \bar{H^\prime}_d e^{-2g_2 W +g_1 B}H_d^\prime = \frac{w^2}{2}\left[ 1- gZ + g_2^2 W^+ W^- +\frac{1}{2}g^2 Z^2 + \cdots  \right]\cr
%K_{uu}^\prime &=& \frac{1}{2}\left(\bar{H}_u e^{-2g_2 W -g_1 B}H_u^\prime +\bar{H^\prime}_u e^{-2g_2 W -g_1 B}H_u\right) = \frac{v_u v^\prime}{2}\left[ 1+ gZ + g_2^2 W^+ W^- +\frac{1}{2}g^2 Z^2 + \cdots  \right]\cr
%K_{dd}^\prime &=& \frac{1}{2}\left(\bar{H}_d e^{-2g_2 W +g_1 B}H_d^\prime +\bar{H^\prime}_d e^{-2g_2 W +g_1 B}H_d\right) = \frac{v_d v^\prime}{2}\left[ 1- gZ + g_2^2 W^+ W^- +\frac{1}{2}g^2 Z^2 + \cdots  \right],\cr
% & & 
%\eea
%various bilinear products of these operators contribute to $T$.  
The simplest such example is
\bea
Y &=& \bar{H^\prime}_u e^{-2g_2 W -g_1 B}H_u^\prime - \bar{H^\prime}_d e^{-2g_2 W +g_1 B}H_d^\prime\cr
&=& \frac{w^2}{2}\left[ 1+ gZ + g_2^2 W^+ W^- +\frac{1}{2}g^2 Z^2 + \cdots  \right] - \frac{w^2}{2}\left[ 1- gZ + g_2^2 W^+ W^- +\frac{1}{2}g^2 Z^2 + \cdots  \right]\cr
& =& gv^{\prime 2} Z +\cdots .
\eea
The effective action for this term takes the form
\bea
\Gamma_u &=& \frac{-M_Z^2 t}{16 g^2 v^{\prime 4} \Lambda^2} \int dV Y^2\cr
& =& \frac{-M_Z^2 t}{16  \Lambda^2}\int dV \left[ Z^2 +\cdots \right] \cr
 &=& \frac{-M_Z^2 t}{16 \Lambda^2}\int d^4 x \left[ \frac{1}{2} Z_\mu Z^\mu +\cdots \right] 
\eea
and provides a contribution  to $T$  given by
\be
\alpha T = \frac{t}{\Lambda^2} ,
\ee
with no contribution to $S$. Fitting to $S$ and $T$ can determine the allowed range of values for the coupling constants $s_{11}, s_{12}, s_{21}, s_{22}, t$ and the dynamical scale $\Lambda$ and thus provides a potent constraint on model building. 

\begin{figure*}
\begin{center}
$\begin{array}{cc}
\includegraphics[scale=0.75]{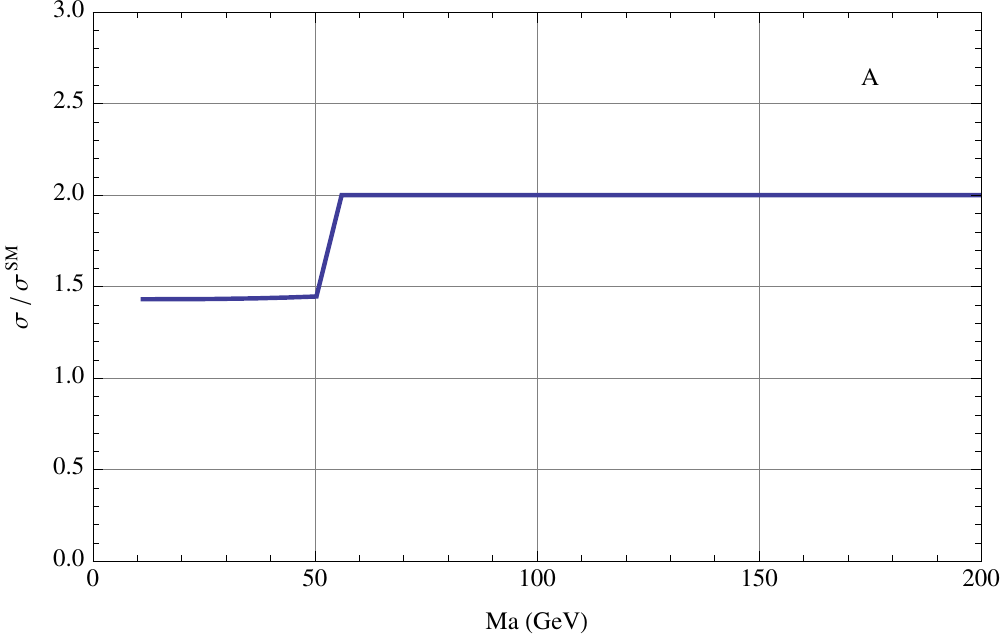} &
\includegraphics[scale=0.75]{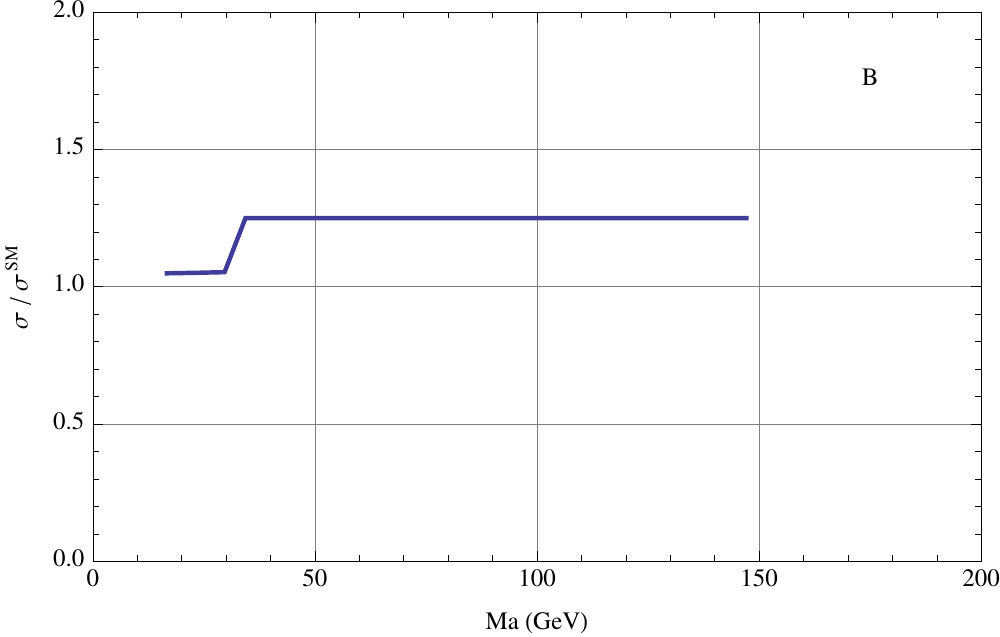} \\
\includegraphics[scale=0.75]{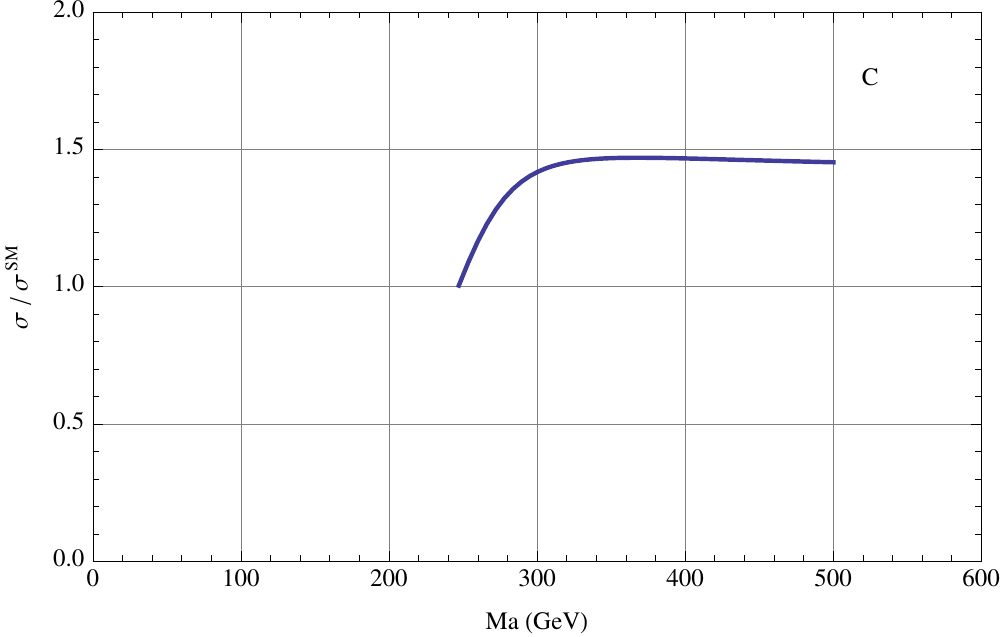} &
\includegraphics[scale=0.75]{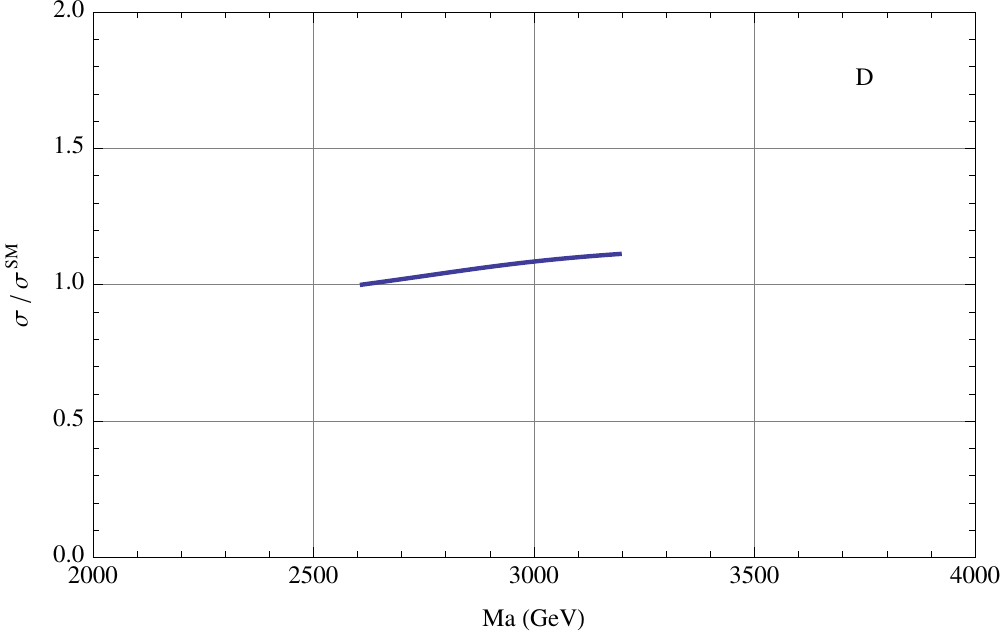} 
\end{array}$
\caption{Ratio of gluon fusion Higgs scalar production cross-section to the standard model result.} 
\label{cross}
\end{center}
\label{fig-cross}
\end{figure*}

\begin{figure*}
\begin{center}
$\begin{array}{cc}
\includegraphics[scale=0.75]{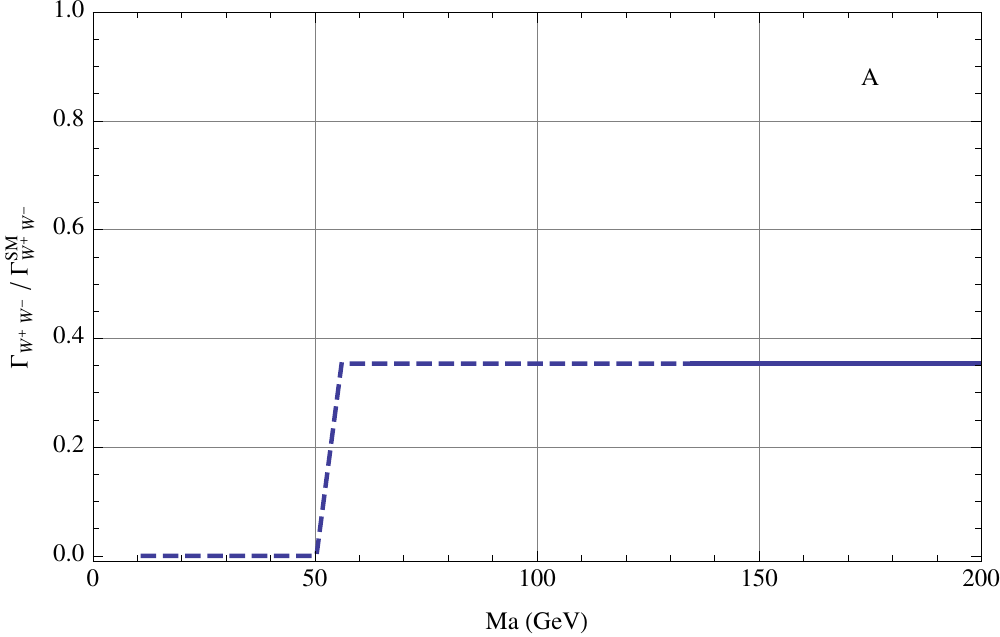} &
\includegraphics[scale=0.75]{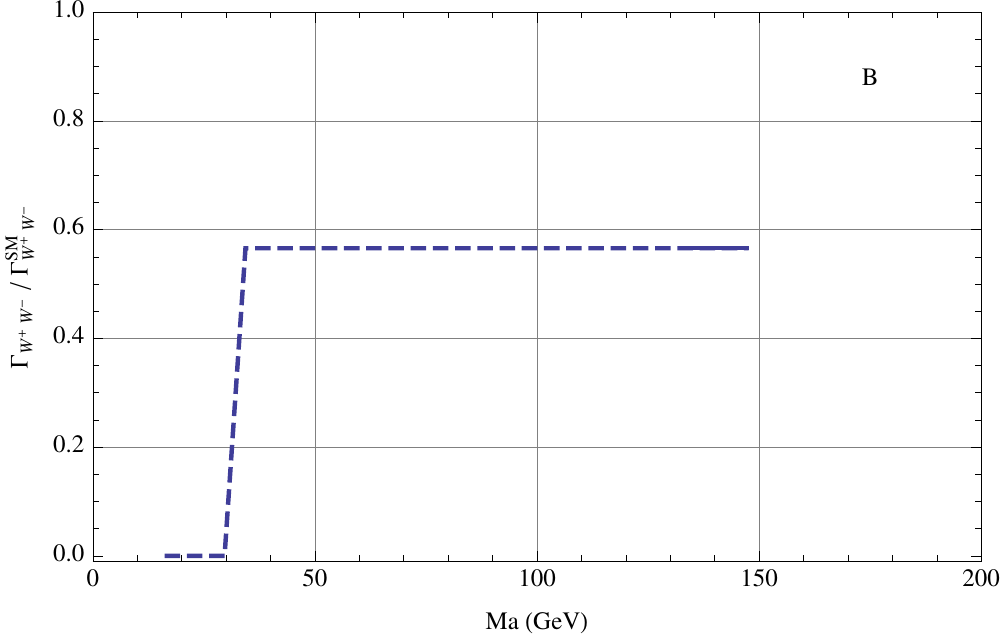} \\
\includegraphics[scale=0.75]{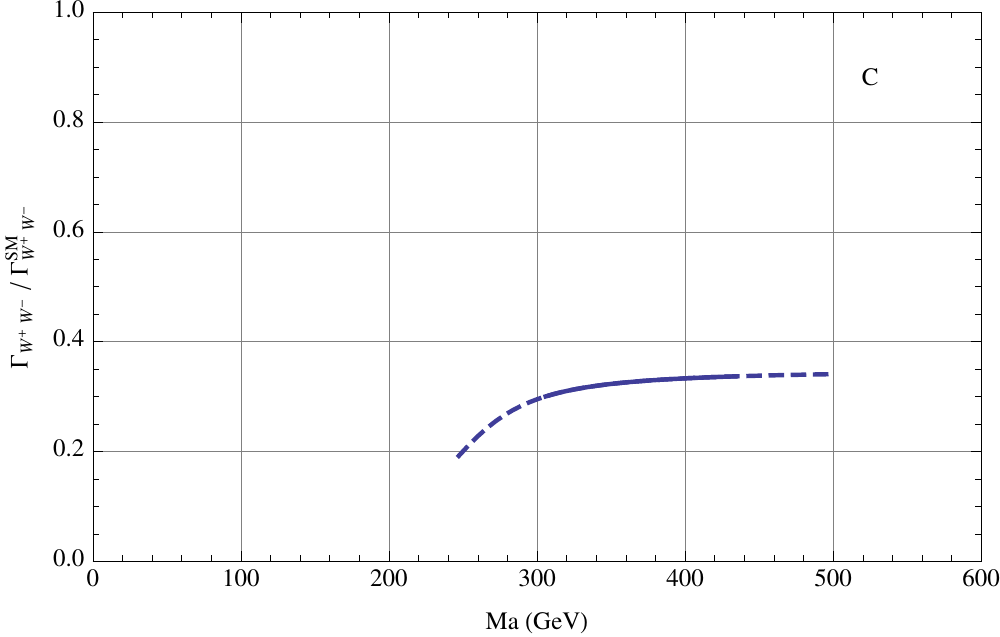} &
\includegraphics[scale=0.75]{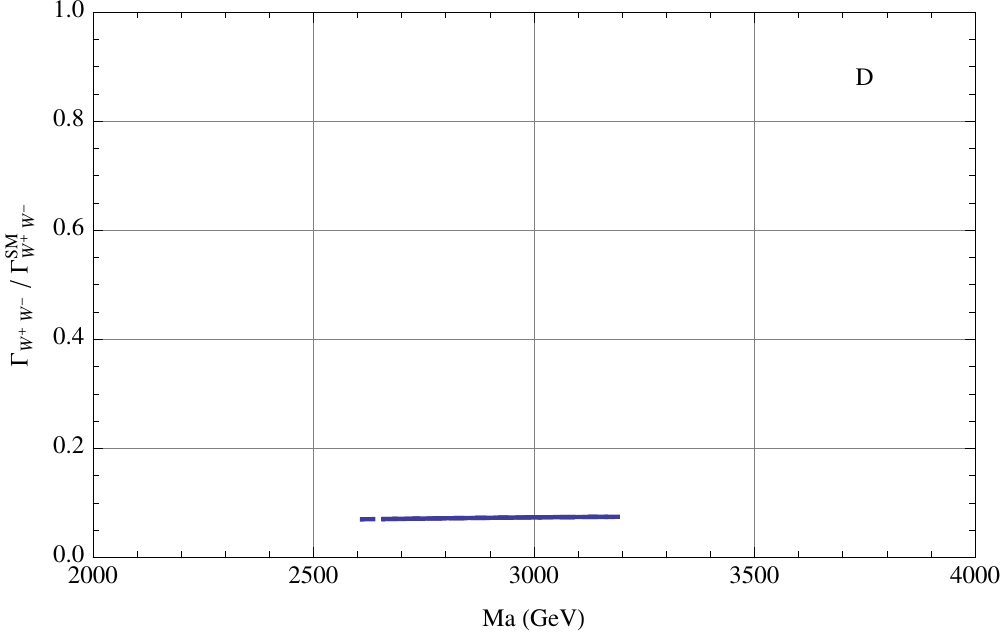} 
\end{array}$
\caption{Ratio of two $W$-boson partial decay width of the Higgs scalar, $h$ to that of the standard model. The dashed line shows the enhancement (suppression) factor over the entire scanned region while solid line corresponds to the  region where the Higgs scalar is sufficiently heavy for the decay to be  kinematically allowed.} 
\label{gammahWW}
\end{center}
\label{fig-gammahWW}
\end{figure*}

\begin{figure*}
\begin{center}
$\begin{array}{cc}
\includegraphics[scale=0.75]{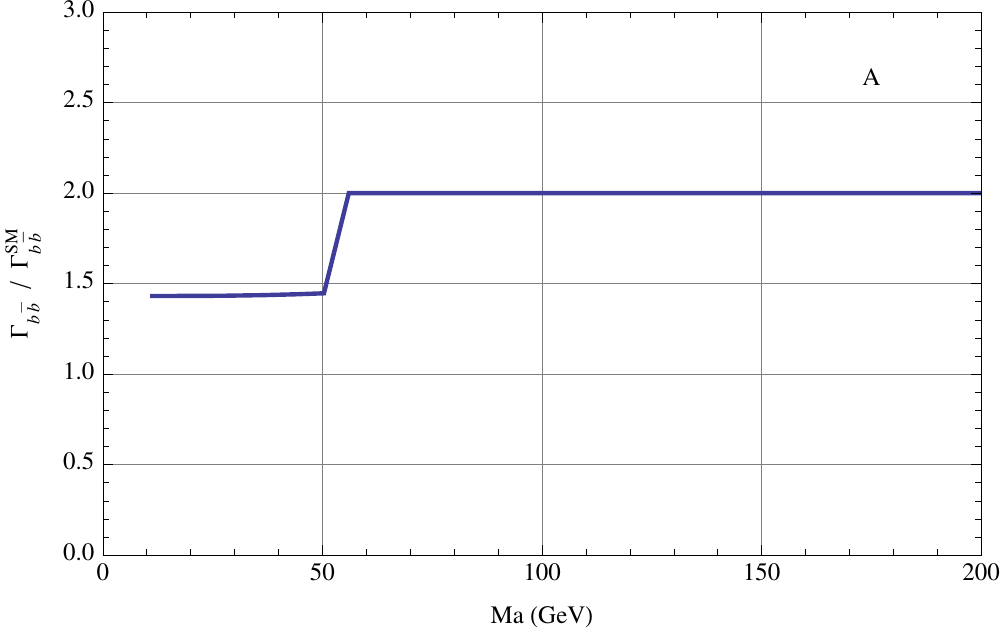} &
\includegraphics[scale=0.75]{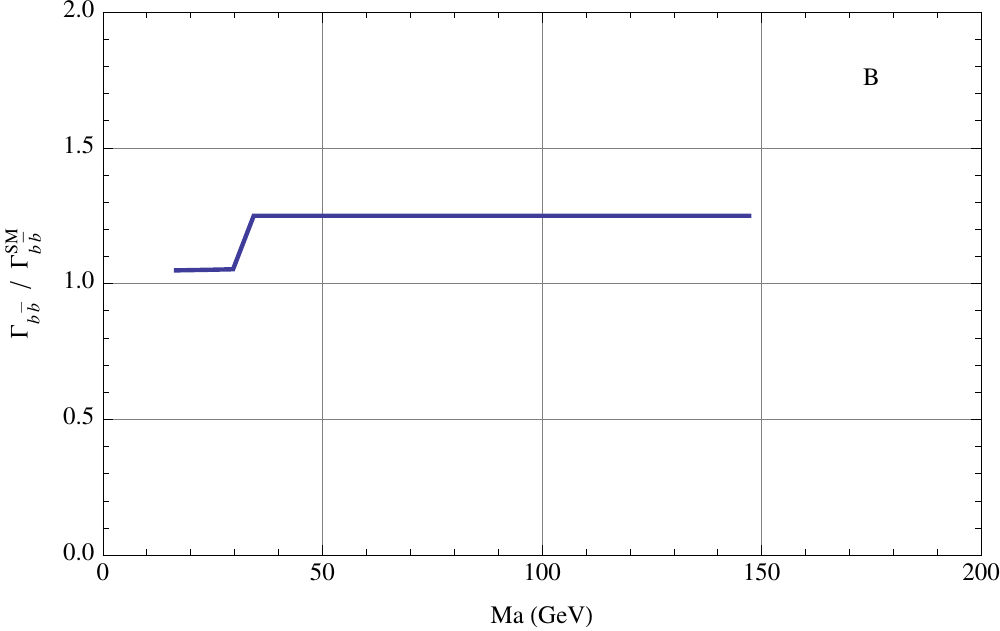} \\
\includegraphics[scale=0.75]{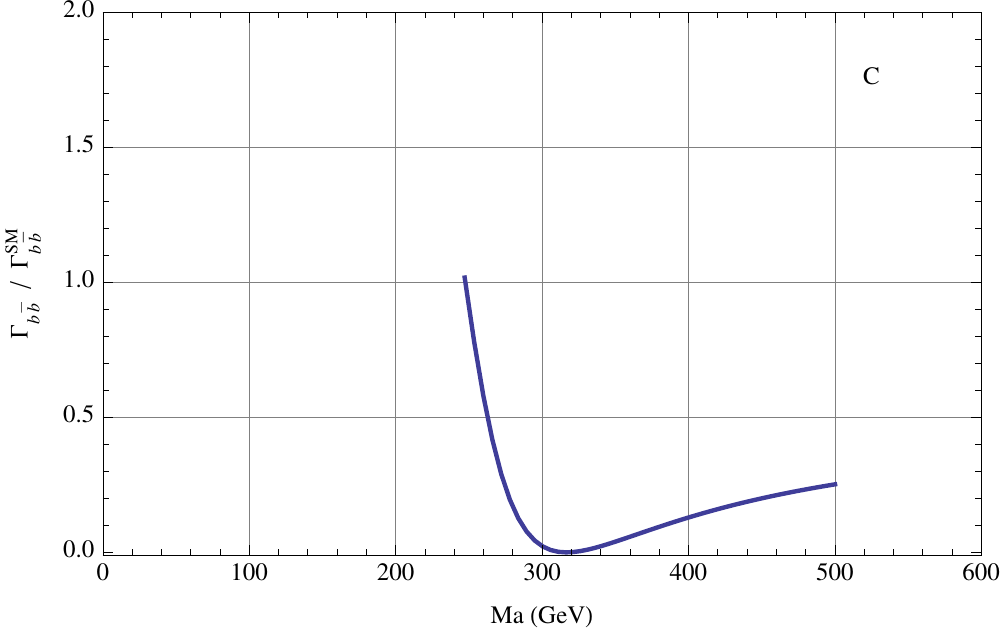} &
\includegraphics[scale=0.75]{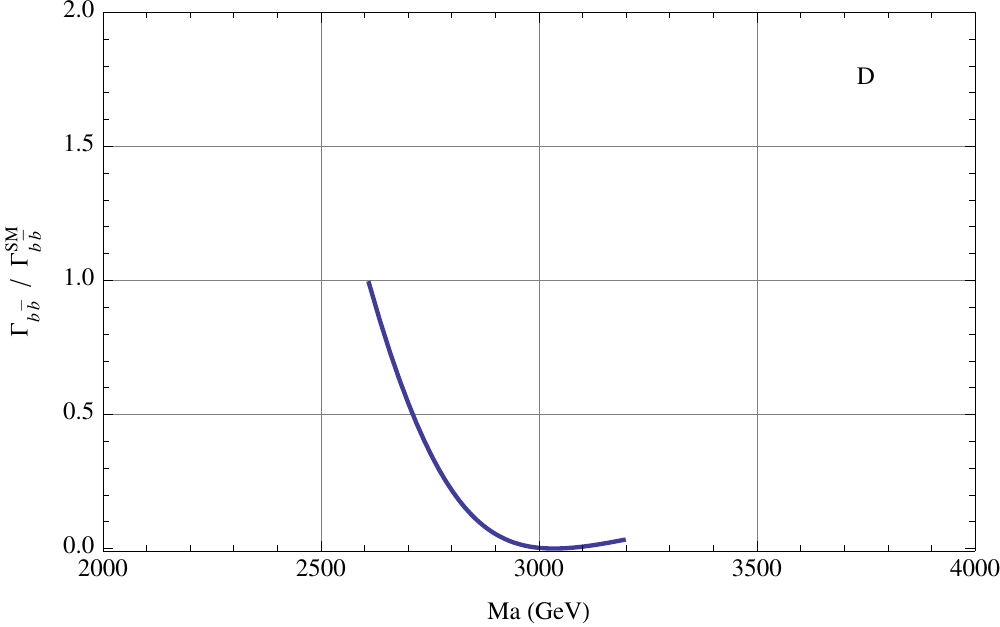} 
\end{array}$
\caption{Ratio of partial width for the decay of the Higgs scalar, $h$,  to two b quarks to that of the standard model.} 
\label{gammavbbbar}
\end{center}
\end{figure*}

As a final topic, we briefly address the modifications to Higgs boson production and decay.  For moderate $\tan{\beta}$ values, the top quark loop gives the dominant contribution to gluon fusion Higgs production at the LHC provided the squark masses are sufficiently high \cite{Spira:1995rr}.  The lightest Higgs boson can be written as a linear combination of the MSSM scalars $S_u$, $S_d$ and nonlinearly transforming scalar $S_\pi$  as  
\be
h = a_u S_u + a_d S_d + a_\pi S_\pi .
\ee
The modulus squares of various amplitudes are presented in Fig. \ref{HiggsContent} for the four regions of parameter space numerically probed in this paper. 
Since the top quark interacts only with the $S_u$ component with the enhanced Yukawa coupling $m_u/(v\sin{\theta}\sin{\beta})$, the tree level gluon fusion production cross section is equal to that of the standard model times an overall factor so that
\be
\sigma =  |a_u|^2 \left(1 + \frac{1}{\tan^2{\theta}}  \right)\left(1 + \frac{1}{\tan^2{\beta}}  \right) \sigma^{\rm SM}.
\ee
Note that the production rate depends on the details of the MSSM Higgs scalar $S_u$ content for the chosen values of parameter space.  It is clear from Fig.~\ref{HiggsContent} that since $S_u$ comprises at least one-half the Higgs scalar, there will be an enhanced gluon fusion  production rate relative to the standard model as seen in Fig. \ref{fig-cross}. Modifications to other Higgs production processes such as Higgsstrahlung off a vector boson or top quark, or in the decay of a heavy charged Higgs boson, can also be considered.

When considering the decay of the Higgs scalar, $h$, differences from the standard model can arise from  both the presence of the mixing angles, $\beta, \theta$, in the vacuum expectation values as well as the various particle content of $h$ mentioned above. Since  $v_u^\prime=v_d^\prime$, the coupling of $S_\pi$ to the $W^+ W^-$ pair identically cancels. Consequently, the process $h \rightarrow W^+ W^-$ proceeds only through the  $S_u$ and $S_d$ field components and the tree level decay rate of a heavy Higgs boson to $W^+ W^-$ is the standard model rate modified by a suppression factor
\be
\Gamma_{W^+ W^-} = \left(\frac{\tan^2{\theta}}{1+\tan^2{\theta}}\right)\left(\frac{1}{1+\tan^2{\beta}}\right)\left| a_u  \tan{\beta} + a_d ~\right|^2 \Gamma_{W^+ W^- }^{\rm SM} .
\ee
Likewise, the decay to $\bar{b}b$ quarks also depends on the $b$-Yukawa enhancement and the constituent fraction of the $S_d$ content of the Higgs field.  This leads to the modified tree level rate given by
\be
\Gamma_{b\bar{b}} = |a_d|^2 \left(1 + \frac{1}{\tan^2{\theta}}  \right)\left(1 + {\tan^2{\beta}} \right) \Gamma_{b\bar{b}}^{\rm SM} .
\ee
and displayed in Fig. \ref{gammavbbbar} using the parameter scans appropriate to the four regions. For regions $A$ and $B$, the $b$-pair partial rate is enhanced relative to that of the standard model, while for regions $C$ and $D$, the rate is suppressed. This suppression is a consequence of the very small admixture of $S_d$ in $h$ for these regions.

\section{Discussion}

A model consisting of a supersymmetric nonlinear sigma model incorporating the low energy effects of an unspecified electroweak symmetry breaking sector and coupled to a supersymmetric version of the standard model was constructed and analyzed. The superpotential coupling of the constrained pair of Higgs doublets to the  MSSM Higgs doublet pair catalyzes a nontrivial vacuum expectation value in the later thus producing an additional contribution to the electroweak symmetry breaking which is in turn communicated to the MSSM matter fields. Supersymmetry breaking was assumed to be a perturbation that does not effect the strong dynamics and was added to the model by the introducing explicit soft supersymmetry breaking parameters.  The tree level particle spectrum of the model was obtained for a variety of model parameters. The MSSM upper limit on the mass of the lightest Higgs scalar  was obviated. Throughout the region of the explored parameter space, the lightest Higgs scalar and the neutralino LSP, which can be identified as a dark matter candidate, was primarily composed of the MSSM fields with only a small admixture of the nonlinear transforming components. Since quarks and leptons were assumed to have direct couplings only to the linearly transforming MSSM Higgs doublets and not to the non-linearly transforming Higgs fields, the Yukawa couplings in the model tend to be larger than in the MSSM and standard model. An initial survey of phenomenological constraints on the Higgs scalar was performed. The main difference from the standard model predictions in both Higgs boson production from gluon fusion and Higgs scalar decay to either $W^+W^-$ or $\bar{b}b$ resulted from the constituent nature of the Higgs scalar and the variant Yukawa couplings. Depending on the process and region of parameter space, these differences could lead to either an enhancement or a suppression. Further phenomenological studies of the model including consequences of radiative corrections  are left for future study as is the possible form of the ultraviolet completion to the nonlinear sigma model supersymetric effective Lagrangian. 

%The low energy behavior of many strongly coupled supersymmetric gauge theories is very well understood. Electroweak symmetry breaking can be modeled by weakly gauging a global $SU(2)$ flavor symmetry of such theories and identifying it with $SU(2)_L$. Coupling of the meson chiral superfields of the strongly interacting theory to the Higgs chiral superfields communicates the electroweak symmetry breaking to the matter and gauge superfields. At low energy the effective Lagrangian should then give an appropriate description of the physics. For example, strongly coupled supersymmetric gauge theory with a quantum modified moduli space  \cite{Choi:1999yaa,Luty:2000fj} and with a dynamically generated superpotential \cite{Harnik:2003rs} have been considered as electroweak symmetry breaking sectors. 

\begin{acknowledgments}
The work of TEC and STL was supported in part by the U.S. Department of Energy under grant DE-FG02-91ER40681 (Theory).  The work of TtV was supported in part  by the NSF under grant PHY-0758073.  
\end{acknowledgments}

\appendix*
\section{Standard Coordinates and $SU(2)_V$ Symmetry \label{appendixA}} 

In this appendix, we address the model limit in which  $m_u^2 = m_d^2 $ and $\mu_{12} =\mu_{21}$ so that the model exhibits an approximate global $SU(2)_L \times SU(2)_R$ symmetry which is spontaneously broken to the diagonal $SU(2)_V$ subgroup with explicit breaking only by the hypercharge gauge coupling $g_1$. This approximate symmetry is the source of the degeneracies and near degeneracies in the spectrum plots presented for $\tan\beta=1$ in the main text. In order to make this approximate symmetry more manifest, it proves convenient to embed the Higgs doublets in covariantly transforming matrix chiral superfields $U$ and $V$ containing the MSSM Higgs superfields and the constrained Higgs superfields, respectively. So doing leads to the  parameterization 
\begin{eqnarray}
U & = & 
\left(
\begin{array}{cc}
H_{d}^0 & H_{u}^+ \\
H_{d}^- & H_{u}^0 \\
\end{array}
\right) =  \frac{1}{\sqrt{2}}
\left(
\begin{array}{cc}
\eta+ i \zeta_3 &  i \zeta_1 + \zeta_2 \\
i \zeta_1 - \zeta_2 & \eta- i \zeta_3
\end{array}
\right) \nonumber \\
 & = & \frac{1}{\sqrt{2}} \left( \eta {\bf 1} + i \vec{\zeta} \cdot \vec{\sigma} \right),
\end{eqnarray}
and
\begin{eqnarray}
V & = & 
\left(
\begin{array}{cc}
H_{d}^{0\prime} & H_{u}^{+\prime} \\
H_{d}^{-\prime} & H_{u}^{0\prime} \\
\end{array}
\right)\cr
& = & \frac{1}{\sqrt{2}}
v^{\prime} e^{i  \frac{\vec{\xi} \cdot \vec{\sigma}}{v^{\prime}}}
=\frac{1}{\sqrt{2}} v^{\prime} \left( \cos\sqrt{\frac{\vec{\xi} \cdot \vec{\xi}}{v^{\prime 2}}}\, \mathbf{1}  + i\frac{\vec{\xi} \cdot \vec{\sigma} }{\sqrt{ \vec{\xi }\cdot\ \vec{\xi} }} \sin\sqrt{ \frac{\vec{\xi} \cdot \vec{\xi}}{v^{\prime 2}}}  \right).
\end{eqnarray}

The relevant supersymmetric part of the action then takes the form
\begin{eqnarray}
\Gamma_S & = & \Gamma_K +\Gamma_W,
\end{eqnarray}
with
\begin{eqnarray}
\Gamma_K & = & \int dV \left\{ \bar{U} e^{-2 g_2 W } U e^{-g_1 B \sigma^3} + \bar{V} e^{-2 g_2 W } V e^{-g_1 B \sigma^3} \right\},
\end{eqnarray}
and
\begin{eqnarray}
\Gamma_W & = & \int dS W + \int d\bar{S} \bar{W},
\end{eqnarray}
where the superpotential is given by
\begin{eqnarray}
W & = & {2} \mu_{11} U U \epsilon \epsilon + 4\mu_{12} U V \epsilon \epsilon, 
\end{eqnarray}
while the constraint reads
\begin{eqnarray}
VV\epsilon\epsilon & = & V_{ia}V_{jb} \epsilon_{ij}\epsilon_{ab} = 2\det{V}   = v^{\prime2}.
\end{eqnarray}
The supersymmetry breaking part of the action takes the form
\begin{eqnarray}
\Gamma_{\rm \rlap{/}{S}} &  =  & \int d^4 x \Bigl\{ \frac{1}{2}M_1 \left( \lambda \lambda +\bar{\lambda}\bar{\lambda} \right)
+\frac{1}{2} M_2 \left( \lambda^i \lambda^i + \bar{\lambda}^i \bar{\lambda}^i \right)  \nonumber \\
 & & - m_u^2 \bar{U} U + \frac{1}{2} \mu_{11} B U U \epsilon \epsilon + \frac{1}{2} \mu_{11} B \bar{U} \bar{U} \epsilon\epsilon \Bigr\}.
\end{eqnarray}

Since in the $SU(2)_V$ limit considered here $v_u = v_d$ ( $\tan\beta=1$), the vacuum expectation values of $U$  and $V$ reduce to 
\begin{eqnarray}
<0| U |0> & = &\frac{1}{\sqrt{2}}
\left(
\begin{array}{cc}
v_u & 0 \\
0 & v_u
\end{array}
\right),
\end{eqnarray}
and
\begin{eqnarray}
<0| V |0> & = &\frac{1}{\sqrt{2}}
\left(
\begin{array}{cc}
v^\prime& 0 \\
0 &v^\prime
\end{array}
\right),
\end{eqnarray}.
Defining  $v^2 = 2 v_u^2 + 2 v^{\prime 2}$,  the potential minimization condition takes the form $m_u^2 =-16 \mu_{11}^2 - 16\mu_{11} \mu_{12} \cot \theta$, where $\tan \theta = v_u/v^{\prime}$. 

It is convenient to split the complex scalar components of the chiral superfields into their real and imaginary parts as
\begin{eqnarray}
\begin{array}{ll}
\vec{P}_\xi = \frac{1}{\sqrt{2}} ( \vec{\xi} +\vec{\bar{\xi}} ),\,\,\,\,\,\,\, & \vec{S}_\xi = \frac{i}{\sqrt{2}}(\vec{\xi} -\vec{\bar{\xi}})  \\
\vec{P}_\zeta = \frac{1}{\sqrt{2}} ( \vec{\zeta} +\vec{\bar{\zeta}} ),\,\,\,\,\,\,\, &  \vec{S}_\zeta = \frac{i}{\sqrt{2}}(\vec{\zeta} -\vec{\bar{\zeta}})  \\
P_\eta = \frac{i}{\sqrt{2}} (\eta - \bar{\eta} ),\,\,\,\,\,\,\,  &  S_\eta = \frac{1}{\sqrt{2}}(\eta + \bar{\eta}).
\end{array}
\end{eqnarray}
The mass terms in the scalar potential then take the form
\begin{eqnarray}
V_{\rm mass} & = & -{8} \mu_{11} \mu_{12} \cot \theta S_\eta^2 - (8\mu_{11} \mu_{12} \cot \theta -  \mu_{11} B) P_\eta^2 \nonumber \\
&& +16(\mu_{11} \mu_{12} \tan \theta + \mu_{12}^2) \vec{S}_{\xi}^2 + \mu_{11} B \vec{S}_\zeta^2 \nonumber \\
&& -\frac{8}{ \cos^2 \theta} (\mu_{11} \mu_{12} \cot \theta -\mu_{12}^2) (\cos \theta \vec{S}_\zeta-\sin\theta \vec{S}_\xi)^2
\nonumber \\
&& -\frac{8}{ \cos^2 \theta} (\mu_{11} \mu_{12} \cot \theta -\mu_{12}^2) (\cos \theta \vec{P}_\zeta-\sin\theta \vec{P}_\xi)^2
\nonumber \\
&& +\frac{1}{2} M_W^2 (\sin\theta \vec{S}_\zeta + \cos \theta \vec{S}_\xi)^2 
\nonumber \\
&& + \frac{1}{2} \sin^2 \theta_W M_Z^2 (\sin \theta S_\zeta^3 + \cos\theta S_\xi^3)^2.
\label{scalarmasses}
\end{eqnarray}
Only the last term in Eq.(\ref{scalarmasses}) breaks the $SU(2)_V$ symmetry into its $U(1)_{\rm EM}$ subgroup. The exact and approximate degeneracies of the tree level mass spectrum appearing  in the spectrum plots in the main text are a consequence of the relatively small value of $M_Z\sin\theta_W $.  The mass matrix in this basis has some diagonal blocks. The scalar $S_\eta$ (labeled h in Fig. \ref{MassSpectrum}) has mass-squared  $-16 \mu_{11} \mu_{12} \cot \theta$ while the pseudoscalar $P_\eta$ (labeled a in Fig. \ref{MassSpectrum})has mass-squared $-16\mu_{11} \mu_{12} \cot \theta + 2 \mu_{11} B$.  One massive pseudoscalar (labeled A in Fig. \ref{MassSpectrum}) and a charged scalar (labeled C2 in Fig. \ref{MassSpectrum})) lie in the triplet $(\cos\theta \vec{P}_\zeta - \sin\theta \vec{P}_\xi)$ and have degenerate mass-squared $-16(\mu_{11} \mu_{12} \cot \theta - \mu_{12}^2)\sec^2 \theta$. The three Nambu-Goldstone bosons lie in the triplet $\sin\theta \vec{P}_\zeta + \cos \theta \vec{P}_\xi$. Two remaining triplets each contain a  massive scalar and a charged scalar ((H1,C1) and (H2,C3) in Fig. \ref{MassSpectrum}) and are mixed. The mass degeneracy within these triplets is slightly lifted by the breaking term and the tree level masses can be calculated by diagonalizing  two by two matrices. The expressions for the eigenvalues are not very illuminating and therefore are not presented here. The supersymmetric limit of the model is recovered by taking $B=0$ and $\tan \theta = - \mu_{12}/\mu_{11}$.

The mass terms for the fermions in the Lagrangian are
\begin{eqnarray}
{\cal L}_{\rm mass} & = &  -{2} \mu_{11} \tilde{\eta} \tilde{\eta}  
-{2} \mu_{11} \tilde{\zeta}_i \tilde{\zeta}_i - 4\mu_{12} \tilde{\zeta}_i \tilde{\xi}_i + {2}  \mu_{12} \tan \theta \tilde{\xi}_i \tilde{\xi}_i \nonumber \\
& &  \frac{1}{2} M_1 \lambda \lambda+ \frac{1}{2} M_2 \lambda_i \lambda_i  + i M_W  \lambda_i \left( \sin \theta \tilde{\zeta}_i + \cos \theta \tilde{\xi}_i \right) 
 \nonumber \\
& & - i M_Z\sin \theta_W   \lambda  \left( \sin \theta \tilde{\zeta}_3 + \cos \theta \tilde{\xi}_3 \right) + {\rm h.c.}
\label{fermionmasses}
\end{eqnarray}
Only the last term in Eq.(\ref{fermionmasses}) breaks the $SU(2)_V$ symmetry. Since $M_Z\sin\theta_W$ is parametrically small, the fermion mass spectrum also shows a large number of near degeneracies. The singlet (neutral) fermion  $\tilde{\eta}$ (labeled by N1 in Fig. \ref{MassSpectrum}) has mass-squared $16\mu_{11}^2$. The remaining fermions fall into an  singlet and three  triplets that are mixed, each containing a neutral fermion and a charged fermion. The degeneracies of the masses of the fermions in each triplet is slightly lifted by the breaking term. In the limit that the explicit breaking can be neglected, the singlet $\lambda$ (labeled N3 in Fig. \ref{MassSpectrum}) has mass-squared $M_1^2$, while the masses of each of the triplets ((N2,C1), (N4,C2) and (N5,C3) in Fig. \ref{MassSpectrum}) can be obtained by diagonalizing a three by three matrix. The supersymmetric limit of the model is recovered by taking $M_1=M_2=0$ and $\tan \theta = - \mu_{12}/\mu_{11}$.

\newpage
\end{document}